\documentclass[a4paper]{aa}         

\usepackage{graphicx}
\graphicspath{{Figures/}}

\usepackage{txfonts}
\usepackage{hyperref}

\DeclareMathOperator{\Dex}{Dex}

\newcommand{\moy}[1]{\ensuremath{\langle #1 \rangle}}
\newcommand{\SiII}{\ion{Si}{II}}
\newcommand{\CaII}{\ion{Ca}{II~H\&K}}
\newcommand{\ew}[1]{\ensuremath{\textup{ew}^{\textup{#1}}}}
\newcommand{\Si}{\ensuremath{\textup{Si}}}
\newcommand{\Ca}{\ensuremath{\textup{Ca}}}
\newcommand{\DCaSi}{\ensuremath{\textup{DCaSi}}}
\renewcommand{\d}{\ensuremath{\textup{d}}}

\usepackage{natbib}
\bibpunct{(}{)}{;}{a}{}{,}      
\defcitealias{Chotard2011}{C11}
\defcitealias{Fitzpatrick1999}{FM99}
\defcitealias{CCM1989}{CCM}

\usepackage{etoolbox} 

\begin{document}

\title{%
  Intrinsic and extinction colour components in SNe~Ia and the
  determination of~$R_V$}

\author{%
  G. Smadja\inst{1}
  \thanks{\email{g.smadja@ipnl.in2p3.fr}}
  \and
  Y. Copin\inst{1}
  \and
  W. Hillebrandt\inst{3}
  \and
  C. Saunders\inst{4}
  \and
  C. Tao\inst{2, 5}
}

\institute{%
  Université de Lyon, Université Lyon 1, CNRS/IN2P3, IP2I, F69622,
  Villeurbanne, France%
  \and%
  Centre de Physique des Particules de Marseille, Aix-Marseille
  Université, CNRS/IN2P3, 163 avenue de Luminy-Case 902, 13288
  Marseille Cedex 09, France%
  \and%
  Max-Planck-Institut fûr Astrophysik, Karl-Schwartzschild-Str. 1,
  D-85748 Garching, Germany%
  \and%
  Princeton University, Department of Astrophysics, 4 Ivy Lane,
  Princeton, NJ 08544, USA%
  \and%
  Tsingua Center for Astrophysics, Tsinghua University, Beijing
  100084, China%
}

\date{Received \today}
\abstract%
{The colour fluctuations of type~Ia supernovae (SNe~Ia) include intrinsic and
  extrinsic components, which both contribute to the observed variability.
  Previous works proposed a statistical separation of these two contributions,
  but the individual intrinsic colour contributions of each SN~Ia were not
  extracted.  In addition, a large uncertainty remains on the value of the
  parameter $R_V$, which characterises the dust extinction formula.}
{Leveraging the known parameterisation of the extinction formula for dust in our
  Galaxy, and applying it to the host galaxy of SNe~Ia, we propose a new method
  of separation ---valid for each SN--- using the correlations between colour
  fluctuations.  This also allows us to derive a well-constrained value of the
  extinction parameter $R_V$ with different, possibly smaller systematic errors.
  We also define a three-dimensional space of intrinsic colour fluctuations.}
{The key ingredients in this attempt at separating the intrinsic and extinction
  colour components for each SN ---and subsequently measuring $R_V$--- are the
  assumption of a linearized dependence of magnitude on the extinction component
  of colour, a one-dimensional extra-intrinsic colour space (in addition to
  \CaII~$\lambda$3945 and \SiII~$\lambda$4131 contributions) over
  four~independent colours, and the absence of correlation between the intrinsic
  and extrinsic variabilities.}
{We show that a consistent solution is found under the previous assumptions, but
  the observed systematic trends point to a (small) inadequacy of the extinction
  formula.  Once corrected, all systematic extinction effects can be cancelled
  by choosing a single scaling of the extinction colour component as well as an
  appropriate value of $R_V = 2.181 \pm 0.117$.  The observed colours are
  described within an accuracy of 0.025~mag.  The resulting magnitude
  variability is 0.13 over all $UBVRI$ bandpasses, and this fluctuation is shown
  to be independent of the bandpass to within 0.02~mag.}
{}


\titlerunning{Intrinsic and extinction colours}
\authorrunning{G. Smadja et al.}
\maketitle

(Accepted in Astronomy \& Astrophysics)
\section{Introduction}
\label{sec:introduction}

Type~Ia supernovae (SNe~Ia) have been used for a long time as standardisable
candles in the extraction of cosmological parameters \citep{Perlmutter1999,
  Riess1998}, and the largest correction in the standardisation scheme is
related to the extinction by the host galaxy.  However, a significant
uncertainty remains on this reddening correction, which affects the potential of
SNe~Ia for the accurate determination of cosmological parameters.  Here, we try
to answer some remaining questions related to the use of the semi-empirical
extinction formulae derived for our galaxy by \cite{Fitzpatrick1999}.  A widely
employed standardisation method using SNe~Ia is provided by SALT2
\citep{Guy2007, Betoule2014}.  This latter incorporates purely empirical
correlation between colour and magnitude with the general form
$m_B = M_0 + \alpha x_1 + \beta c,$ where $x_1$ characterises the shape of the
light curve, and $c \sim (m_B-m_V)$ the colour of the supernova (SN).  The
parameters $\alpha$ and $\beta$ are tuned so as to improve the standardisation
of the SNe~Ia in a given sample.  This description is simple, but explicitly
abstains from separating the intrinsic and reddening contributions to the
observed colour.  The reddening is caused by a variable mix of gaseous
contributions (H$_2$, H, CH, etc.) and dust grains of variable sizes.  Such a
complex mixture would not be expected a priori to lead to a simple extinction
law, but detailed work by several authors extending over 50 years
(\citealt{Rieke1985}; \citealt[hereafter \citetalias{CCM1989}]{CCM1989};
\citealt[hereafter \citetalias{Fitzpatrick1999}]{Fitzpatrick1999}) led to
universal formulae ---depending on a single parameter $R_V$--- related to the
cross sections of the diffusion centres (molecules or dust grains) for photons.
$R_V$ is operationally defined as the ratio of the extinction in the $V$
bandpass $A_V$ to the reddening indicator $E^{\prime} \sim E(B-V)$, either
measured or adjusted by a fit to the observations.

A detailed study by \cite{Schlafly2016, Schlafly2017} confirmed a directional as
well as distance variability of $R_V$ in our galaxy (from 3 to 3.75), and
provided confirmation of the results of \citetalias{Fitzpatrick1999}.  All
attempts to determine the parameter $R_V$ from SNe assume that the extinction
formulae derived for our galaxy apply directly to all host galaxies, leaving the
possibility of different values of $R_V$, as the averaging which occurs when
considering a sample of host galaxies differs from that performed when deriving
the extinction formulae from stars of the Milky Way.  The same universal
extinction formula is assumed to be valid for the host galaxies of the SNe~Ia,
but previous investigations actually relied on a combination of photometric
measurements with slit spectroscopy, which lacked the spectrophotometric
information provided by the Nearby Supernova Factory
\citep[SNfactory,][]{Aldering2020} collaboration.  The contribution of the
intrinsic variability was not directly monitored, and the applicability of the
extinction formula to other galaxies could not be checked accurately.  All the
groups restrict their extinction analysis to redshifts larger than 0.01 in order
to avoid a significant contribution of peculiar velocities.

\section{Previous investigations}

There have been many attempts to separate the intrinsic and extinction colour
components of SNe, and we describe a few of them here.  In many instances, an a
priori distribution of the extinction is assumed; for instance an exponential
distribution with a cutoff.  This introduces a bias in the determination of the
extinction of each SN, as the reddening involves an average over many different
geometrical configurations of the SN with respect to its host.  The exponential
behaviour is a simple but unrealistic assumption.  In the present work, an
extinction colour and an intrinsic colour are extracted for each SN.  The
intrinsic colour contribution to $E(B-V)$ is frequently derived from the
light-curve shape, whether it is the SALT2 $x_1$ or the stretch parameter.  In
our case, an extra intrinsic component is introduced (for each SN).

\cite{Lira1998} and later \cite{Phillips1999} noticed that the colour evolution
of all SNe~Ia in the 30--90 days post $V$-maximum is universal, which allowed
them ---by selecting SNe in E or S0~galaxies (without dust)--- to derive a
universal intrinsic SN~Ia colour $E(B-V)_0$ at any reference date in the
interval from 30 to 90~days; the magnitude error quoted is~0.05.  The galaxy
reddening $E(B-V)_{\text{tail}}$ at late epochs is then found by subtracting the
intrinsic colour from the observed colour.  After selecting a sample of SNe with
low extinction ($E(B-V)_{\text{tail}} < 0.06$), a time-dependent correction
provided by the light curve allows them to evaluate the intrinsic value
$E(B-V)_{\max,0}$ at maximum $B$~luminosity.

Another photometric technique was used by \cite{Wang2003}, who analysed the
colour--magnitude diagrams using the data from~\cite{Hamuy1996}
and~\cite{Riess1998}.  The $B$ magnitude varies linearly as a function of colour
in the 10 to 30 day period post $B$ maximum.  The scatter of the residuals from
the straight line is typically 0.05.  It is claimed that extinction does not
affect the shape of these diagrams (an approximation, as the spectra of the SNe
evolve).  The magnitude $M'_{BV}$ of the SN~Ia at colour $B-V = 0.6$ is used as
a reference value for the standardisation.  To obtain a `reference' magnitude at
maximum, the linear behaviour is extended to $B-V = 0$, although some SNe~Ia
show a `bump' departing from linearity.  In the absence of a such a bump, the
comparison with the observed $M^B_{\max}$ allows a value of the reddening
$E(B-V)$ to be derived, which has a dispersion of about 0.1~mag with respect to
the determination of \cite{Phillips1999} for the same SN~Ia.

\cite{Jha2007} also relied on photometric bandpasses and K corrections, but took
into account the colour evolution along the light curve of the SNe using the
MLCS2k2 software (Multicolor Light Curve Shapes).  A preliminary fit extracts
the intrinsic Gaussian fluctuation (with $\sigma = 0.049$, and an average value
of zero) and the exponential reddening distribution at date $+35$~days (with a
colour decay constant of $\tau_E = 0.138$).  In each $UBVRI$ bandpass, the same
reddening distribution is used in the modelling of the light curve.  Each of the
five light curves is described by five parameters: date of peak $B$ luminosity,
distance modulus, time evolution, a unique reddening scale ($A_V$) (with the
$B-V$ shape found in the previous step), and $R_V$.  A prior $R_V = 3.1$ with
$\sigma = 0.4$ is assumed for the \citetalias{CCM1989} law (the parameter $R_V$
is needed to convert $A_V$ into an extinction in different filters).  These
latter authors do not claim to describe the intrinsic distribution at maximum
luminosity, nor to measure~$R_V$.

The multi-colour light curves of a sample of 80~SNe measured by different groups
are analysed by \cite{Nobili2008}.  The light curve shape is characterised by
its stretch, that is, the time dilatation factor $s(sn)$ bringing the mean
light-curve width to the SN light-curve width.  The stretch is similar to the
$x_1$ variable mentioned earlier for SALT2.  Given two bandpasses, $X$ and $Y$,
and stretch $s$: $X-Y = b_{XY}(t) + a_{XY}(s-1) + c_{XY}E(B-V)$.  The data are
K-corrected for the changes of rest-frame bandpasses with redshifts, and
$E(B-V)$ is the average over all epochs of the $B-V$ colour curve of each~SN.
$R_V$ is first extracted from the measured coefficients $c_{XY}$ found
(e.g. $c_{BV} = R_B - R_V$).  The best fit is found (with the
\citetalias{CCM1989} law) for $R_V = 1.01 \pm 0.25$.  When the excess dispersion
of the residuals along the light curve after correction for the extinction is
assigned to the intrinsic colour contribution, the same analysis finds
$R_V = 1.75 \pm 0.27$ using \citetalias{CCM1989}.  This difference strongly
suggests that the intrinsic colour components should be taken into account.

The `twin' SNe SN2014J and (unextincted) SN2011fe were compared in the optical
and near-infrared (NIR) range by \cite{Amanullah2014} from data collected by the
HST (UV bands), at Mauna Kea (NIR), and NOT (optical).  The comparison of the
12~light curves was performed using the extinction formula of
\citetalias{Fitzpatrick1999} .  \cite{Amanullah2014} find
$E(B-V) = 1.37 \pm 0.03$, $R_V = 1.4 \pm 0.1$.

The Carnegie Supernova Project \citep{Burns2014} emphasises the use of a
`pseudo-stretch' $s_{BV}$, which describes the time dependence of the $B-V$
colour instead of the usual `stretch', which relates directly to the light
curve.  This pseudo-stretch allows a prediction of the maximum $B-V$ colour.
The analysis proceeds to analyse the eight~observed colours $B - m_i$ in
bandpasses $(u,g,r,i,V,Y,J,H)$ of each SN as a sum
$c_i(sn) = P(s_{BV}) + \Delta A_i(E(B-V),R_V),$ where $P(s_{BV})$ is a
second-order polynomial, and $\Delta A_i$ is the predicted reddening for colour
$c_i$ from the extinction formula.  From a set of 75 SNe, \citet{Burns2014} find
$R_V = 2.14 \pm 0.16$.  With one intrinsic colour and one (measured) host
extinction per SN, and without any assumption on the distribution of the
extinction, their method and their result are relatively similar to ours.  In
Sect.~\ref{sec:noCaSi}, we suggest using the SALT2 variable $x_1$ in the way
$s_{BV}$ is used by these authors.

\cite{Amanullah2015} include photometric measurements of seven~SNe~Ia between
0.2 and 2~microns from date $-10$ to $+50$.  A mean spectral template is used at
each date to evaluate the extinction correction in all filters with a standard
\citetalias{CCM1989} extinction law where $E(B-V)$ and $R_V$ are parameters to
be adjusted to the observed colour.  \cite{Amanullah2015} assume an intrinsic
$U-V$ dispersion of 0.1~mag (compatible with our own {spectrophotometric}
measurements), and a larger dispersion of 0.3~mag for wide-band filters at lower
wavelengths.  The different intrinsic colours have a mean value of zero and are
supposed to be uncorrelated at any given date, but fully correlated at different
phases.  The light curves are then fitted in all filters as in
\cite{Amanullah2014} to extract $R_V$ and $E(B-V)$ for each~SN.  The values of
$R_V$ for the different SNe vary from $1.4 \pm 0.1$ to~$3.8 \pm 1.5$.

\cite{Mandel2017} analysed a sample of 250 nearby SNe~Ia ($0.01 < z < 0.10$).
They adapted the SALT2 formalism \citep{Betoule2014} at maximum $B$ light to
include an additional intrinsic contribution of the intrinsic colour
$c^{\text{int}}$ to the $B$ magnitude as well as an explicit reddening $R_BE_s$.
The intrinsic B band absolute magnitude is given by
$M^{\text{int}} = M_0^{\text{int}} + \alpha x^{\text{int}} +
\beta_{\text{int}}c^{\text{int}} + \epsilon$ with a contribution of the
intrinsic colour $c^{\text{int}}$ to the $B$ magnitude in addition to the
reddening $R_BE_s$.  The intrinsic light-curve shape parameter $x$, similar to
SALT2 $x_1$, is assumed to have a normal distribution
$x^{\text{int}} \approx \mathcal{N}(-0.40, \sigma = 1.2)$, the intrinsic colour
is $c^{\text{int}} = c_0^{\text{int}} + \alpha_c^{\text{int}} x^{\text{int}}$
with $c_0^{\text{int}}\approx \mathcal{N}(-0.06,\sigma = 0.06)$, and the extra
(grey) magnitude dispersion $\epsilon \approx \mathcal{N}(0., \sigma = 0.1)$ is
introduced.  The `grey' fluctuation is the dispersion in the magnitude of SNe
that have identical or quasi-identical spectral shapes after reddening
correction; it is bound to be achromatic by definition, and its origin is not
understood.  The intrinsic parameters of SN $s$ are thus
$(M_s^{\text{int}}, c_s^{\text{int}}, x^{\text{int}})$.  The host galaxy
reddening $E_s$ is positive, with an exponential dependence on colour
$\exp(-E_s/\tau)$, as in \cite{Jha2007}, with a magnitude scale defined by
$R_B$.  The model is adjusted to the outcome of a SALT2 fit $(m_B, c, x_1)$; the
Hubble residuals are~0.16~mag.  The 11~parameters of the model are obtained by a
hierarchical Bayesian method.  The observed values, which include reddening and
distance modulus, are $(m_s, c_s^o, x_s)$.  The main results of the fit in
\cite{Mandel2017} are: $c_0^{\text{int}} = -0.061 \pm 0.012$,
$\sigma^{\text{int}} = 0.100$ (intrinsic magnitude dispersion) ,
$\sigma_C^{\text{int}} = 0.065\pm 0.008$ (intrinsic colour dispersion),
$R_B = 3.73 \pm 0.31$, $\tau_E = 0.069$ (extinction decay rate).  The dependence
of the magnitude with intrinsic colour is observed with
$\beta^{\text{int}} = 2.25 \pm 0.25$.  The dispersion of the intrinsic B-V
colour found by these latter authors is four times larger than our result.

\cite{Thorp2021} used photometric data to investigate the potential dependence
of the reddening as a function of host mass.  Their model includes the Gaussian
grey magnitude fluctuation as well as a spectral variability function (largely
spectral lines) with an amplitude scale as a parameter.  The $R_V$ values of
individual SNe are drawn from a truncated Gaussian distribution ($R_V > 0.5$).
The Hubble residuals are improved from 0.16 (SALT2) to 0.12.  \cite{Thorp2021}
finds no significant difference in $R_V$ between the large galactic mass and
low-mass samples.  The full sample averages to $R_V =2.70 \pm 0.25$.

As in \cite{Jha2007}, \cite{Brout2021} describe the observed colour as a sum of
intrinsic and extinction colours, with a Gaussian distribution of the intrinsic
colour, and an exponential shape of the dust.  Intrinsic and extinction colour
components are added into the observed colour.  Their model also includes a
Gaussian $R_V$ distribution.  When splitting their sample into low-mass and
high-mass galactic hosts, \cite{Brout2021} find two different values of $R_V$,
with $R_V = 2.0 \pm 0.25$ (low mass) and $R_V = 3.0 \pm 0.4 $ (high mass).

A recent study by \cite{Wojtak2023} also investigates the impact of the host
mass on the standardisation of SNe~Ia within a Bayesian approach using
multi-filter light curves of SNe~Ia.  These authors introduce a shape parameter
for the extinction distribution rather than an exponential with a cutoff and
find strong evidence for two populations of SNe.  The two populations, with
probabilities 0.38 and 0.62, differ in their intrinsic colour and their
reddening distribution.  \cite{Wojtak2023} introduce an (unaccounted) intrinsic
magnitude dispersion of 0.11~mag.  The authors also define two intrinsic colour
dispersions for the two populations: $\sigma_{\text{int}}^1 = 0.076 \pm 0.011$
and $\sigma_{\text{int}}^2 = 0.043 \pm 0.013$.  Both these values are larger
than our own determination, which includes more spectroscopic information.

The only spectrophotometric data available are provided by the SNfactory with
the SNIFS instrument.  A first attempt by \citet[hereafter
\citetalias{Chotard2011}]{Chotard2011}, analysed the magnitudes of 76~SNe~Ia at
B maximal luminosity in the five synthetic top-hat filters $UBVRI$.  The
magnitudes were corrected for the \SiII{} and \CaII{} equivalent widths, and the
extinction correction was then derived from the \citetalias{CCM1989} law with
scale $A_V$.  A value of $R_V = 2.8 \pm 0.30$ was obtained.  The extinction
scale $A_V$ was a free parameter adjusted to the data for each SN in this latter
study.  The magnitude uncertainties were estimated from the measurement errors
(dominated by the flux calibration uncertainty) without including the
(unaccounted) grey fluctuation of 0.11~mag.  Colours are better measured than
magnitudes as some systematic uncertainties cancel out, such as the absolute
flux calibration uncertainty of 0.03~mag, and (largely) the error on the
redshift (measurement and peculiar velocity).  The present work is an extension
of this latter investigation, but with an increased sample and several
improvements.

\cite{Huang2017} uses the twin SNe SN2012cu (with a significant reddening) and
SN2011fe (without extinction) measured by the SNfactory to derive the parameter
$R_V$ by comparing their magnitudes at different epochs.  Although the spectra
are very similar after extinction correction (the two SNe are `twins'), some
spectral bins near absorption lines differ.  These bins are averaged by a
Gaussian convolution, and deweighted so that the $\chi^2$ in these regions is
unity (a `floor' error of 0.03 is added); the residual fluctuations of the
deweighted residuals are of the order of 0.01 mag, and the value of $R_V$ is
$R_V = 2.952 \pm 0.081$ for SN2012cu.  \cite{Leget2020} and \cite{Boone2021}
used the SNfactory spectrophotometric data to show that the largest spectral
features occur with a remarkable {intrinsic} variability in particular around
\CaII{} and \SiII.

While \cite{Leget2020} singles out the correlation of spectral lines with
magnitudes, \cite{Boone2021} introduced three intrinsic variables $\xi_i$, which
provide an accurate description of the full spectra for 173 SNe~Ia (after
selection).  The formula of \citetalias{Fitzpatrick1999} is used to correct for
reddening.  As in \cite{Huang2017}, the parameter $R_V$ is found by deweighting
the spectral ranges with large intrinsic variability by a factor
$1/\sigma(\lambda)$ at each wavelength $\lambda$, with the result
$R_V = 2.40 \pm 0.16$ (the method is called Read Between The Lines).  The
amplitude of the residual intrinsic variability of magnitudes away from spectral
lines after reddening correction is generally of the order of 0.08~mag, but it
is as low as 0.02 mag between 6600 and 7200~\AA.

A large uncertainty remains today on the (mean) value of the parameter $R_V$,
which plays a substantial role in the standardisation of SNe.  Part of the
discrepancy arises from underestimated systematic measurement errors on the path
from detector to magnitude: the absolute calibration, the bandpass wavelength
dependence, the redshift accuracy, the peculiar velocities, and the corrections
for galactic reddening (host and ours).  Another source of variability lies in
different definitions of $R_V$ as a parameter of an extinction formula, or as
$A_V/E(B-V)$, where $A_V$ and $B-V$ depend on the spectral distribution of the
stars used (HD stars and SNe) and on the rest-frame bandpasses selected.
Peculiar velocities are part of the redshift accuracy and influence the rest
frame bandpasses as well as the absolute magnitude.

The spectrophotometric quality of the SNfactory data \citep{Aldering2020} helps
alleviate some of these uncertainties, and the difference in $R_V$ between
\cite{Huang2017} and \cite{Boone2021} ---who average over 173~SNe--- as well as
the results of \cite{Amanullah2015} ---who use a somewhat less homogeneous
sample--- may well point to an actual physical distribution of the value of
$R_V$.  Here, to quantify the smooth spectral evolution of the reddening, we
favour the use of colours rather than magnitudes, as the latter exhibit an
`unaccounted' grey fluctuation of 0.08 to 0.10~mag from one SN to another, and
the evaluation of the ratio of the correlated to uncorrelated colour
variabilities, including the intrinsic contribution, is not straightforward.

We assume in the present work that the reddening of SNe by the host galaxy can
be factored out and consider the observed colour fluctuations (from one SN to
another) as a sum of intrinsic and extinction contributions, as done previously
by \cite{Jha2007, Mandel2017} or \cite{Brout2021}.  As \SiII~$\lambda$6355~\AA{}
and \CaII~$\lambda$3945~\AA{} are dominant components of the spectral intrinsic
variability, we first consider their contribution ---as in
\citetalias{Chotard2011}---, but substituting \SiII~$\lambda$4131~\AA{} for
\SiII~$\lambda$6355~\AA{} as the dependence of the $B$ magnitude on the
equivalent width of \SiII~$\lambda$6355~\AA{} is quadratic.  We shall assume in
addition that there is a single dominant source of intrinsic variability on top
of the \CaII{} and \SiII{} absorption lines, so that the intrinsic variation of
the different `colours' is linked by three `intrinsic' couplings.  For {each}
SN, we can then extract the extinction and intrinsic colour components using the
extinction formula as leverage, and we revisit the determination of $R_V$ using
the correlations of these colour variations.

There are three advantages to using the particular data sample chosen here: the
high-quality spectrophotometric measurement, with well-measured spectral lines
and subpercent errors on colours, and the sizeable homogeneous sample of
165~SNe, which allows us to use small colour fluctuations with respect to a mean
template rather than the absolute values of these colours.

\section{Sample of supernovæ}
\label{sec:sample-sn}

The data used were gathered by the SNfactory collaboration using their SuperNova
Integral Field Spectrograph \citep[SNIFS,][]{Lantz2004}, an automated instrument
optimised for the observation of point sources on a structured background, with
a spectral resolution of 0.25 to 0.30~nm and a good efficiency from 330 to
900~nm.  SNIFS consists of a multi-filter photometric channel used to monitor
the transmission in non-photometric nights and provide an image for guiding, a
lenslet integral field spectrograph covering a field of view of
$6\farcs4 \times 6\farcs4$ with a grid of $15 \times 15$ spaxels, and an internal
calibration unit (continuum and arc lamps).  A more complete description of
SNIFS, its operation, and the data processing can be found in
\cite{Aldering2006}; updated in \cite{Scalzo2010}.  The measured wavelength
range in the SN rest-frame extends from 3300 to 8400~\AA.  It is divided into
five logarithmically distributed top-hat filters $UBVRI$, with central
wavelengths at 363.9, 438.6, 528.7, 637.4, and 768.4~nm.  The light curves are
reconstructed in these five synthetic rest-frame bandpasses, and the present
study is limited to the spectrum closest in time (within a window of
$\pm 2.5$~days) to the maximal $B$ luminosity, as found from a SALT2 fit
\citep{Betoule2014} to the light curves.  The redshift range covered by our
measurements extends from $z = 0.02$ to $z= 0.11$.  Although there are only four
independent colours, we consider all pairs of filters in the present analysis,
which leads to ten possible filter combinations.  The initial sample consists of
172 SNe~Ia, which were selected as those yielding an acceptable SALT2 fit
\citep{Betoule2014}, allowing us to define the date of maximum light: at least
5~nights of observations, nMAD\footnote{Normalised median absolute deviation:
  $\text{nMAD}(x) = 1.486\times \text{median}(|x-\text{median}(x)|$).}  of
residuals $< 0.12$~mag, and a satisfactory phase coverage: at least four epochs
from $-10$ to $+35$ days from maximum light, with one epoch between $-10$ and
$+7$ days, and one between $+7$ and $+20$~days.  In addition, we required in
this work one spectrum within 2.5~days of maximum light.  We eliminated seven
SNe~Ia where the subtraction of the host galaxy signal in our version of the
data processing left a brightness gradient in either the $B$ channel of the
spectrograph (up to a wavelength of 5000~\AA) or the $R$ channel (above
5000~\AA) over the 225~micro-lenses larger than 0.05~mag.  The analysis is
performed on the remaining 165 SNe~Ia from the SNfactory.  This sample is almost
the same as that described in \cite{Aldering2020}.  As explained in the
following section, the SALT2 magnitudes for our five synthetic bandpasses are
{not} used; only the shape of the light curve near maximum light is used, as
provided by the SALT2 model after the determination of $x_0$ and $x_1$ from the
data.

\subsection{Magnitudes at maximum light and errors}
\label{sec:magnitudes-errors}

We require the magnitudes produced by the SALT2 fit in each bandpass to be
consistent with the data near peak.  From each rest-frame spectrum within a
$\pm 2.5$~day window around the $B$-band maximum light, the distance-corrected
magnitudes (brought back to an arbitrary common redshift of 0.05) are derived by
integrating the photon count in each of the five $UBVRI$ top-hat bandpasses
described in Table~\ref{tab:bandpasses}.  The index $T$ in
Table~\ref{tab:bandpasses} is used to distinguish the bandpasses from Bessell
filters, and is dropped later in the text.  The magnitude in each filter is
rescaled to an interpolation of the actual observations as described in
\citetalias{Chotard2011}.  The SALT2 fit extracts the date of the $B$ maximum,
the values of $x_0$ (absolute magnitude scale) and $x_1$ ---which relate to the
light curve---, and the colour $c$, which is very close to $B-V$ at maximum $B$
light.  These parameters are then used to integrate the SALT2 spectral templates
over our bandpasses, and the corresponding SALT2 light curves are reconstructed
at the observation dates, for each SN, according to the values of $M^{0S}_F$,
$x_1$ and $c$ parameters found in the light-curve fit.

\begin{table}
  \caption{SNfactory top-hat filter bandpasses (rest-frame).}
  \label{tab:bandpasses}
  \centering
  \begin{tabular}{c c c c c c}
    \hline\hline
    Filter & $U_T$ & $B_T$ & $V_T$ & $R_T$ & $I_T$ \\
    $\lambda_{\min}$ [\AA] & 3300.0 & 3978.0 & 4795.3 & 5780.6 & 6968.3 \\
    $\lambda_{\max}$ [\AA] & 3978.0 & 4795.3 & 5780.6 & 6968.3 & 8400.0 \\
    \hline
  \end{tabular}
\end{table}

In each bandpass, the shape of the SALT light curve near maximal $B$ luminosity
is then used as an interpolating curve in {each} filter and scaled as described
in Eqs.~\ref{eq:chi2_epsilon} and \ref{eq:magmax} to fit our synthetic
photometry from spectra within 5~days of maximum luminosity, as in
\citetalias{Chotard2011}.  The same integration over the bandpass is applied to
the SALT2 spectral template so that the corresponding SALT2 light curves are
reconstructed at the observation dates, for each SN, according to the values of
$x_0$, $x_1$, and $c$ parameters found in the light curve fit.  The SALT
magnitudes $m^{S}_{F,p}$ reconstructed from the light curve in each bandpass $F$
are then shifted at all dates $p$ according to Eq.\ref{eq:magmax} so as to match
the observed distance-corrected magnitude $m_{F,p}$.  The shift $\epsilon_F$ is
found by minimising
\begin{equation}
  \label{eq:chi2_epsilon}
  \chi^2_F =
  \sum_{p=1}^N \frac{(m_{F,p} - m^S_{F,p} - \epsilon_F)^2}{\sigma_{F,p}^2}.
\end{equation}
The values of $\epsilon_F$ are typically about 0.01~mag within the 2.5~day
window around maximum light $t_{\max}$, with a spread of 0.02 to 0.05~mag around
the SALT2 light curve for $UBVR$, and somewhat larger for $I$.  The rescaled
magnitudes $m_{F}$ at maximum~$B$ luminosity are then obtained by shifting the
SALT maximum $m^S_{F,\max}$ by this (averaged) $\epsilon_F$:
\begin{equation}
  \label{eq:magmax}
  m_F = m^S_{F, \max}  + \epsilon_F.
\end{equation}
Our magnitudes are then largely independent of the SALT colour law and of the
SALT light-curve parameters.  Given the measurement errors of the magnitudes for
the different filters at each date, and the covariance of the measurement errors
between the different bandpasses (mostly a grey absolute calibration factor and
a redshift uncertainty from peculiar velocities), we can then estimate the
covariance matrix of the magnitudes in the different filters at maximum $B$
luminosity.  The rescaled SALT2 magnitudes at maximum are corrected for the
extinction in our galaxy.

\section{Methodology}
\label{sec:methodology}

The data analysis presented here starts from the processing described in
\cite{Aldering2020}.  It is followed by an evaluation of the magnitudes at
maximum luminosity in each bandpass, and a measurement of the equivalent widths
of the significant spectral features as performed in \citetalias{Chotard2011}
and \cite{ChotardPhD}.  The first step of the analysis, as in the earlier work
of \citetalias{Chotard2011}, is to subtract in Sect.~\ref{sec:CaSi} the
intrinsic colour variability correlated to \SiII~$\lambda$4131~\AA{} and
\CaII~$\lambda$3945~\AA{} signals using their equivalent widths \ew{Si} and
\ew{Ca}.  What is referred to as `colour' in our study ---except in this first
step--- is actually the `colour difference' between the measured colour and the
average value of the same colour over our sample of 165 SNe after subtraction of
the contribution of these two lines.  It differs from the standard colour
excess, measured with respect to a sample of SNe without extinction by a
constant, and can be negative.  This constant offset (different for each filter)
has no impact on any of our results.  Each colour difference $e_i$ is the sum of
two components, an intrinsic part $I_i$ and an extinction part $X_i$.  In our
method, $X_i$ is also an extinction variation with respect to its average value
and may assume negative values.  For each filter, it also differs from the full
extinction by an (irrelevant) constant.  We then express the ten colour
differences observed (after correction for \ew{Si} and \ew{Ca}) in terms of the
following relations ({without any summation of the indices}):
\begin{equation}
  \label{eq:method}
  e_i(n) = I_i(n) + X_i(n) + \epsilon_i (n),\qquad i= (1 \ldots 10),
\end{equation}
and for each colour $i$:
\begin{equation}
  \label{eq:method_delta}
  \begin{aligned}
    X_i(n) &= \delta_{ij}(R_V)\,X_j (n) \\
    I_i(n) &= \gamma_{ij}\,I_j(n).
  \end{aligned}
\end{equation}
The coefficients $\gamma_{ij}$ and $\delta_{ij}$ are common to all SNe, labelled
by index $n$; $\epsilon_i$ is the residual associated with the description of
the colour difference $e_i$ within the present model.  While $\gamma_{ij}$
relates intrinsic colours and does not depend formally on $R_V$, its
determination will be weakly dependent on the value of $R_V$ assumed.  The
previous relations do not involve convolutions with measurement errors, as these
`colour' errors are small enough with respect to the physical quantities $I_i$
and $X_i$ to be ignored at this stage.

Our notations are summarised in Table~\ref{tab:definitions}.  The colours are
numbered from zero to 9 in the sequence $(U-B$, $U-V$, $U-R$, $U-I$, $B-V$,
$B-R$, $B-I$, $V-R$, $V-I$, $R-I$), and the differences in their measurement
accuracies is ignored.  The coefficients $\delta_{ij}(R_{v})$ relating the
extinction colour components $X_{i}$ are derived from the extinction formula as
described in Appendix~\ref{sec:Annex}, while the 45 $\gamma_{ij}$, which relate
the intrinsic colour variations $I_i$, can be expressed as a function of three
of them, selected (arbitrarily) to be $\gamma_{10}$, $\gamma_{20}$,
$\gamma_{30}$.  Thus, the proposed model is characterised at this stage by
four~parameters: $R_V$, and the three~intrinsic coefficients $\gamma_{10}$,
$\gamma_{20}$, $\gamma_{30}$.  However, below, we find it necessary to introduce
four~corrections to the extinction formula in the bandpasses $BVRI$, which
results in a total of eight~parameters.  For presentation purposes, an extra
parameter (not required by our analysis, where it is unity) will be introduced
for the extinction scale.  For each SN, two coordinates are derived from the
data using the previous parameters: $X_0(n)$ and $I_0(n)$.  We call them
coordinates rather than parameters as they are computed with a fixed algorithm
from the measured spectra, without any tuning, once the parameters involved in
our modelling of the full sample have been found.  The extinction analysis
proceeds in eight stages, with the successive application of different fits.  To
assess the validity of the model used, we have chosen to control each successive
stage and apply judgement, rather than combine the different $\chi^2$, which
occur along the path into a single $\chi^2$ or a likelihood:
\begin{enumerate}
\item We subtract the contributions of the silicon and calcium equivalent widths
  \ew{Si} and \ew{Ca} to the colour fluctuations to ensure they are small in
  Sect.~\ref{sec:CaSi}, allowing a linear approximation.

\item The extinctions $A'_F$ in filter $F$ derived from the extinction formula
  \ref{eq:extinctionlaw} are rescaled to $A_F$ (Sect.~\ref{sec:rescaling}) to
  ensure mathematical consistency.  The colour coefficients $\delta_{ij}$ of
  Eq.~\ref{eq:method} are invariant in this rescaling.

\item For a given set of $\gamma_{ij}$, the intrinsic ($I_i)$ and extinction
  ($X_i$) colour components are found by a least square fit to the observed
  colours (Sect.~\ref{subsec:chsq}).

\item The values of the intrinsic colour correlation coefficients $\gamma_{ij}$
  are found by iteration, and these are tuned in
  Sect.~\ref{subsec:Intrinsic_gamma} by minimising the scatter of the output
  ratios of the $10\times 9/2 = 45$ pairs of colour-ratios found from the colour
  fit in Eq.~\ref{eq:chsq}.

\item The extinction formula (\citetalias{Fitzpatrick1999}) is modified so as to
  suppress any correlation between the residuals $\epsilon_{i}$ derived from the
  fit to the extinction colour.  This is achieved in Sect.~\ref{sec:offsets} by
  minimising $\chi_{\Dex}^2 = \sum_i (\d\epsilon_i/\d X_i)^2$.  The consistency
  rescaling is then applied to the corrected coefficients.

\item Iterations are performed over steps 4 and 5 on the values of the filter
  extinction corrections and the intrinsic coefficients $\gamma_{ij}$, until
  neither $\chi^2_{\Dex}$ nor the scatter can be improved by a change of
  $10^{-3}$ of the extinction corrections or the coefficients
  $(\gamma_{10}, \gamma_{20}, \gamma_{30})$.

\item The previous steps do not depend on an overall scale factor applied to the
  (filter-colour) extinction coefficients $\delta_{Fi}$.  For each value of
  $R_V$, this scale factor is unity within our model, but it is introduced in
  almost all extinction corrections under the name of $A_V$, where the
  extinction at wavelength $\lambda$ is $A_V \phi(\lambda)$, with the function
  $\phi$ standing for the extinction formula.  To compare with models that do
  not enforce the normalisation of the extinction, we introduce in
  Sect.~\ref{sec:RVslopes} a (illegitimate) scale parameter $s$ multiplying the
  extinction magnitude $A_F$, so that the extinction in bandpass $i$ will now be
  $s\,\delta_{i4}\,X_4$, where colour $4 \equiv B-V$ (with our top-hat filters).

\item It is then observed in Sect.~\ref{sec:RVslopes} that there is a unique
  value of $R_V$ for which the extinction-corrected magnitudes in the other
  filters $UBRI$ do not depend on the extinction colour.  This provides our
  determination of~$R_V$.
\end{enumerate}

\begin{table*}
  \caption{Definitions.}
  \label{tab:definitions}
  \centering
  \begin{tabular}{r l}
    \hline\hline
    Term & Description\\
    \hline
    $\cdot_{F}$ & Filter index \\
    $\cdot_{i}$ & Colour index \\
    $\moy{\cdot}$ & Average over all SNe \\
    $c'_i$ & Colour $i$ before \ew{Si} and \ew{Ca} corrections \\
    $c_i$  & Colour $i$ after \ew{Si} and \ew{Ca} corrections  \\
    $e_i$  & Excess $e_i = c_i - \moy{c_i}$ \\
    $I_i$  & Intrinsic component of colour $i$ \\
    $X_i$  & Extinction component of colour $i$ \\
    $\epsilon_{i}$ & Residual of the colour description by the model \\
    $A'_F$ & Magnitude increase from extinction by host (from the
             extinction formula) in bandpass $F$ \\
    $A_F$
         & Same as previous after modification
           of the extinction formula value \\
    $\delta_{ij}(R_V)$
         & Relates extinctions in colours $j$ and $i$ using the
           extinction formula (Sect.~\ref{sec:offsets}) \\
    $R'_{Fj}(R_V)$
         & Ratio of the magnitude extinction in bandpass $F$ to the extinction
           component of colour $j$ (Eq.~\ref{eq:extinctionfilter}) \\
    $R_{Fj}(R_V)$
         & Same as previous after rescaling (Sect.~\ref{sec:rescaling}) \\
    $\gamma_ {ij}$
         & Relates the intrinsic colours $j$ and $i$
           (free parameters to be found) \\
    $\sigma_i$
         & Partial derivative of colour $c_i$ on \ew{Si},
           $\partial c_i/\partial\ew{Si}$ \\
    $\kappa_i$
         & Partial derivative of colour $c_i$ on \ew{Ca},
           $\partial c_i/\partial\ew{Ca}$ \\
    \hline
\end{tabular}
\end{table*}

This derivation of $R_V$ relies on the suppression of the (well-measured) colour
dependencies on $R_V$, rather than on the fit of magnitudes involving a $\chi^2$
(subject to the `grey' fluctuation of 0.08 to 0.12~mag and other correlated
effects, such as modelling and calibration).  There is no attempt at any stage
in the previous sequence to minimise the colour residuals or the residuals of
magnitudes in the Hubble diagram.

\subsection{Calcium and silicon contributions}
\label{sec:CaSi}

\cite{Nugent1995} showed that the impact of \CaII{} and \SiII{} absorption lines
can be characterised by their equivalent widths and that they correlate strongly
with magnitudes, colours, and light-curve shapes.  This section addresses a
mathematical issue: how to suppress the part of the colour variation that
depends on \ew{Si} and \ew{Ca} in the presence of a correlation between these
two widths.  \citetalias{Chotard2011}, and other earlier works referenced
therein, observed that there is a strong correlation between the equivalent
widths \ew{Si} of \SiII~$\lambda$4131~\AA, \ew{Ca} of \CaII~$\lambda$3945~\AA,
and the different colours.  The equivalent widths \ew{Si} and \ew{Ca} are
measured as in \cite{Bronder2008}, using an algorithm described in
\citetalias{Chotard2011}, and in more detail in \cite{ChotardPhD}: following
\cite{Nugent1995}, a spectral range centred on the spectral line is defined from
$\lambda_1$ to $\lambda_2$, and a reference level $f_c(\lambda)$ is defined by a
straight line connecting the two spectral values associated to $\lambda_1$ and
$\lambda_2$.  The actual spectrum observed is $s(\lambda)$.  The equivalent
width is
\begin{equation}
  \label{eq:1}
  \ew{} = \int_{\lambda_1}^{\lambda_2} \left(
    1 - \frac{s(\lambda)}{f_c(\lambda)}
  \right) \d\lambda.
\end{equation}

The errors on the equivalent widths are derived using a simulation that takes
into account the photon statistics as well as the algorithm used to select the
boundaries in the integration over the line width.  As these two Ca and Si
spectral features have a strong colour contribution over the whole spectral
range, it is advantageous to treat them separately from the extra-intrinsic
colour and take advantage of the detailed spectral information.  The analysis
described in the following sections refers to the intrinsic colour component
remaining {after} subtraction of the effect of \SiII~$\lambda$4131~\AA{} and
\CaII~$\lambda$3945~\AA{} on the ten~different colours\footnote{To recover the
  full intrinsic colours, the suppressed intrinsic colour fluctuations
  associated to Si and Ca contributions should be added back, as in
  Sect.~\ref{sec:noCaSi}.}.  The small (formal) variations $\delta c'_i$ of the
`observed' rest-frame colour $c'_i$ around its mean value are assumed to depend
linearly on the variations $(\delta \ew{Si}, \delta \ew{Ca})$ of the equivalent
widths (with other contributions treated as noise).

The subtraction of the \ew{Si} and \ew{Ca} contribution to the colour
variability is performed sequentially, taking into account their correlation.
Let the rest-frame colour $c'_i (\ew{Si}, \ew{Ca})$ be a function of the two
equivalent widths with $\sigma_i = \partial c'_i/\partial \ew{Si}$ and
$\kappa_i = \partial c'_i/\partial \ew{Ca}$.  The small fluctuations (from SN to
SN) are defined as $\delta \ew{} = \ew{} - \moy{\ew{}}$ (where $\moy{\cdot}$
denotes the averaging over all SNe).  The associated change in `colour
difference' $\delta c'_{i} = e_i$ is given by the partial derivatives:
\begin{equation}
  \label{eq:colour_vs_Ca-Si}
  e_i = \delta c'_{i} = \sigma_i\, \delta \ew{Si} + \kappa_i\, \delta \ew{Ca}.
\end{equation}
As we are only interested in the derivatives, the subtraction of the averages is
unnecessary.  The coefficients $\sigma_i$ and $\kappa_i$ are not directly
observed from the data as a consequence of the correlation between the two
equivalent widths, which needs to be removed, and can be described by two linear
relations:
\begin{equation}
  \label{eq:8}
  \begin{aligned}
    \delta\ew{Si}
    &= a_{\Si/\Ca}\, \delta\ew{Ca} + \text{uncorrelated fluctuations,} \\
    \delta\ew{Ca}
    &= a_{\Ca/\Si}\, \delta\ew{Si} + \text{uncorrelated fluctuations}.
  \end{aligned}
\end{equation}
The coefficients $a_{\Si/\Ca}$ and $a_{\Ca/\Si}$ are {not} inverse of each other
as a consequence of measurement errors and uncorrelated physical fluctuation.
They are found directly from the observed equivalent widths by two different
linear fits to the observed distribution in the $(\ew{Si}, \ew{Ca})$ plane.  Two
$\chi^2$ are minimised:
\begin{equation}
  \label{eq:10}
  \begin{aligned}
    \chi^2_{\Ca/\Si}
    &= \sum_i \left(%
      \ew{Ca}_i - a_{\Ca/\Si} \ew{Si}_i + b_{\Ca/\Si}
      \right)^{2}, \\
    \chi^2_{\Si/\Ca}
    &= \sum_i \left(
      \ew{Si}_i - a_{\Si/\Ca} \ew{Ca}_i + b_{\Si/\Ca}
      \right)^{2}.
  \end{aligned}
\end{equation}
The dependence of \ew{Si} on \ew{Ca} gives $a_{\Si/\Ca} = 0.0379 \pm 0.0771$,
and the dependence of \ew{Ca} on \ew{Si} gives $a_{\Ca/\Si} = 0.513 \pm 0.286$.
The observed values are not weighted by the measurement errors, which do not
include the intrinsic variability of the equivalent widths.  The correlation
$a_{\Si/Ca}$ of \ew{Si} as a function of \ew{Ca} is compatible with zero, and
ignoring it would not have a significant impact on the corrections.  We
nevertheless take it into account to ensure that when the procedure described
below is applied, there is no residual dependence whatsoever of colours on
\ew{Si} or \ew{Ca}.  We stress that we do not claim that either $a_{\Si/\Ca}$ or
$a_{\Ca/\Si}$ represents the physical correlation between the equivalent widths,
which is described by a single number
$ a^{\text{phys}}_{\Ca/\Si} = 1/a^{\text{phys}}_{\Si/\Ca}$.  Extracting
$a^{\text{phys}}_{\Ca/\Si}$ would require a full understanding of the
convolution of measurement errors and variability, which we do not actually
need.

\begin{figure}%
  \includegraphics[width=\columnwidth]{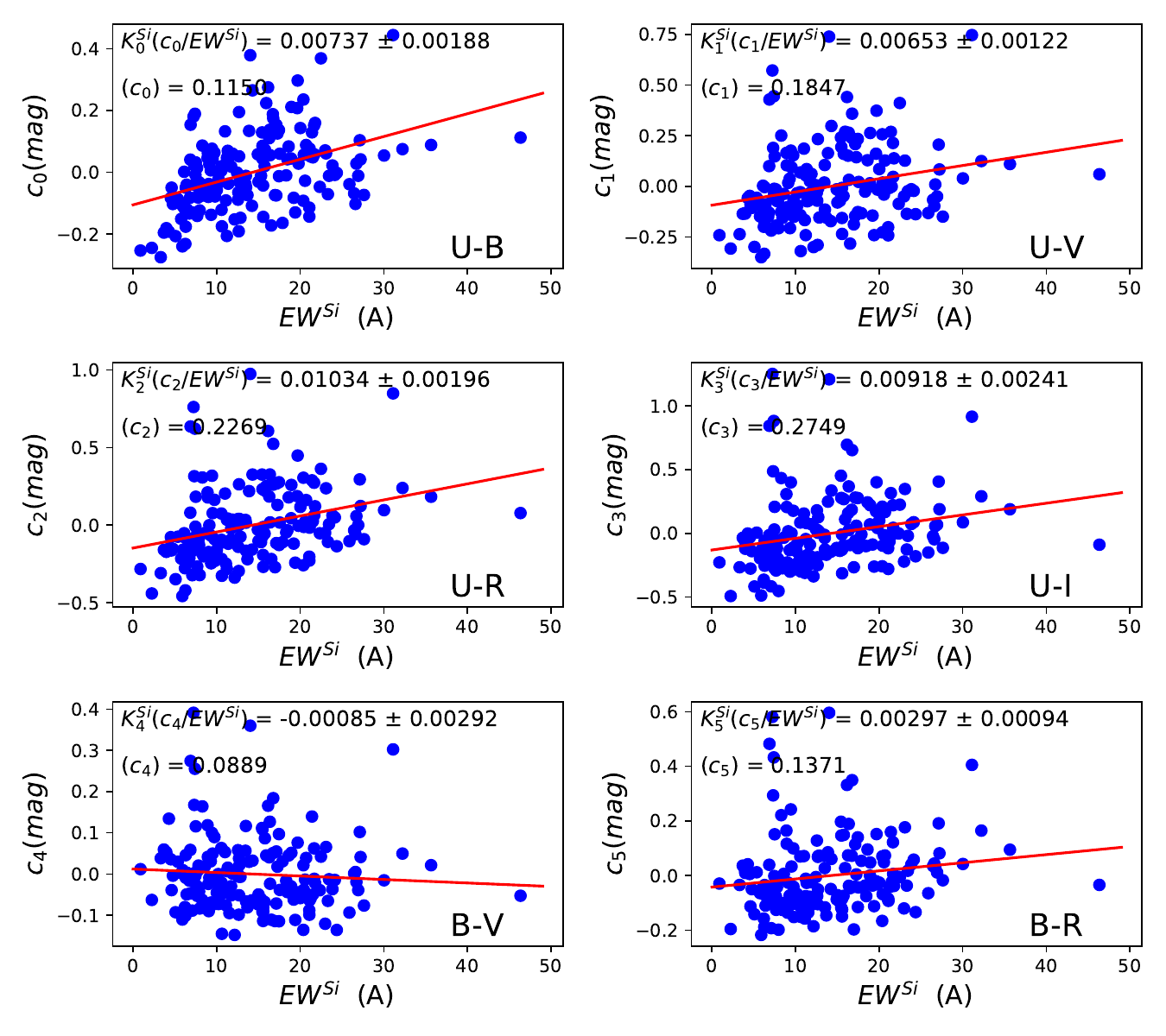}
  \caption{Correlations between the first five colours $c_0' \equiv U-B$,
    $c_1' \equiv U-V$, $c'_2 \equiv U-R$, $c'_3 \equiv U-I$, $c'_4 \equiv B-V$,
    $c'_5 \equiv B-R$ (mag) and \ew{Si} (\AA).  The slopes of the straight line
    fits (red lines) give the value of~$K^{\Si}$ directly.  The correlations of
    the colours can be derived from the first~four.}
  \label{fig:col_ewsi4000}
\end{figure}

\begin{figure}
  \includegraphics[width=\columnwidth]{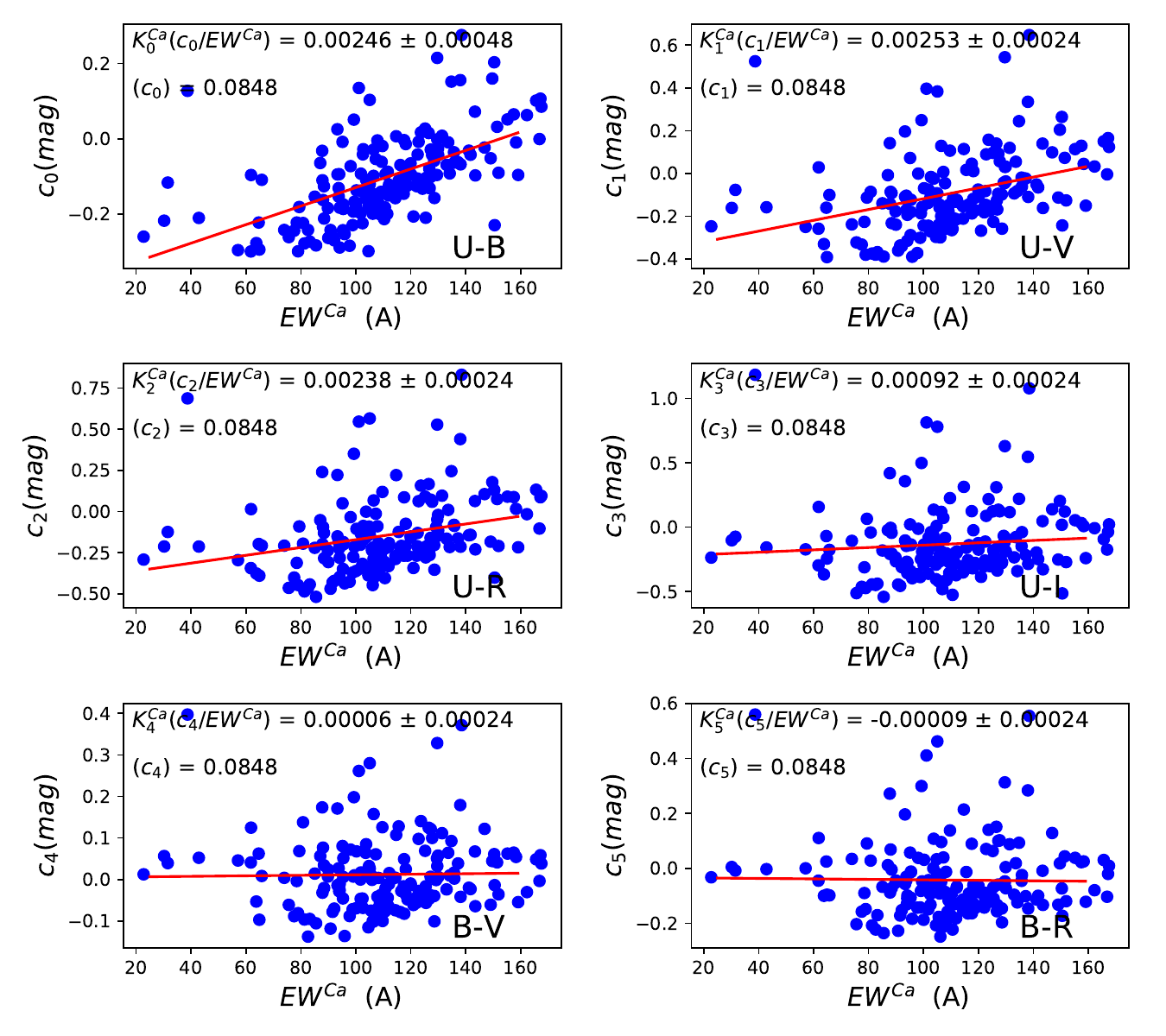}
  \caption{Correlations between the same colours as Fig.~\ref{fig:col_ewsi4000},
    corrected for the \ew{Si} correlation (mag), as a function of \ew{Ca} (\AA).
    The slopes of the straight line fits (red lines) give the value
    of~$K'^{\Ca}$.}
  \label{fig:col_Si_ewCaHK}
\end{figure}

The large difference between the two coefficients reflects the large difference
in the equivalent widths of the two lines.  The observed dependencies of colours
on the variations of Ca and Si equivalent widths are the total derivatives of
the rest-frame colours with respect to \ew{Si} and \ew{Ca}, $K^{\Si}_i$ and
$K^{\Ca}_i$, which can be expressed in terms of the partial derivatives
$\sigma_i = \partial c'_i/\partial \ew{Si}$ and
$\kappa_i = \partial c'_i/\partial \ew{Ca}$:
\begin{equation}
  \label{eq:CaSi}
  \begin{aligned}
    \frac{\d\,c'_i}{\d\,\ew{Si}} = K^{\Si}_i
    &= \sigma_i + \kappa_i\, a_{\Ca/\Si}, \\
    \frac{\d\,c'_i}{\d\,\ew{Ca}} = K^{\Ca}_i
    &= \kappa_i + \sigma_i\, a_{\Si/\Ca}.
  \end{aligned}
\end{equation}
The total derivatives $K^{\Si}$ and $K^{\Ca}$ are obtained from $\sigma_i$ and
$\kappa_i$ in Eq.~\ref{eq:CaSi}, once the coefficients $a_{\Ca/\Si}$ and
$a_{\Si/\Ca}$ are known.  Inversely, the coefficients $\sigma_i$ and $\kappa_i$
can then be derived in Eq.~\ref{eq:9} from the observed dependencies of the
colours on the equivalent widths, as seen in Figs.~\ref{fig:col_ewsi4000} and
\ref{fig:col_Si_ewCaHK}:
\begin{equation}
  \label{eq:9}
  \begin{aligned}
    \sigma_i &=
    \frac{K^{\Si}_i - a_{\Ca/\Si}\,K^{\Ca}_i}{1 - a_{\Ca/\Si}\,a_{\Si/\Ca}}, \\
    \kappa_i &=
    \frac{K^{\Ca}_i - a_{\Si/\Ca}\,K^{\Si}_i}{1 - a_{\Ca/\Si}\,a_{\Si/\Ca}}.
  \end{aligned}
\end{equation}
The denominator is zero when $a_{\Ca/\Si} = 1/a_{\Si/\Ca}$, but as explained
previously, this would only occur if there were no fluctuations in the
correlation between the two equivalent widths, and the issue addressed in this
section would vanish.

We now use Eq.~\ref{eq:9} to correct for the dependence of colours on the
equivalent widths in the data.  If the \ew{Si}-corrected colour is first
applied, the observed dependence with \ew{Ca} will be
$K'^{\Ca}_i = \kappa_i(1 - a_{\Si/\Ca}\,a_{\Ca/\Si})$.  The values of $K^{\Si}$
and $K'^{\Ca}$ are directly obtained from the correlation between the colours
and the equivalent widths observed in Figs.~\ref{fig:col_ewsi4000} and
\ref{fig:col_Si_ewCaHK}.  The \ew{Si} and \ew{Ca} contributions to each colour
$c'_i$ are then subtracted to yield
$c_i = c'_i - \sigma_i\, \ew{Si} - \kappa_i\, \ew{Ca}$.  In all the following
steps of the analysis, we use the colour differences $e_i = c_i - \moy{c_i}$,
where $\moy{\cdot}$ represents ---as above--- the mean over all the SNe~Ia of
our sample\footnote{The simulation in Sect.~\ref{sec:simulation}, which ensures
  the consistency of the analysis, starts from the $\Si$- and $\Ca$-corrected
  colours.}.  To preserve the algebraic relations between the ten colours, the
\ew{Si} and \ew{Ca} corrections are performed on the first four colours, $U-B$,
$U-V$, $U-R$, and $U-I$, while the others are obtained by combining them.  The
coefficients $K^{\Si}$, $K'^{\Ca}$, $\sigma$, and $\kappa$ are summarised in
Table~\ref{tab:Ca-Si_coeff} for the first four~colours; the other colours can
then be obtained by subtraction from the appropriate pair.

\begin{table*}
  \caption{Colour correction from \ew{Si} and \ew{Ca}.  The first four colours
    allow the prediction of the coefficients $K^{\Si}$ and $K^{\Ca}$ for the
    other ones.  The coefficients $\sigma$ and $ \kappa$ are then derived via
    Eq.~\ref{eq:9}.}
  \label{tab:Ca-Si_coeff}
  \centering
  \begin{tabular}{c l c c c c}
    \hline\hline
    & colour & $K^{\Si}$ & $K'^{\Ca}$ & $\sigma_i$ & $\kappa_i$ \\
    \hline
    && (mag/\AA)  & (mag/\AA) & (mag/\AA) & (mag/\AA) \\
    0 & $U-B$ & $ 0.00737 \pm 0.00122 $ & $0.00246 \pm 0.00027$ & $0.00608 \pm 0.00123 $ & $0.00251 \pm 0.000295$ \\
    1 & $U-V$ & $ 0.00653 \pm 0.00195 $ & $0.00253 \pm 0.00049$ & $0.00520 \pm 0.00196 $ & $0.00258 \pm 0.000501$ \\
    2 & $U-R$ & $ 0.01034 \pm 0.00241 $ & $0.00238 \pm 0.00063$ & $0.00910 \pm 0.00243 $ & $0.00242 \pm 0.000687$ \\
    3 & $U-I$ & $ 0.00918 \pm 0.00292 $ & $0.00092 \pm 0.00079$ & $0.00870 \pm 0.00295 $ & $0.00934 \pm 0.000806$ \\
    \hline
  \end{tabular}
\end{table*}

As shown in Sect.~\ref{sec:noCaSi}, we found that the correction of colours for
spectral features was necessary.  The parameters $\sigma_i$ and $\kappa_i$ could
be considered as additional parameters of the model implemented, though they are
not tuned but measured; we show in Sect.~\ref{sec:noCaSi} that the Ca-Si colour
correction is closely related to the SALT variable $x_1$.

\subsection{Evaluation of extinction coefficients $\delta$}
\label{sec:delta}

\subsubsection{From extinction formula to coefficients $\delta$}
\label{subsec:subsdelta}

This section describes the contribution of the reddening to the colour
variations.  Our reddening indicator is the extinction part $X_4$ (defined in
Eqs.~\ref{eq:method}, \ref{eq:method_delta}) of $e_4 = E(B-V) -\moy{E(B-V)}$
after the \ew{Ca} and \ew{Si} corrections have been applied; its mean
$\moy{X_i}$ is zero, and it can have a positive or negative sign, but its
variations are the same as the standard one (up to the intrinsic contributions
of \ew{Si}, \ew{Ca}, and $I_i$, which we do want to exclude from the extinction
correction).  The SNfactory top-hat filters are defined in
Table~\ref{tab:bandpasses}, and the extinction formula is (at first) assumed to
be perfectly known, as given by \citetalias{Fitzpatrick1999}, once a value of
$R_V$ is chosen, which we assume until Sect.~\ref{sec:RVslopes}.

The reddening corrections $A'_{F}$ and $R'_{Fi} $ of
Eq.~\ref{eq:extinctionfilter} for the bandpass $F$ are then evaluated for each
value of $R_V$ in a grid, and the analysis described in this section is repeated
over the grid.  The difficulty that we address is that the extinction indicator
$E_n(B-V)$ (with or without subtraction of the mean) is always tied to the
choice of a wide bandpass, top-hat (our case), or Bessell.  The reddening
corrections at each wavelength depend on wavelength and $R_V$, but also on the
choice of wide band filter, and on the spectral shapes of the selected stars
through the wide-band integration.  The rest-frame photon-count spectrum
$s_{n}(\lambda)$ for SN $n$ is
\begin{equation}
  \label{eq:extinctionlaw}
  s_{n}(\lambda) = s_0(\lambda)\,10^{-0.4 \phi(\lambda)\,E_n(B-V)}.
\end{equation}
With the colour difference definition $E_{n}(B-V) = (B-V)_{n} - \moy{B - V}$,
the spectrum $s_0(\lambda)$ at null colour difference is the mean SN~Ia
photon-count spectrum; the function $\phi(\lambda)$ is provided by
\cite{CCM1989, ODonnell1994, FM05}.  This extinction indicator $E_{n}(B-V)$, or
`colour difference', differs by a constant term from the standard $E(B-V),$
which is referenced to the zero extinction value rather than the mean; the
extinction indicator $E_n(B-V)$ can be positive or negative.  This constant
shift (bandpass-dependent) affects colours, but not their variations, with which
we deal exclusively.  Equation~\ref{eq:extinctionlaw} is a slight extension of
the standard extinction formula of \citetalias{Fitzpatrick1999} on two counts:
the extinction formula relies on a specific set of blue stars, not SNe, and the
bandpasses are defined by Bessell filters, rather than top-hat filters as
described in Table~\ref{tab:bandpasses}.  This issue is addressed in
Sect.~\ref{sec:rescaling}.  Hereafter, we use the extinction component of
$E_n(B-V)$, that is, $X_4(n) = X_{B-V}(n)$, as a substitute for the colour
difference $E_n$ in Eq.~\ref{eq:extinctionlaw}.  It is shown in
Appendix~\ref{sec:Annex} that, in the linear approximation, the extinction
correction $A'_{F}$ in filter $F$ is given for a SN with photon-count spectrum
$s(\lambda)$ by:
\begin{align}
  \label{eq:extinctionfilter}
  A'_F(n) &= m_{F}^{\text{obs}}(n) - m_{F}^{\text{true}}(n) \nonumber\\
          &\simeq R'_{F4} X_{4}(n) =
            \frac{\int_{F} s_0(\lambda)\phi(\lambda)\,\d\lambda}{%
            \int_F s_0(\lambda)\,\d\lambda} X_{B-V}(n).
\end{align}
The coefficient $R'_{F4}$ is the exact derivative of the reddening correction
$A'_F(n)$ with respect to the extinction colour component $X_4$ at $X_4 = 0$.
We show in Table~\ref{tab:linearcoeff} that the relative error resulting from
this approximation remains smaller than 0.01 over the full range of values of
$X_4$.  The reddening correction coefficient $A'_{F4}$ should actually be
weighted by the spectrum of each SN according to Eq.~\ref{eq:extinctionfilter},
but as we only consider a linear reddening correction, the SN dependence would
bring a negligible second-order correction.  The term $X_{B-V}(n) = X_{4}(n) $
in Eq.~\ref{eq:extinctionfilter} is the extinction content of the colour
difference $e_4 = E_{n}(B-V)$ for SN $n$.  As noted previously, the extinction
colour $X_i(n)$ can be positive or negative, but should be a better reddening
indicator than $E_n(B-V)$ as the intrinsic colour variation has been removed.

For each bandpass $F = UBVRI$, the extinction coefficient
$R'_{F4} = A'_F/X_{B-V}(n)$ is obtained from Eq.~\ref{eq:extinctionfilter}.  The
associated colour differences are $e_i = {c_i} -\moy{c_{i}}$.  The extinction
coefficients $R'_{Fj}$ ($j \neq 4$) are then derived from the integrated
extinction formula.  The value of~$R'_{Fi}$ for another colour $i$ can be
derived from $R'_{F4}$; for example $R'_{F3}$ (with $e_3 \equiv U-I$) is given
by:
\begin{equation}
  \label{eq:extinction-coeff}
  R'_{F3} = \frac{R'_{F4}}{R'_{U4} - R'_{I4}}.
\end{equation}

The colour--colour coefficients $\delta'_{ij}$ relate the (extinction) colour
excesses of all colours; they are obtained from the two bandpasses defining each
colour, from the analogue of Eq.~\ref{eq:extinction-coeff}; for example, for
$e_2 \equiv U-R$ and $e_6 \equiv B-I$, one has
$X_2 = \delta_{24}X_{4} =\delta_{26} X_{6}$ with
\begin{equation}
  \label{eq:colour-delta}
  \delta_{26} = R'_{U6}- R'_{R6} = \frac{R'_{U4} - R'_{R4}}{R'_{B4} - R'_{I4}}.
\end{equation}
The extinction correction for all the other colours is similar.  The colour
indicators used in this study is actually the extinction component $X_{B-V}$ of
$B-V$ computed in Sect.~\ref{subsec:chsq} rather than the measured value
$e_{B-V}$.  For instance, the correction to $e_5 \equiv B-R$ using
$X_6 \equiv B-I$ is
\begin{equation}
  X_{5} = \delta_{56}(R_V)\, X_{6} =
  \frac{R'_{B4} - R'_{R4}}{R'_{B4} - R'_{I4}} X_{6} =
  \delta_{54}(R_V)\, X_{4}.
  \label{eq:3}
\end{equation}
Using for instance the extinction part $X_4$ of colour $4 \equiv B-V$ or the
extinction part $X_6$ of colour~$6 \equiv B-I$:
\begin{align}
  \label{eq:extinctiondef}
  m_{F}(n)^{\text{corr}} & = m_{F}(n)^{\text{obs}} - A'_{F}(n)(R_V) \\ A'_F(n,
  R_V) &= \delta'_{F4}(R_V)\,X_4 (n) = \delta'_{F6}(R_V)\,X_6(n).
\end{align}
(We recall that $m_F^{\text{obs}}(n) $ has been brought to a common reference
redshift.)

\subsubsection{Rescaling of extinction coefficients: from $\delta'$ to $\delta$}
\label{sec:rescaling}

The extinction formulae, such as those of \cite{CCM1989} or
\citetalias{Fitzpatrick1999}, have not been tuned for SN spectra, nor to our set
of top-hat bandpasses.  In addition, the extinction law was established for a
sample of mostly HD stars, where the extinction of each spectral bin is weighed
by a spectrum that differs from the SNe~Ia spectrum.  It should not be assumed
that the extinction formulae provided by \cite{CCM1989, ODonnell1994, FM05}
---with different filters (Bessell) and different stars (HD)--- apply directly
to SNe~Ia, given the presence of the wide band term $E_n(B-V)$ or $X_4(n)$ in
Eq.~\ref{eq:extinctionlaw}.  The minimal correction proposed here is a rescaling
of the formula so as to ensure consistency.  If we apply
Eq.~\ref{eq:extinctionfilter} to $e_4 = E_n(B-V),$ we find
$e_4 = (R'_{B4} - R'_{V4})e_4$ (or $X_4 = [R'_{B4} - R'_{V4}]X_4$).  We must
expect
\begin{equation}
  \label{eq:consistency1}
  R'_{B4} - R'_{V4} = 1.
\end{equation}
If the extinction in the $V$ bandpass is used ---which is a frequent
occurrence--- a different consistency condition should be implemented:
$R'_{V4} = 1$ to ensure that $A_V$ is the actual extinction in this bandpass.
Equation~\ref{eq:consistency1} is sometimes written \citep[e.g.][]{Mandel2017,
  Jha2007} as $R_B - R_V = 1$.  However, we find (for $R_V = 2.25$) with the
previously defined top-hat bandpasses $(R'_{B4} - R'_{V4})_{FM99}= 0.864$, while
with \cite{ODonnell1994} extinction parameters, $(R'_{B4} - R'_{V4})= 0.814$.
Most of this difference is due to the Bessell bandpasses assumed in the
extinction formula, while we are using top-hat filters in the present study.
Indeed, when using Bessell filter weighting \citep{Bessellfilters},
$R'_{B4} - R'_{V4} = 1.0329$, which is closer to but still incompatible with
unity.  We show in Eq.~\ref{eq:extinctionfilterAnnex} of the Appendix that the
expression of the coefficient $R'_{F4}$ in Eq.~\ref{eq:extinctionfilter} is the
{exact} derivative of the reddening correction with respect to the extinction
colour at $X_{4} = X_{B-V} = 0$, and in Table~\ref{tab:linearcoeff} that the
linear approximation is accurate to within $5\,10^{-3}$~mag over the full range
of extinctions.  Another contribution to this mismatch may be the different
spectra of stars used in \citetalias{Fitzpatrick1999} and in the present work
(SNe~Ia).  We assume that an acceptable correction ensuring the `consistency
condition' in Eq.~\ref{eq:consistency1} is an overall rescaling of the
extinction formula at all wavelengths, so as to force $R_{B4} - R_{V4} = 1$.
All the coefficients $R'_{Fi}$ are rescaled:
\begin{equation}
  \label{eq:consistency2}
  R_{Fi} = R'_{Fi}/(R'_{B4}- R'_{V4}).
\end{equation}

The extinction coefficient of each filter is then increased by the factor
$1/(R'_{B4} - R'_{V4}) = 1/0.864$.  No such global rescaling was applied in
\citetalias{Chotard2011} where an extinction parameter $A_V$ was fit
independently for each SN.  As mentioned above, Eq.~\ref{eq:extinctionlaw} is
only an approximation.  Even if one accepts that different molecular and grain
compositions can be summarised by a single coefficient $R_V$ (related to their
cross section for light), its value could differ in the halo and in the disc (if
there is one), and the observations of \cite{Schlafly2016} in our own galaxy
suggest this possibility.  The $\delta_{ij}$ coefficients of
Eq.~\ref{eq:method_delta} verify that $\delta_{ij} = 1/\delta_{ji}$.  The five
values of the extinction coefficients $\delta_{F4}(R_V),$ which provide the
extinction in filters for $F = UBVRI$ from $X_4 = X_{B-V}$, are given in
Table~\ref{tab:extinctioncoeff}.  The extinction coefficients $\delta_{F}(R_V)$
are subject to a first correction ($\Dex_F$), which cancels the correlations of
the colour residuals with the extinction components as discussed in
Sect.~\ref{sec:offsets}; they are then rescaled for algebraic consistency as
described in Eq.~\ref{eq:consistency2}.  Finally, for arbitrary values of $R_V$,
the suppression of the correlation of extinction-corrected {magnitudes} with the
extinction colour in the $V$ bandpass requires an extra overall rescaling $s,$
which is a function of the extinction parameter $R_V$, as explained in
Sect.~\ref{sec:RVslopes}.  Such a scaling destroys the consistency condition in
Eq.~\ref{eq:consistency1}, but is introduced in practice by almost all authors
\citep{Boone2021, Chotard2011, Amanullah2015, Saunders2018, Thorp2021} who
substitute
$m_F^{\text{corr}}(\lambda) = m_F^{\text{obs}}(\lambda) - A_V\,\phi(\lambda,
R_V)$ where $A_V$ is arbitrary and $\phi(\lambda, R_V)$ stands for one of the
reddening formula.  The analysis of the extinction is performed over a grid of
nine~values of $R_V$: $1.95, 2.05, 2.15, 2.20, 2.25, 2.35, 2.50, 2.60$, and
$3.10$.  We show below that for our value of $R_V$, the scaling factor $s$ is
indeed compatible with unity, which confirms the consistency of the model, as
well as the value of~$R_V$ obtained.

\begin{table*}
  \caption{Extinction coefficients from \citetalias{Fitzpatrick1999} and $UBVRI$
    corrections to the extinction formula for the case $R_V = 2.25$.  In the
    second line, the extinction formula values are modified as explained in
    Sect.~\ref{sec:offsets}.}
  \label{tab:extinctioncoeff}
  \centering
  \begin{tabular}{l c c c c c}
    \hline \hline
    $R_{F4}$ & $U$ & $B$ & $V$ & $R$ & $I$ \\
    \hline
    \citetalias{Fitzpatrick1999} $R'_{F4}$
             & 3.9232 & 3.2477 & 2.3832 & 1.6914 & 1.2167 \\
    Modified (Sec.~\ref{sec:offsets}) $R_{F4}$
             & 3.9907 & 3.2062 & 2.3397 & 1.7910 & 1.1161 \\
    Scaled (Eq.~\ref{eq:consistency2})
             & 4.6057 & 3.7003 & 2.7003 & 2.0671& 1.2881 \\
    \hline
  \end{tabular}
\end{table*}

\section{Extraction of intrinsic and extinction colour components}
\label{sec:colour-components}

\subsection{Intrinsic colour coefficients $\gamma$}
\label{subsec:gamma}

As mentioned in Eq.~\ref{eq:method}, the colour difference $e_i(n)$ of SN~$n$
(after correction for \ew{Si} and \ew{Ca}) is the sum of two components
$e_i(n) = I_i(n) + X_i(n) + \epsilon_i(n)$, and the intrinsic component $I_i$ of
colour $i$ is assumed to belong to a one-dimensional space of intrinsic colour
fluctuations, as suggested by \cite{Leget2020} and \cite{Boone2021}.  These are
then interrelated by a set of coefficients $\gamma_{ij}$ common to all
supernovæ~$n$ as introduced in Eq.~\ref{eq:method_delta}.  As shown below, small
corrections to the extinction formula are required.  An initial value for the
coefficients $\gamma_{ij}$ can be derived directly from the requirement that
intrinsic and extrinsic colours be uncorrelated, though this value can differ
significantly from the result of the $\chi^2$ minimisation discussed later.  For
any pair of leading ($i$) and auxiliary ($j$) colour, two equations link
$X_{i}, X_{j}$ and $I_{i}, I_{j}$:
\begin{align}
  \label{eq:colour_i}
  e_i &= I_{i} + X_{i}  + \epsilon_i =
        \gamma_{ij}I_{j} + \delta_{ij}X_{j} + \epsilon_{ij}, \\
  \label{eq:colour_j}
  e_j &= I_{j}  + X_{j} +\epsilon_j.
\end{align}

The initial values of $\gamma_{i0}$ can be obtained from the previous equations,
assuming the absence of correlations between the $I_i$ and the $X_i$, and
neglecting the contribution of the residuals $\epsilon_i$ and $\epsilon_{ij}$,
\begin{align}
  \label{eq:var_1}
  \moy{e_1^2} &= \gamma_{10}^2\moy{I_0^2} + \delta_{10}^2 \moy{X_{0}^2}, \\
  \label{eq:var_0}
  \moy{e_0^2} &= \moy{I_0^2} + \moy{X_0^2},\\
  \label{eq:var_01}
  \moy{e_0\,e_1} &= \gamma_{10}\moy{I_0^2} + \delta_{10}\moy{X_0^2}.
\end{align}
Here, $\delta_{ij}$ is the ratio of the extinction components $X_i/X_j$, as
defined in Table~\ref{tab:definitions} ---and is not the Kronecker~$\delta$---,
and the index $n$ of the SN is dropped.

Combining Eq.~\ref{eq:var_0} ($\times\delta_{10}$) and \ref{eq:var_01}, one gets
$\moy{e_0\,e_1} - \delta_{10} \moy{e_0^2} = (\gamma_{10} - \delta_{10})
\moy{I_0^2}$.  Similarly,
$\moy{e_1}^2 - \delta_{10} \moy{e_0\,e_1} = \gamma_{10} (\gamma_{10} -
\delta_{10}) \moy{I_0^2}$; dividing the two previous relations one obtains an
initial value of $\gamma_{10}$ (with large errors).  More generally, an initial
value for any pair of colours would be:
\begin{equation}
  \label{eq:variance}
  \gamma_{ij} =
  \frac{\moy{e_i^2} - \delta_{ij} \moy{e_i\,e_j}}{%
    \moy{e_i\,e_j} - \delta_{ij}\moy{e_j^2}}.
\end{equation}

The initial values of the three coefficients $\gamma_{10}$, $\gamma_{20}$, and
$\gamma_{30}$ are respectively 0.671, 0.759, and 0.401.  The statistical errors
on these initial values are relatively large, and an improved determination is
obtained from the iterative fits described in the following section.  (We recall
that all colour differences have zero average, and that intrinsic and extinction
components are assumed to be uncorrelated.)

Given the relation $\delta_{ij} = 1/\delta_{ji}$, we find that (as expected)
$\gamma_{ij} = 1/\gamma_{ji}$.  The coefficients $\gamma_{ij}$ can be expressed
according to Eq.~\ref{eq:method_delta} in terms of three of them, which have
been selected as $\gamma_{10} = I_{(U-V)}/I_{(U-B)}$,
$\gamma_{20} = I_{(U-R)}/I_{(U-I)}$, and $\gamma_{30} = I_{(U-I)}/I_{(U-B)}$.
The remaining 42 coefficients can be obtained from the following relations,
which result from the algebraic relations between colours:
\begin{align}
  \label{eq:gammaij}
  \gamma_{40} &= -1 + \gamma_{10},
  &\gamma_{50} &= -1 + \gamma_{20},
  &\gamma_{60} &= -1 + \gamma_{30},  \nonumber\\
  \gamma_{70} &= -\gamma_{10} + \gamma_{20},
  &\gamma_{80} &= -\gamma_{10} + \gamma_{30},
  &\gamma_{90} &= -\gamma_{20} + \gamma_{30}, \\
  &&\gamma_{ik} &= \gamma_{i0}/\gamma_{k0}. \nonumber
\end{align}

\subsection{$\chi^2$ for intrinsic and extinction colour components}
\label{subsec:chsq}

The model described by Eq.~\ref{eq:colour_i} should reproduce the rest-frame
colours.  We consider the colour difference for SN $n$:
$e_i(n) = c_i(n) - \moy{c_i}$.  The two components $(I_i(n), X_i(n))$ can be
found by requiring that all colours $e_j(n)$ are properly described by
Eqs.~\ref{eq:colour_i}--\ref{eq:colour_j}, and the coordinates
$I_j(n) = \gamma_{ji}I_i(n)$, $X_j(n) = \delta_{ji}X_i(n)$.  Colour components
$X_2(n)$ and $I_2(n)$ are, for example, obtained by minimising ---for each
SN~$n$--- the $\chi^2_2(n)$ in Eq.~\ref{eq:chsq} as a function of the
`coordinates' $I_2(n)$ and $X_2(n)$.  The reddening correction $A_V(n)$ in
bandpass $V$ $A_V(n) = R_{V4} X_4(n)$, which varies from SN to SN, plays no role
in the minimisation: the extinction correction is $A_{F4}(n) = R_{F4} X_4(n)$
for filter $F$ and SN $n$ ($= A_{Fi}(n) = R_{Fi} X_{i}(n)$ using another
colour):
\begin{align}
  \label{eq:chsqepsilon}
  \epsilon_{2} &= X_{2} + I_{2} - e_{2},  \\
  \label{eq:chsqepsilonk}
  \epsilon_{k2} & = (\gamma_{k2}I_2 + \delta_{k2}X_2 - e_k),\\
  \label{eq:chsq}
  \chi^2_2 & = \epsilon_{2}^2 + \sum_{k \in Aux} \epsilon_{k2}^2,
\end{align}
as a function of $I_2$ and $X_2$, under the constraint
\begin{equation}
  \label{eq:decorrel}
  \sum_{\text{all SNe}} I_2(n)\,X_2(n) = 0.
\end{equation}

Equation~\ref{eq:decorrel} imposes the decorrelation of the extra-intrinsic
colour from extinction.  The index $n$ is omitted in Eq.~\ref{eq:chsqepsilon}.
For each colour $i$ and each SN, the residual $\epsilon_{i}(n)$ defined by
Eq.~\ref{eq:chsqepsilon} should not be confused with the magnitude shift at
maximum light defined in Eq.~\ref{eq:magmax}.  We are aware that
Eq.~\ref{eq:decorrel} could be criticised under certain circumstances, such as
in the presence of circumstellar matter with a composition differing from the
average host galaxy.  Such matter, as pointed out by \cite{Borkowski2009,
  Ferretti2017}, might invalidate this assumption, but if the circumstellar dust
is created by a non-degenerate companion, as suggested in one case by
\cite{Nagao2017}, it is not necessarily correlated to the SN~Ia properties.
While not rejecting the possibility, we wanted to explore the accuracy that
could be reached within the assumption of decorrelation.

Each pair $(I_i(n), X_i(n))$ of components allows the prediction of all the
other colours of the SN.  When defining $\chi^2(n)$ in Eq.~\ref{eq:chsq}, each
colour $i$ enters with equal weight, which implicitly assumes that the errors on
all colours are similar.  The solution of Eq.~\ref{eq:chsq} in terms of
$(X_2(n), Y_2(n))$ is found by requiring
$\partial \chi_2^2/\partial X_2 = \partial \chi_2^2/\partial I_2 = 0$:
\begin{equation}
  \label{eq:XandI}
  \left(%
    \begin{array}{c}
      Y_0 \\
      Y_1
    \end{array}
  \right) = \left(
    \begin{array}{c c}
      M_{00} & M_{01} \\
      M_{10} & M_{11}
    \end{array}
  \right) \left(%
    \begin{array}{c}
      I_2\\
      X_2
    \end{array}
  \right)
\end{equation}
with
\begin{align}
  M_{00} &= 1 + \sum_{k \in Aux} \gamma_{k2}^2,
  & M_{01} &= 1 + \sum_{k \in Aux} \gamma_{k2} \delta_{k2} - \lambda \nonumber,\\
  \label{eq:Matrix}
  M_{10} &= M_{01},  & M_{11} &= 1 + \sum_{k \in Aux} \delta_{k2}^2, \\
  Y_{0} &= e_2(k) + \sum_{k \in Aux}\gamma_{k2}\, e_k,
  & Y_{1} &= e_2(k) + \sum_{k \in Aux} \delta_{k2}\, e_k. \nonumber
\end{align}
The colours excesses $e_1(n)$ and $e_2(n)$ are measured, meaning that $Y_0$ and
$Y_1$ are known.  The intrinsic and extinction components $I_2(n)$, $X_2(n)$ are
obtained by inverting Eq.~\ref{eq:XandI}.  The Lagrange multiplier $\lambda$
arises from the assumption of uncorrelated $I_i$, $X_i$, and is found from
Eq.~\ref{eq:decorrel}; it depends on the colour, but is the same for all SNe.
We introduce the set $Aux$ of auxiliary colours used to extract the intrinsic
and extinction components of each colour.  In the optimisation of the intrinsic
couplings $\gamma_{ij}$ below, the full set of ten colours cannot be used in
view of the appearance of algebraic constraints, as explained in the following
section.  Table~\ref{tab:map} presents the actual list of auxiliary colours used
for each colour.  There is an arbitrariness in the choice of the participating
colours in $Aux$; we selected the one that yielded the smallest residuals among
five~trials.

For each SN $n,$ this minimisation yields ten~intrinsic colours $I_i(n)$, and
ten~extinction colours $X_i(n)$, once the intrinsic colour coefficients
$(\gamma_{01}, \gamma_{02}, \gamma_{03})$ are known.  Equation~\ref{eq:chsq}
does not take into account the correlation between colours, but as the solution
gave acceptable residuals for all colours, we kept this simplified expression
for $\chi^2$.  We verified that including the obvious algebraic correlations
between the different colours does not change the solution, but it does add
significant complications.  As the mean of $e_i$ is zero (by definition), the
mean values of $I_i$ and $X_i$ obtained by linear combinations of $e_i$ will
also be zero.

\begin{table}
  \caption{Map of the auxiliary colours used in Eq.~\ref{eq:chsq} and
    Eq.~\ref{eq:Matrix}.  The first column is the measured colour being adjusted
    by the model.  Colours 5 and 7, which have a small intrinsic part, are not
    included in the evaluation of other colours.  The sample of auxiliary
    colours with the smallest mean residual was selected.}
  \label{tab:map}
  \centering
  \begin{tabular}{r l}
    \hline\hline
    Colour index &  Auxiliary colours\\
    \hline
    0 & 1, 2, 3, 4, 6, 9 \\
    1 & 0, 2, 3, 4, 6, 8 \\
    2 & 0, 1, 3, 4, 6  \\
    3 & 0, 1, 2, 4, 6, 8, 9 \\
    4 & 0, 1, 2, 3, 4, 6 \\
    5 & 0, 1, 2, 6, 8, 9 \\
    6 & 0, 2, 3, 4, 8, 9 \\
    7 & 1, 2, 3, 4, 6, 8, 9 \\
    8 & 1, 3, 4, 6, 9 \\
    9 & 2, 3, 4, 6, 8 \\
   \hline
  \end{tabular}
\end{table}

\subsection{Determination of intrinsic components and intrinsic couplings}
\label{subsec:Intrinsic_gamma}

The colour differences which we now use were subtracted from the contributions
of Ca and Si, but leave room for an extra-intrinsic component $I_i$ of each
colour, with all of them dependent on three parameters ($\gamma_{10}$,
$\gamma_{20}$, $\gamma_{30}$) as required by the assumption of a single
extra-intrinsic contribution.  Up to this stage, the $\gamma_{ij}$ are obtained
from their initial values in Eq.~\ref{eq:variance}.  The solution of
Eq.~\ref{eq:XandI} is only acceptable if the ratio of the solutions of
Eq.~\ref{eq:XandI} $I_i(n)$ to $I_j(n)$ for two colours $i$ and $j$ is equal to
$\gamma_{ij}$ for all SNe and all pairs of colours.  Similarly, the ratio of
$X_i(n)$ to $X_j(n)$ should be equal to $\delta_{ij}$.

\begin{figure}
  \includegraphics[width=\columnwidth]{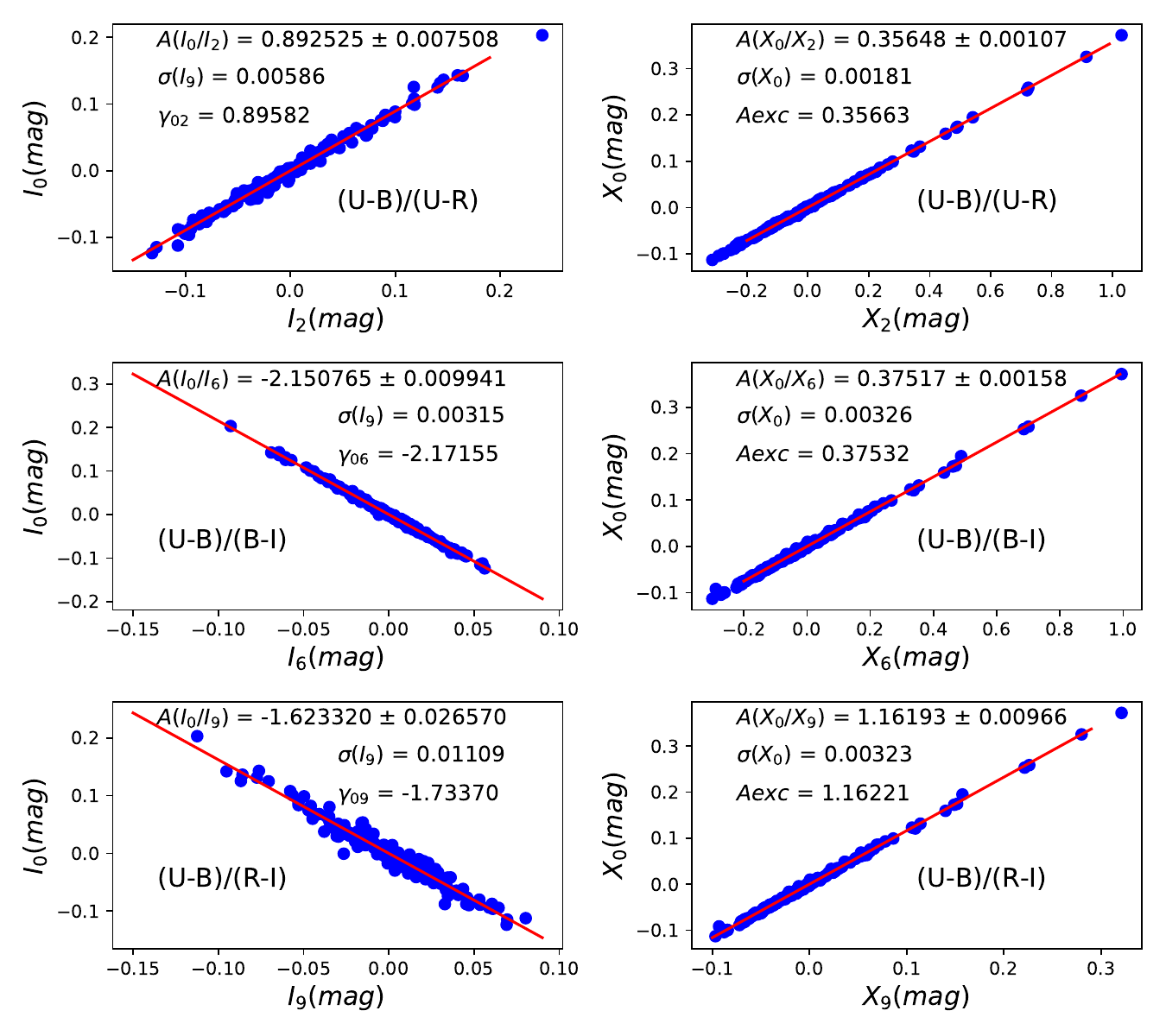}
  \caption{Intrinsic $\gamma^{\text{out}}_{0i} = I_0/I_i$ and extinction
    $\delta^{\text{out}}_{0i} = X_0/X_i$ ratios as defined in Eq.~\ref{eq:XandI}
    $(i=2, 6, 9)$ compared to the input $\gamma^{in}_{0i}$ and calculated
    $\delta_{0i}$ for $R_V = 2.25$.  The compatibility is at the level of 0.001
    for the extinction, and the relative accuracy of 0.05 for the (smaller)
    intrinsic colour (worst case).  The red lines are linear fits to the data,
    and $\gamma^{\text{out}}, \delta^{\text{out}}$ are the values of the linear
    coefficients.}
  \label{fig:intrinsic-extinction_coeff}
\end{figure}

This is achieved by a sequence of iterations.  At each iteration, and for each
colour, the input coefficients $\gamma_{ij}^{\text{in}}$ are the parameters
introduced in Eqs.~\ref{eq:method_delta} and \ref{eq:chsqepsilonk}, while the
linear coefficients of the straight line fits in the plane $(I_i, I_j)$ as in
Fig.~\ref{fig:intrinsic-extinction_coeff} are the output value
$\gamma_{ij}^{\text{out}}$ of the intrinsic couplings.  The output values of
($\gamma_{10}, \gamma_{20}, \gamma_{30}$) are reinjected into Eq.~\ref{eq:chsq}
until Fig.~\ref{fig:intrinsic-extinction_coeff} is satisfactory.  The scatter is
seen to range from 0.002 mag to 0.01 mag depending on the colours, and the
measured values of the ratios $\gamma_{ij}^{\text{out}} =I_i/I_j$ are within
0.01 of the corresponding set of $\gamma_{ij}^{\text{in}}$, as obtained from
$(\gamma_{10}, \gamma_{20}, \gamma_{30})$ using Eq.~\ref{eq:gammaij}.  The
$\delta_{ij}$ coefficients are left unchanged in this sequence, as their values
are fixed by the extinction formula (at this stage).

\begin{figure}
  \includegraphics[width=\columnwidth]{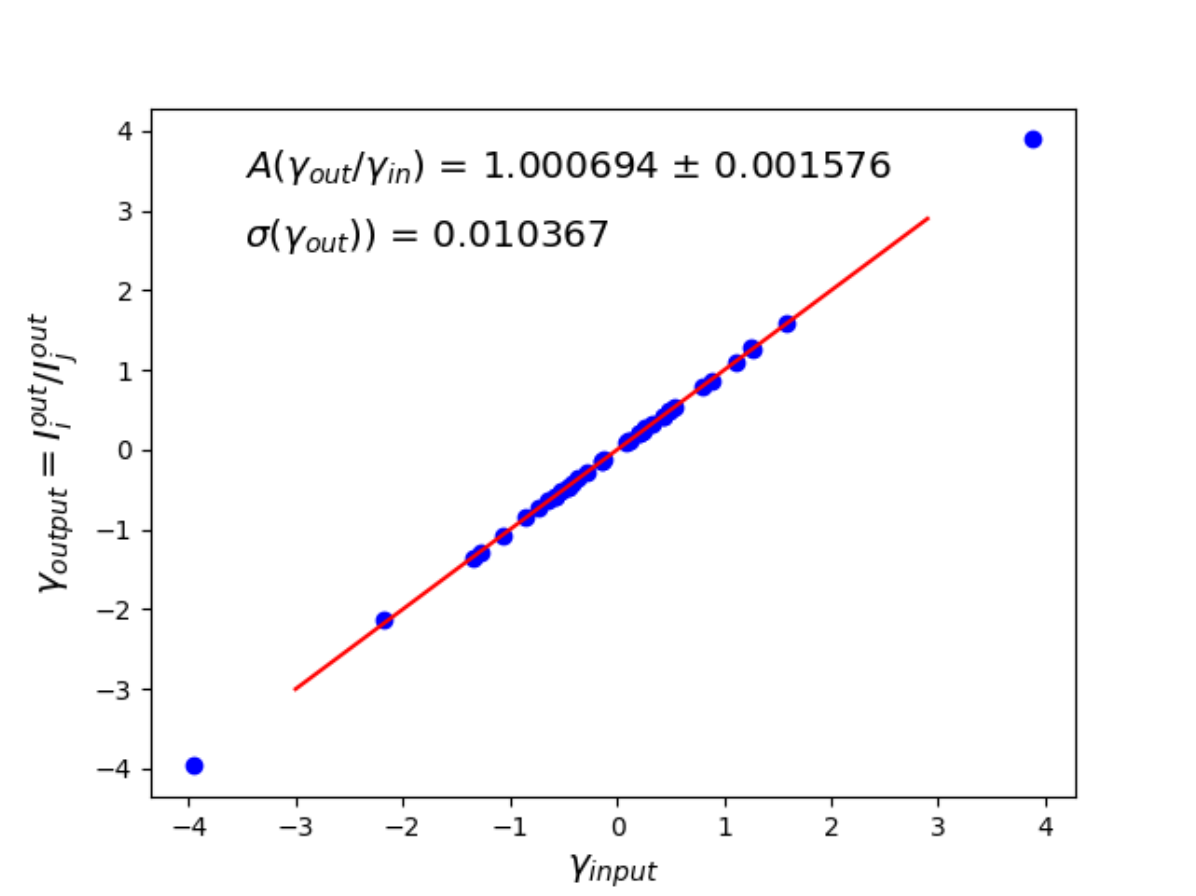}
  \caption{Relation between input and output ($\moy{I_i/I_j}$) values for the
    coefficients $\gamma_{ij}$.  The overall agreement for all $\gamma_{ij}$
    (excluding those involving colours $5 \equiv (B-R)$ and $7 \equiv (V-R)$) is
    within~0.01.}
  \label{fig:gamma_vs_gamma}
\end{figure}

We then turn to Fig.~\ref{fig:gamma_vs_gamma}, which compares the full set of
input values $\gamma_{ij}^{\text{in}}$ in Eq.~\ref{eq:XandI} in the
second-to-last iteration to the output values, that is, of $I_i/I_j$ from the
linear fits in Fig.~\ref{fig:intrinsic-extinction_coeff}.  The final values of
the three coefficients ($\gamma_{10}^{\text{out}}$, $\gamma_{20}^{\text{out}}$,
$\gamma_{30}^{\text{out}}$) are found by minimising the scatter (RMS) of the
output coefficients $\sigma(\gamma_{ij}^{\text{out}})$ with respect to the input
coefficients (horizontal scale).  This last step brings in the constraint of
consistency with Eq.~\ref{eq:gammaij} for the {whole} set of $\gamma_{ij}$, and
modifies the result of the previous sequence of iterations; however, as seen in
Fig.~\ref{fig:intrinsic-extinction_coeff}, the outcome is still satisfactory
with respect to the convergence of the solution for each $\gamma_{ij}$

The relation between $\gamma_{ij}^{\text{out}}$ and $\gamma^{\text{in}}_{ij}$ is
shown in Fig.~\ref{fig:gamma_vs_gamma} at the last iteration.  The linear
coefficient is expected to be close to unity if convergence has been achieved,
while we find 1.00069, with an RMS of 0.0104.  This dispersion arises from a
combination of measurement errors, Si-Ca correction errors, and modelling
errors.  Two precautions were implemented to extract the three intrinsic
couplings from Eq.~\ref{eq:XandI} and Fig.~\ref{fig:gamma_vs_gamma}: we
eliminated the auxiliary colours $4 \equiv B - V$ and $7 \equiv V - R$ that have
very small intrinsic components (cf. Table~\ref{tab:colour_error}) from the
$10\times 9/2 = 45$ $\gamma_{ij}$ couplings, 28 remain after the elimination of
colours 5 and 7.  The 28 output values of the parameters
$\gamma^{\text{out}}_{ij} = I_i/I_j$ at the last iteration are compared in
Fig.~\ref{fig:gamma_vs_gamma} to the input values derived from ($\gamma_{10}$,
$\gamma_{20}$, $\gamma_{30}$) using Eq.~\ref{eq:gammaij}.  The equality of input
and output values was not forced into the optimisation process.  The ratios
$\delta_{ij} = X_i/X_j$ are also seen to have a small scatter of 0.002~mag (in
Fig.~\ref{fig:intrinsic-extinction_coeff}), with averaged output values close to
the input coefficients, supporting the model
Eqs.~\ref{eq:chsqepsilon}--\ref{eq:chsq}.

Nevertheless, the RMS in Fig.~\ref{fig:gamma_vs_gamma} fails as an indicator
when the same set of colours (auxiliary and main) is used to find the intrinsic
content, as the algorithm defined by Eq.~\ref{eq:XandI} yields
$\sigma(\gamma^{\text{out}}_{ij}) = 0$.  For each colour, we eliminated at least
one extra auxiliary colour to avoid such duplicated sets.  The choice of the
auxiliary colours has an impact on the values found for the three intrinsic
couplings.  We selected the one that gave the smallest mean residual scatter
(RMS) in Fig.~\ref{fig:gamma_vs_gamma}.  The scatter of
$\gamma_{ij}^{\text{out}}$ in Fig.~\ref{fig:gamma_vs_gamma} is 0.0103, and the
ratio $\gamma^{\text{out}}_{ij}/\gamma^{\text{in}}_{ij}$ is within 0.001 of
unity.  After the sequence of iterations described in Sect.~\ref{sec:offsets},
the optimal values in Fig.~\ref{fig:gamma_vs_gamma} for $R_V = 2.20$ are given
in Table~\ref{tab:gammaij}.

\begin{table}
  \caption{Coupling constants of the intrinsic colours.}
  \label{tab:gammaij}
  \centering
  \begin{tabular}{c c c c }
    \hline\hline
    $R_V$ & $\gamma_{10}$ & $\gamma_{20}$ & $\gamma_{30}$\\
    $2.20$ & $1.2687 \pm 0.0586$ & $1.1248 \pm 0.0134$ & $0.5579 \pm 0.0286$ \\
    $2.25$ & $1.2647 \pm 0.0586$ & $1.1163 \pm 0.0134$ & $0.5395 \pm 0.0286$\\
    \hline
  \end{tabular}
\end{table}

The errors are derived at this stage from the spread of the results in four
different choices of auxiliary colours and do not include the statistical error.
The comparison of our intrinsic colours to previous results is postponed to
Sect.~\ref{sec:noCaSi}, as the contribution of the equivalent widths \ew{Ca} and
\ew{Si} must be reintroduced.

\subsection{Modifications to the extinction formula}
\label{sec:offsets}

\begin{figure}
  \includegraphics[width=\columnwidth]{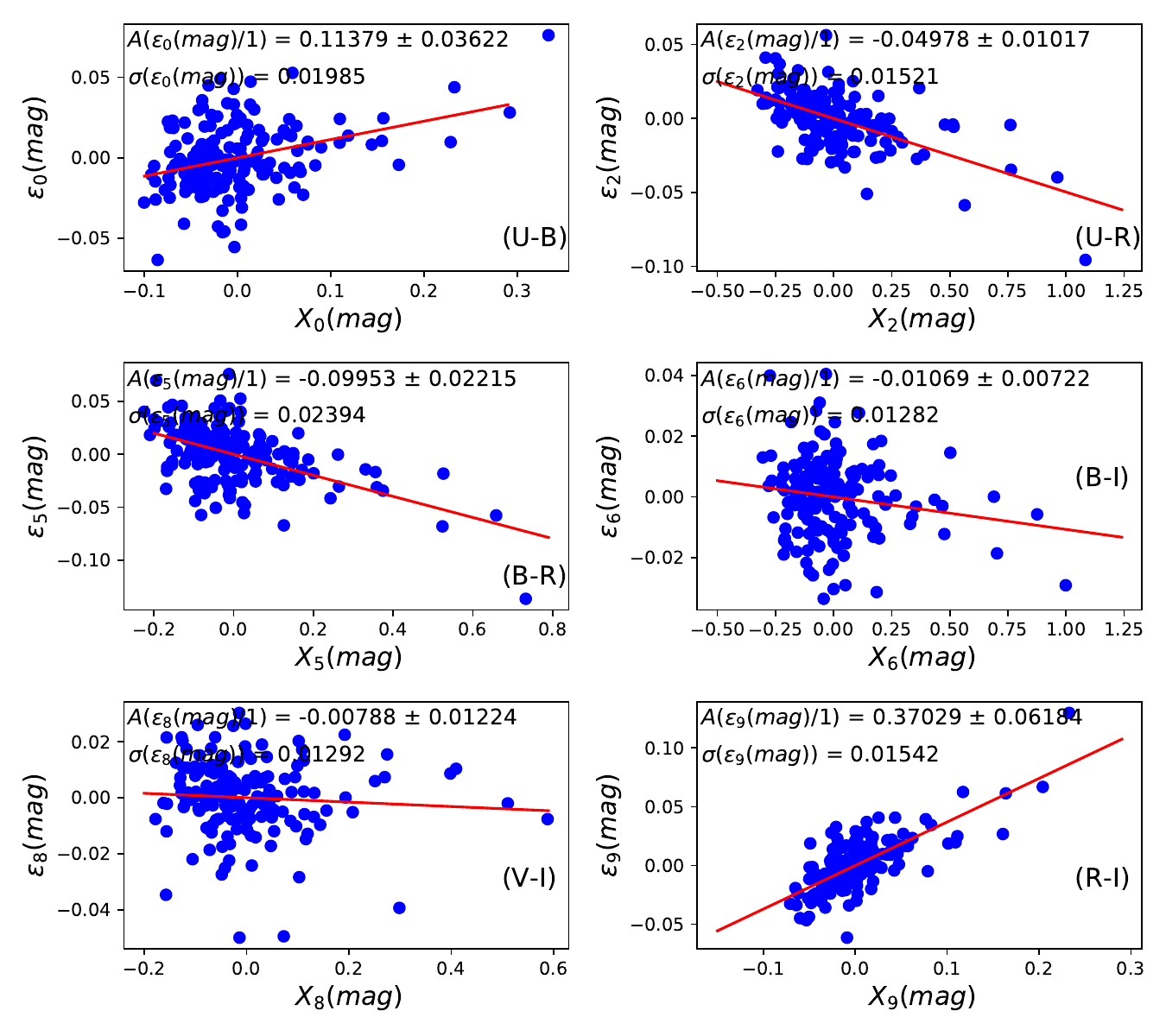}
  \caption{Correlation of the colour residuals (mag) for $R_V = 2.25$ with the
    extinction colour for colour differences $e_0, e_2, e_5, e_6, e_8, e_9$ (no
    extinction formula correction).  The strong correlations observed would
    forbid a safe evaluation of $R_V$.}
  \label{fig:diffcol_vs_colx_nocorr}
\end{figure}

For all values of $R_V$, the colour residuals $\epsilon_i$ described in the
previous section (after Eq.~\ref{eq:chsq}) show a strong correlation with the
extinction colour component when the \citetalias{Fitzpatrick1999} extinction
formula is used.  This (unexpected) correlation is apparent in
Fig.~\ref{fig:diffcol_vs_colx_nocorr}, and there is no value of $R_V$ that
suppresses this effect.  As our subsequent determination of $R_V$ relies on the
absence of any correlation between the extinction-corrected magnitudes and the
extinction, it is crucial to suppress the correlation observed in
Fig.~\ref{fig:diffcol_vs_colx_nocorr}.  This suppression depends on the
coefficients $\gamma_{ij}$ and $\delta_{ij}$ in Eq.~\ref{eq:Matrix} but in no
way on the scale of the reddening correction (nor on the rescaling applied in
Eq.~\ref{eq:consistency2}).

We can cancel the effects observed in Fig.~\ref{fig:diffcol_vs_colx_nocorr} by
introducing corrections $\Dex_F$ to the extinction coefficients $R_{F4}$ of the
five $F= UBVRI$ bandpasses in Eq.~\ref{eq:extinctionfilter}.  The minimised
quantity is
\begin{equation}
  \label{eq:extcorrection}
  \chi_{\Dex}^2 = \sum_{i = 0}^{9} (sl_{i})^2
  \qquad\text{with}\qquad
  sl_i = \partial \epsilon_{i}/\partial X_{i}.
\end{equation}

\begin{figure}
  \includegraphics[width=\columnwidth]{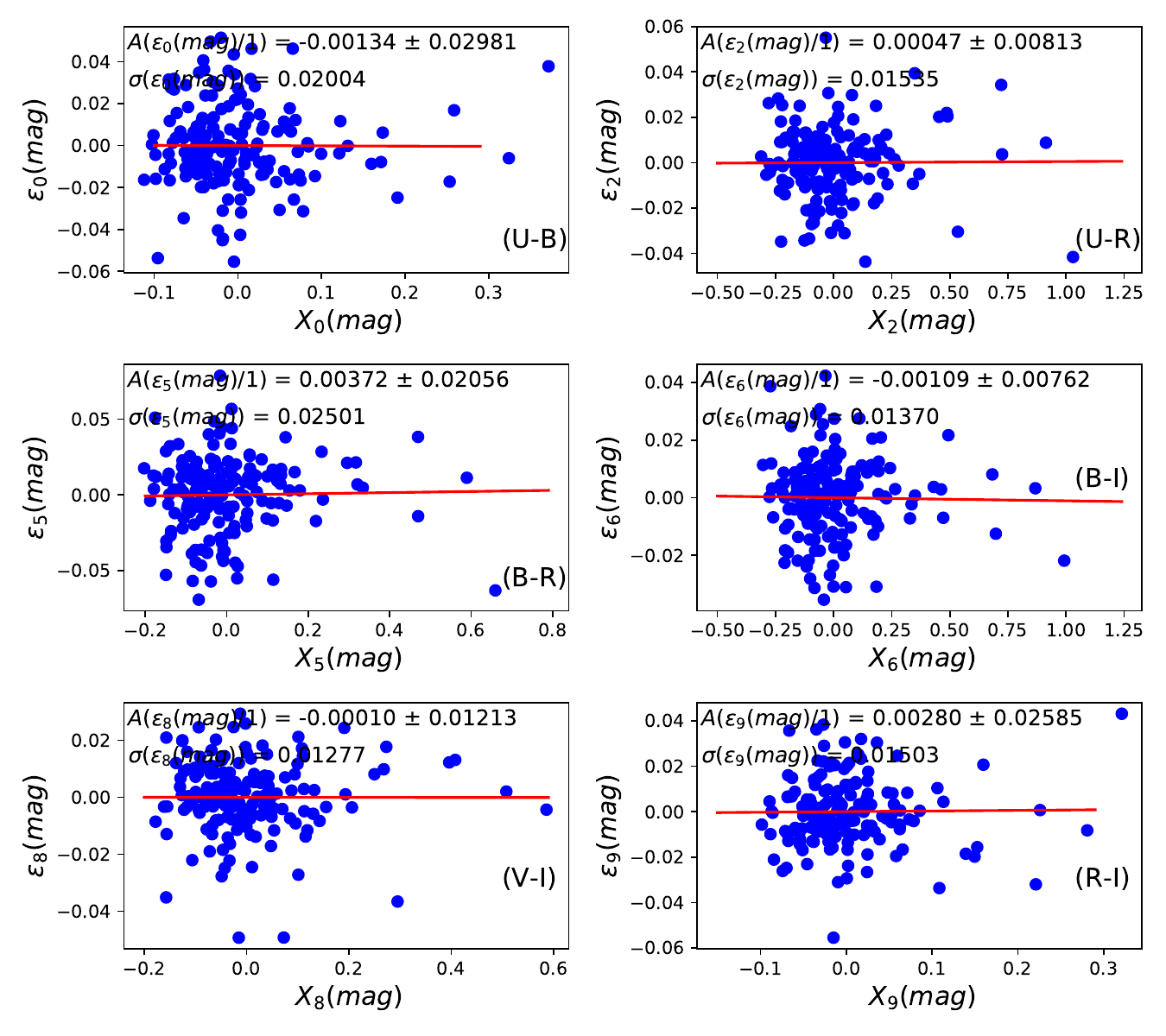}
  \caption{Correlation of residuals of colour differences (mag)
    $(e_0, e_2, e_5, e_6, e_8, e_9)$ with extinction after correcting the
    extinction formula with $R_V = 2.25$.  There are only four independent
    colours, meaning that the four extinction corrections allow us to cancel the
    correlations of all ten~colours.}
   \label{fig:diffcol_vs_colxcorr}
\end{figure}

The extinction corrections $\Dex_F$ in the five filters are applied to the
values of $R'_{F4}$ from Eq.~\ref{eq:extinctionfilter}:
$R'_{F4} \to R'_{F4} + \Dex_F$ and fitted for each value of $R_V$ so as to
minimise $\chi_{\Dex}^2$, and cancel the coefficients $sl_i$ in
Fig.~\ref{fig:diffcol_vs_colx_nocorr}.  The result is given in
Table~\ref{tab:extinctionoffsets}.  As there is an overall scale degeneracy of
the extinctions in the colour analysis, we arbitrarily set the $U$~bandpass
correction to be 0.0675.  At each value $R_V$ of the grid, the corrections are
evaluated.  As these corrections (partially) compensate for the distortions
introduced by the difference between the `optimal $R_V$' and the value used in
the grid, they tend to blow up when $R_V$ is close to the limits of the range
considered in this work, namely as $R_V \to 1.95$ or $R_V \to 3.10$.  The
offsets are applied in the second line of Table~\ref{tab:extinctioncoeff}.  The
impact of these corrections on the correlation between the residuals and the
extinction is shown in Fig.~\ref{fig:diffcol_vs_colxcorr}, and for all colours
in Table~\ref{tab:extinction_diff}.  The large value of $sl_9$ observed
($e_9 \equiv R-I$) in Fig.~\ref{fig:diffcol_vs_colx_nocorr} reflects the
significant corrections to the extinction formula in bandpasses $R$ and $I$,
with opposite signs.  The errors given in these figures are only indicative, as
somewhat arbitrary `floor' errors of 0.006~mag are imposed on the determination
of the extinction colour.

\begin{table*}
  \caption{$UBVRI$ corrections (mag) to the extinction formula.  The number of
    decimals describes the convergence criterion, not the (lower) accuracy of
    the results.  The choice of the $U$ correction is arbitrary.}
  \label{tab:extinctionoffsets}
  \centering
  \begin{tabular}{l c r r r r c c c}
    \hline\hline
    $R_V$ & $\Dex_{U}$ & $\Dex_{B}$ & $\Dex_{V}$ & $\Dex_{R}$ & $\Dex_{I}$
    & $\gamma_{10}$ & $\gamma_{20}$ & $\gamma_{30}$ \\
          & (mag) & (mag) & (mag) & (mag) & (mag) & & & \\
    \hline
    2.20 & 0.0675 & $-0.03679$ & $-0.03923$ & 0.1013 & $-0.11126$ & 1.2687 & 1.1248 & 0.5579 \\
    2.25 & 0.0675 & $-0.04147$ & $-0.04351$ & 0.1015 & $-0.11090$ & 1.2647 & 1.1163 & 0.5395 \\
    \hline
  \end{tabular}
\end{table*}

The corrections to the extinction formula are extremely unlikely to be caused by
the limitations of our model: The linear approximation involved is numerically
correct to better than 0.005 mag in the whole range of $e_4 = E(B-V)$
considered.  There could be a second intrinsic colour involved; that is, rather
than one-dimensional (beyond Ca and Si), as in Eq.~\ref{eq:method_delta}, the
intrinsic colour space could be two-dimensional, with a second set of
$\gamma_{ij}$ coefficients, but intrinsic colours are uncorrelated (or are at
best weakly correlated) to extinction, and this new intrinsic colour would have
to be larger than the one already introduced to make up for the 0.10 mag
discrepancies observed in Fig.~\ref{fig:diffcol_vs_colx_nocorr}.  There is no
room in the residuals for such a large extra-intrinsic component.  On the other
hand, the need for an adaptation of the extinction formula to SNe and Top-hat
filters is shown in Sect.~\ref{sec:rescaling}, where the presence of a rescaling
factor is first introduced.

The numerical value of the corrections are dependent on the specific bandpasses
involved, on the spectral template, and also possibly on the dust properties of
our galaxy, which may differ from the averaged dust extinction.
Figure~\ref{fig:diffcol_vs_colxcorr} shows that, after correction, the remaining
correlations of the residuals $\epsilon_i$ with extinction are negligible.  This
can be achieved regardless of the value of~$R_V$.  The largest of these effects
amounts to a residual linear dependence of 0.006 for $B-V$.  The coefficients of
all other colours are smaller than 0.0036.  The statistical errors on the
offsets away from the extinction formula for our sample of SNe will be estimated
from a simulation in Sect.~\ref{sec:errors_in_simulation}.

It seems unnatural to constrain the $\gamma_{ij}$ coefficients with
$\chi^2_{\Dex}$, although they are (weakly) sensitive to the value of the
extinction corrections.  Our choice is instead to optimise the dispersion
$\sigma(\gamma)$ displayed in Fig.~\ref{fig:gamma_vs_gamma}, as stated in
Sect.~\ref{subsec:Intrinsic_gamma}.  At each value of the parameters ($\Dex_F$
or $\gamma_{i0}$), the $I_i$ and $X_i$ are obtained from Eq.~\ref{eq:XandI}, and
the $\gamma_{i0}$ are retuned in order to minimise $\sigma(\gamma)$ in
Fig.~\ref{fig:gamma_vs_gamma}.  This sequence is repeated four or five times to
reach the final values of the seven unknown parameters (four~$\Dex_F$ and
three~$\gamma_{i0}$).  Within each cycle (extinction corrections $\Dex_F$ or
$\gamma_{i0}$ fixed), convergence is reached when $\chi^2_{\Dex}$ or
$\sigma(\gamma)$ increase whenever any one of the seven parameters is changed by
$10^{-3}$.  As the implemented model is still imperfect, the minima of
$\chi^2_{\Dex}$ and $\sigma(\gamma)$ are not perfectly compatible; the
differences between the two minima are nevertheless approximately ten times
smaller than the quoted errors.

\begin{table*}
  \caption{Residual (mag) dependence on the extinction colour $R_V= 2.25$ (after
    correction of the extinction formula).  The corrections are introduced in
    Table~\ref{tab:extinctionoffsets} and the residuals are displayed in
    Fig.~\ref{fig:diffcol_vs_colxcorr}.}
  \label{tab:extinction_diff}
  \centering
  \begin{tabular}{c c r c c r}
    \hline\hline
    Index & Colour & $\d\epsilon/\d X_3$ & Error ($\sigma$) & Residual ($\sigma$)
    & $\d\epsilon/\d X_3$ \\
    && no corr. & on slope (no corr.) & with corr. & with corr. \\
    \hline
    0 & $U-B$ &  0.1138   & 0.0235 & 0.0201  & $-0.0023$ \\
    1 & $U-V$ &  0.0319   & 0.0059 & 0.0114  & $-0.00127$ \\
    2 & $U-R$ & $-0.0498$ & 0.0054 & 0.0153  & 0.00400  \\
    3 & $U-I$ &  0.0194   & 0.0029 & 0.0089  & 0.00021 \\
    4 & $B-V$ &$-0.0273$  & 0.0272 & 0.0294 & 0.00493 \\
    5 & $B-R$ &$-0.0995$  & 0.0125 & 0.0250 & 0.00915 \\
    6 & $B-I$ &$-0.0107$  & 0.0050 & 0.0137 & $-0.00177$ \\
    7 & $V-R$ &$-0.2293$  & 0.0286 & 0.0248 & 0.02164  \\
    8 & $V-I$ &$-0.0079$  & 0.0086 & 0.0127 & $-0.00035$ \\
    9 & $R-I$ &  0.0370   & 0.0271 & 0.0150 & 0.01268 \\
    \hline
  \end{tabular}
\end{table*}

\section{Quality of the colour reconstruction}
\label{sec:quality}

\begin{figure}
  \includegraphics[width=\columnwidth]{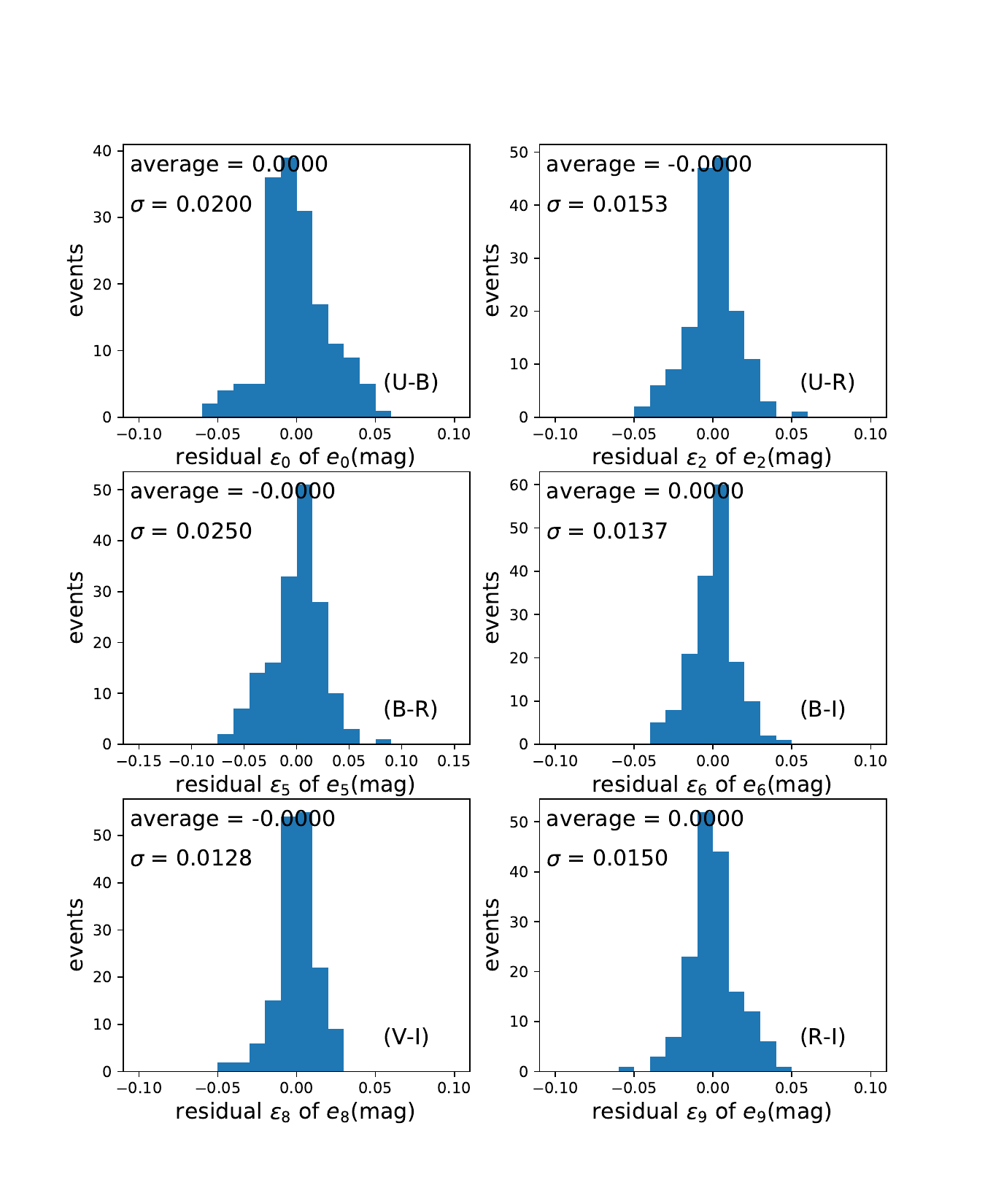}
  \caption{Residuals $\epsilon_i$ (mag) after the reconstruction of six colours
    $(c_0, c_2, c_5, c_6, c_8, c_9)$.  The full list of residuals is given in
    Table~\ref{tab:extinction_diff}.}
  \label{fig:hist_diffcol_all}
\end{figure}

The quality of the reconstruction of the colour differences $e_i$ from its
intrinsic $I_i$ and extinction $X_i$ components is shown in
Table~\ref{tab:colour_error} and in Fig.~\ref{fig:hist_diffcol_all}, where the
distribution of the difference $\epsilon_i = c_i -(I_i + X_i)$ is given for six
selected colours.  The average colour residual over the ten colours is 0.0176.
As the colour measurement errors do not exceed 0.006, this figure (together with
Table~\ref{tab:colour_error}) suggests that the dominant contributions arise
from modelling error and from the subtraction of the \ew{Si} and \ew{Ca}
contributions.  Given the accuracy reached, the phase of the spectrum might also
contribute.

\section{Determination of $R_V$}
\label{sec:RVslopes}

\subsection{Reddening correction of magnitudes}

\begin{figure}
  \includegraphics[width=\columnwidth]{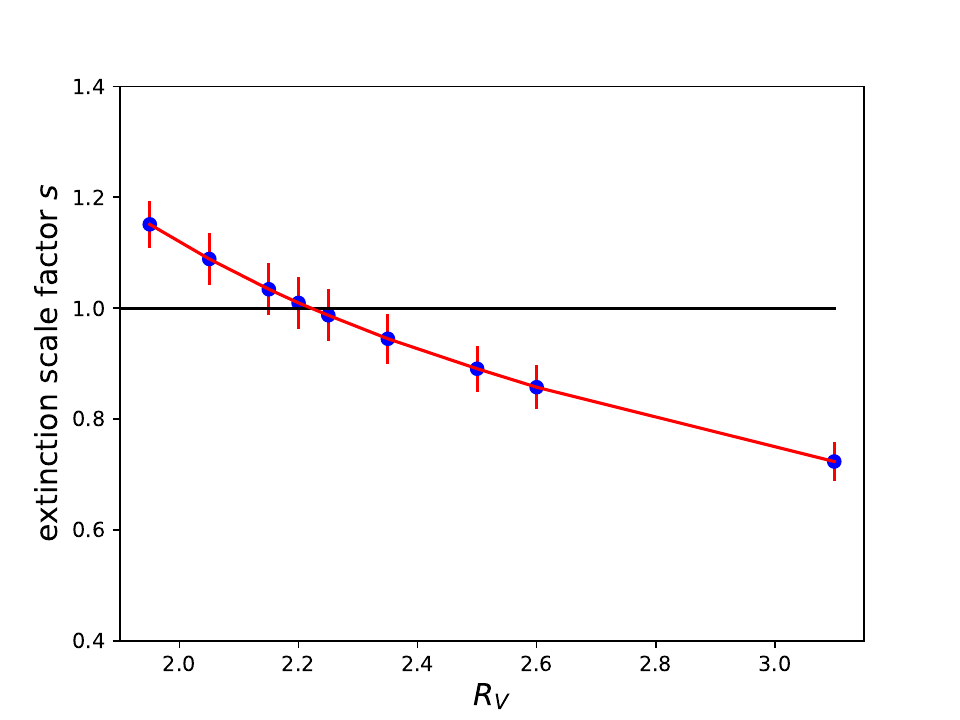}
  \caption{Extinction scale factor $s$ as a function of $R_V$ for
    $\d m_V/\d X_3 = 0$.  For each value of $R_V$, there is a value of the
    extinction scale factor $s$ that cancels the dependence of the $V$ magnitude
    on extinction.}
  \label{fig:scale_vs_RV}
\end{figure}

\begin{figure}
  \includegraphics[width=\columnwidth]{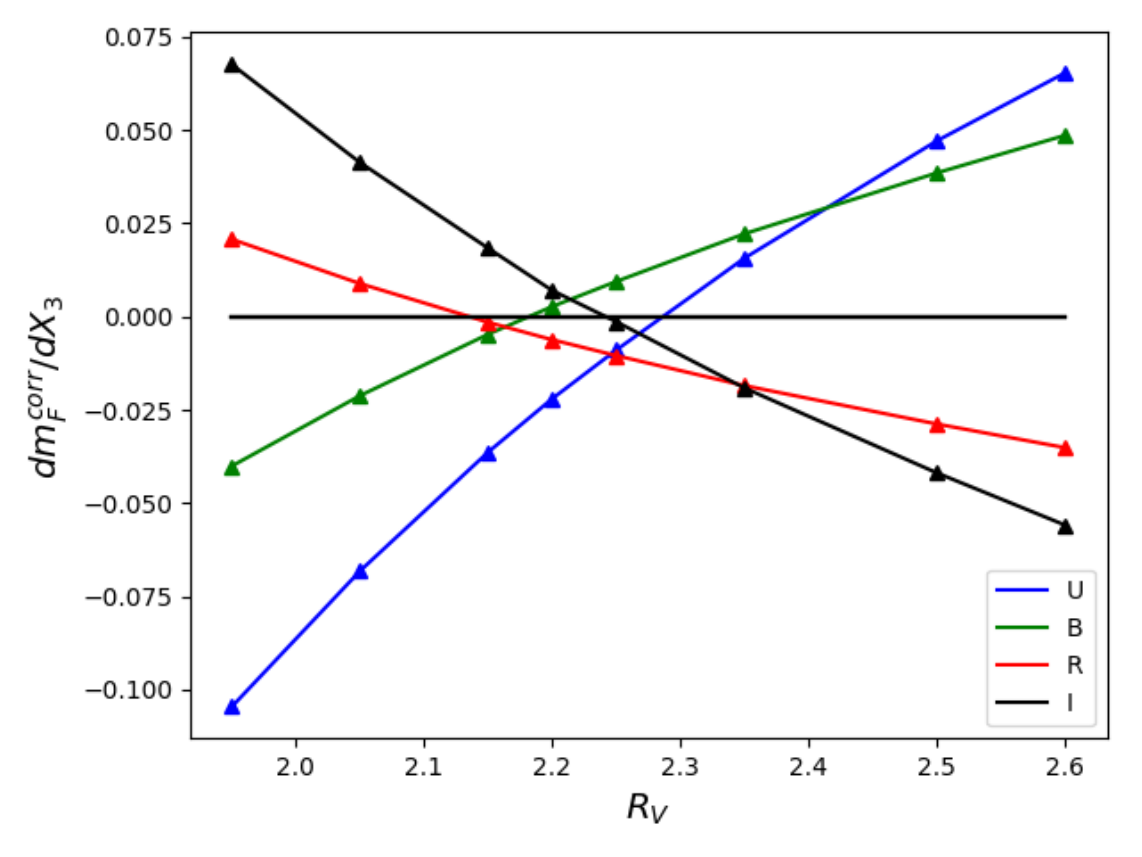}
  \caption{$\d m_{F}^{\text{corr}}/\d X_3$ for the $UBVRI$ bandpasses as a
    function of $R_V$.  Although the dependence of the $V$ bandpasses as a
    function of extinction has been cancelled for each value of $R_V$, the
    extinction in other bandpasses is not adequately corrected.}
  \label{fig:slope_vs_RV_DATA}
\end{figure}

\begin{figure}
  \includegraphics[width=\columnwidth]{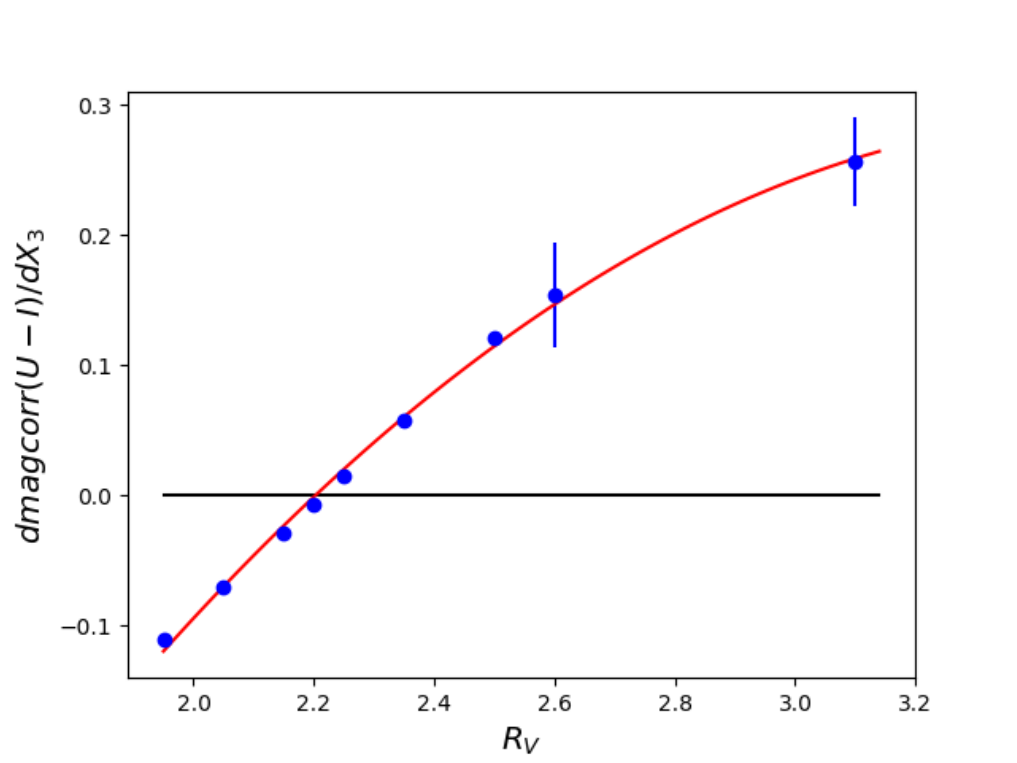}
  \caption{$\d (U-I)^{\text{corr}}/\d X_3$ as a function of $R_V$.  The
    difference between $U$ and $I$ magnitudes (colour~3) is the most sensitive
    to the choice of $R_V$.  The error bar is the dispersion found in the
    simulation.}
  \label{fig:slUI_vs_RV_DATA}
\end{figure}

\begin{figure}
  \includegraphics[width=\columnwidth]{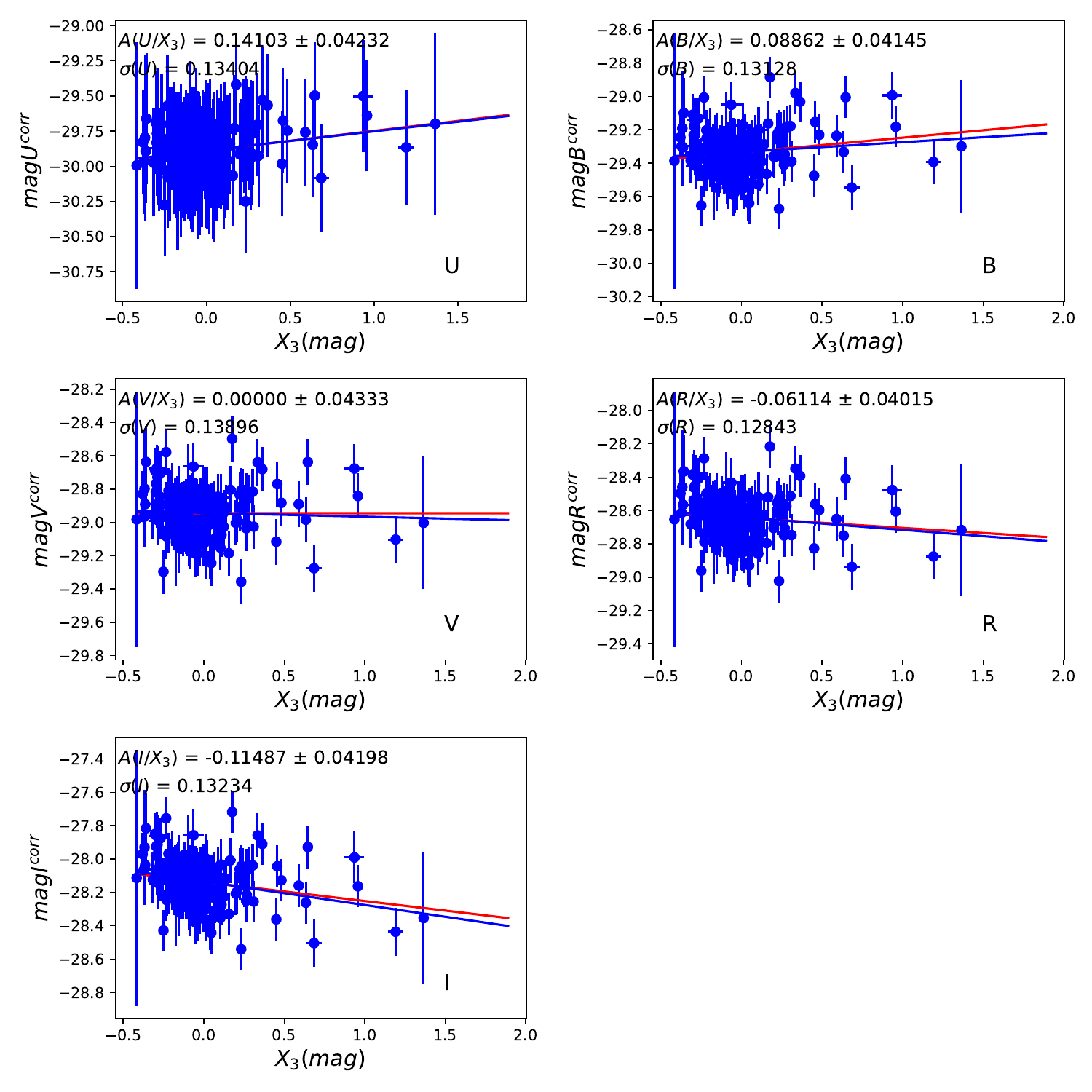}
  \caption{$UBVRI$ magnitudes corrected for $I_3$, $X_3$ with $R_V = 3.100$,
    $s = 0.73344$.  The derivative $\d\text{mag}_F^{\text{corr}}/\d X_3$ is
    compatible with zero in each bandpass, but the variation from $U$ to $I$ is
    quite significant, taking correlations into account.  The red line is
    weighted by errors and the blue straight line is unweighted.}
  \label{fig:mag_Int3corr310}
\end{figure}

In Sect.~\ref{sec:offsets}, we ensure that the colour residuals are independent
of the extinction component for each selected value of $R_V$.  However, after
correction for reddening, the magnitudes and the colours themselves may still
depend on the extinction ---even though the dependences in data and model must
be the same after the corrections of Sect.~\ref{sec:offsets}.  For each choice
of $R_V$, the magnitude $m_F$ found from the data at a reference redshift of
0.05 is corrected for its dependence on \ew{Si} and \ew{Ca} ---as carried out in
Sect.~\ref{sec:CaSi} for colours--- in order to obtain
$m_{F}^{\Ca\Si} = m_F - \alpha_{\Si} \ew{Si} - \beta_{Ca} \ew{Ca}$.  The
dependence of $m_{F}^{\Ca\Si}$ on the intrinsic component $I_i$ with the
(measured) coefficient $z_{F1}$ is then taken into account, and we allow for a
single overall scaling factor $s(R_V)$, which mimics the arbitrary $A_V$ used in
most analyses (it should be unity in our framework) to define the
extinction-corrected magnitude of each SN in the bandpass $F$ using colour~$i$:
\begin{equation}
  \label{eq:mag_Intcorr}
  m_F^{\text{corr}} = m_{F}^{\Ca\Si} - z_{F1} I_1 - s(R_V) \,R_{F1}(R_V) X_1.
\end{equation}
Once the extinctions have been scaled according to Eq.~\ref{eq:consistency2}, as
explained in Sect.~\ref{sec:rescaling}, the magnitude $m_F^{\text{corr}}$ should
not depend on $X_i$.  The extinction scale $s(R_V)$ in Eq.~\ref{eq:mag_Intcorr}
is chosen so as to cancel the dependence of the corrected {data} magnitude in
the $V$ (top-hat) bandpass, and we expect $s(R_V) = 1$ for the physical value of
$R_V$.  It is seen in Fig.~\ref{fig:scale_vs_RV} that this is true when
$R_V \approx 2.25$.  The extinction in filter $F$ is now
$A_F = s(R_V) \times R_{F3} \times X_3$.  However, if we turn to the other
filters, we see in Fig.~\ref{fig:slope_vs_RV_DATA} that even after the extra
scaling factor $s(\lambda)$ has been introduced in Eq.~\ref{eq:mag_Intcorr} to
ensure $\d\text{mag}_V^{\text{corr}}/\d X_3 = 0$, the derivatives of the other
bandpasses are finite when $R_V$ lies away from the $2.15-2.3$ range.  The
dependence of $\d m_F^{\text{corr}}/\d X_3$ as a function of $R_V$ is shown for
the bandpasses $UBRI$ in Fig.~\ref{fig:slope_vs_RV_DATA}.  We find the
extinction scale $s$ to be $s(R_{V} = 2.20) = 1.0269$ and
$s(R_{V} = 2.25) = 1.0014$ (close to our `effective' value of $R_V$).  The
errors in Figs.~\ref{fig:scale_vs_RV} to \ref{fig:mag_Int3corr310} are strongly
correlated over the range of wavelengths, and for different values of $R_V$ as
the same events are used.  The error is evaluated from the dispersion of the
scale factor in the ten samples of the simulation.  From sample to sample, the
scale parameter moves up and down in Fig.~\ref{fig:scale_vs_RV}, but the smooth
behaviour is preserved as a function of $R_V$ as the same events are involved
for a given sample.

Each bandpass provides a value of the parameter $R_V$ that cancels the
derivative $\d m_F^{\text{corr}}/\d X_3$.  For $UBRI$, these values are
respectively 2.283, 2.183, 2.137, and 2.240.  We average the $U$ and $I$
measurements ---which have a stronger $R_V$ dependence--- and obtain (with the
error from the simulation):
\begin{equation}
\label{eq:average_U-I}
  R_V = 2.262 \pm 0.126.
\end{equation}
The scale factor $s$ for this value of $R_V$ is $s = 1.001 \pm 0.045$, which is
additional independent confirmation of the result for $R_V$.  We used magnitudes
rather than colours to derive the previous result, but it can be reframed as a
colour measurement: Figure~\ref{fig:slope_vs_RV_DATA} shows that the largest
mismatch in the value of $\d m_F^{\text{corr}}/\d X_3$ is obtained by comparing
the $U$ and $I$ bandpasses.  We take advantage of this observation in
Fig.~\ref{fig:slUI_vs_RV_DATA} by imposing the condition that the derivative of
the $U-I$ colour with respect to the extinction colour $X_3$ be zero after
application of the extinction correction.  The value of~$R_V$ obtained in
Fig.~\ref{fig:slUI_vs_RV_DATA} is now $R_V = 2.265$, which is almost the same as
in Eq.~\ref{eq:average_U-I}, but the error is slightly smaller, as expected; the
contributions from flux calibration error (0.03~mag), the redshift error, and
the `grey' fluctuation are suppressed.  With this determination, the
corresponding scale factor for the `extinction scale' $s = 1.000$.  If we turn
to the usual `operational' definition of $R_V$, as $ A_V/E(B-V)$, it is seen in
Table~\ref{tab:extinctioncoeff} that $R_V = A_V/X_4 = R_{V4} = 2.7$.  However,
this numerical value is specific to SNe and our top-hat filters.

\begin{figure}
  \includegraphics[width=\columnwidth]{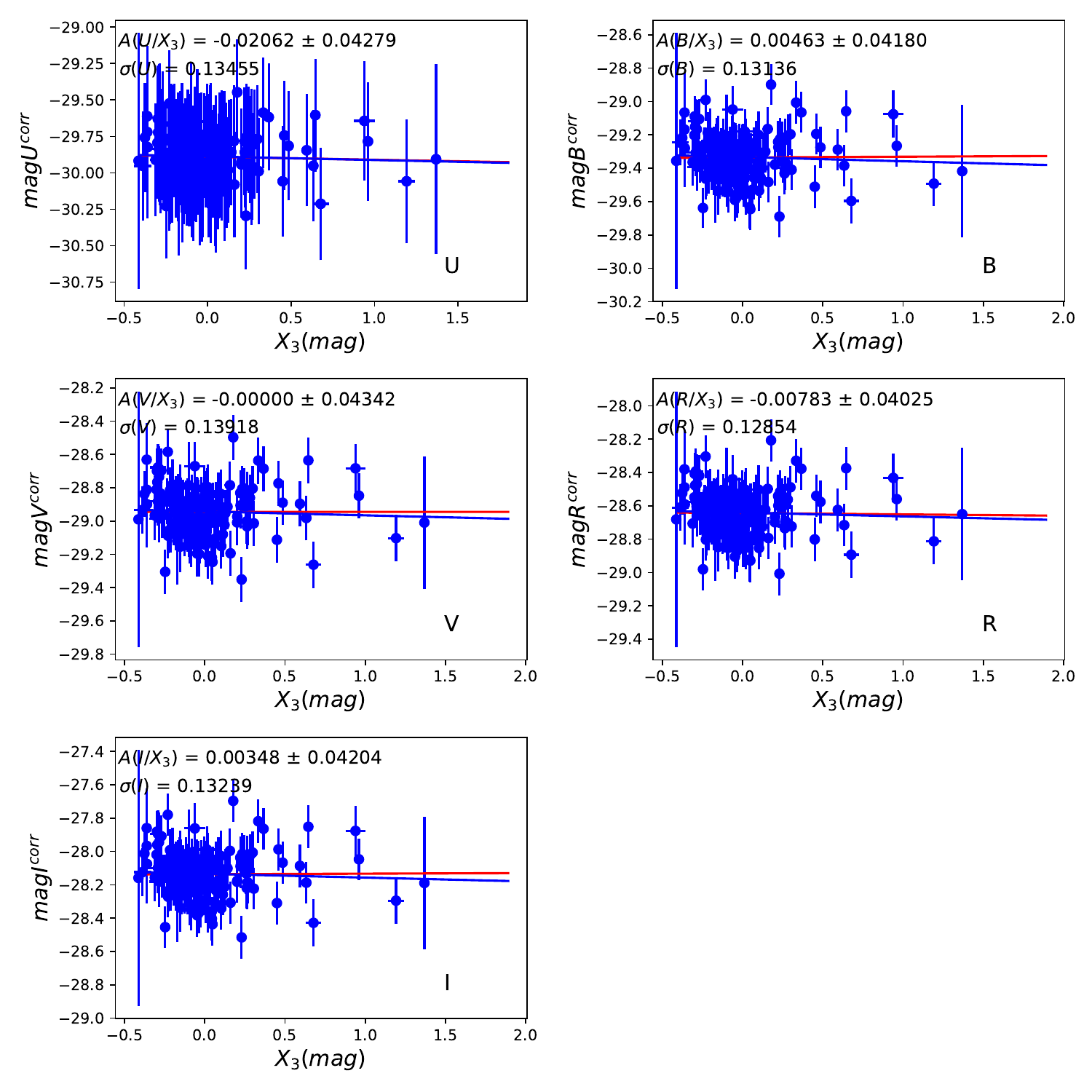}
  \caption{$UBVRI$ magnitudes corrected for $I_3, X_3$ with $R_V = 2.25$,
    $s = 1.0024$.  As for the previous figure, the red straight line is weighted
    by errors.}
  \label{fig:mag_Intcorr}
\end{figure}

The corrected magnitudes in the $UBVRI$ bandpasses are shown in
Fig.~\ref{fig:mag_Intcorr} for the value $R_V = 2.25$.  As expected, no residual
dependence of the (corrected) magnitude on the extinction is observed in any
filter.  The RMS (which was not minimised), has a value of 0.13~mag in all
filters, which is smaller than the scatter of 0.15~mag typically seen in the
SALT2 analyses of our sample \citep{Saunders2018}, but larger than the result
from refined analyses designed to minimise this fluctuation, as in
\cite{Boone2021}.  The blue straight line in Fig.~\ref{fig:mag_Intcorr} is a
weighted fit and the red line is unweighted; the two slopes are almost
identical.  The magnitude fluctuation is `grey', that is, it is almost identical
over all bandpasses to a remarkable accuracy.  For comparison, we show the
magnitude correlations obtained for $R_V = 3.1$ in
Fig.~\ref{fig:mag_Int3corr310}.  The scale factor must now be set to $s = 0.857$
to ensure that the derivative $\d m^{\text{corr}}_V/\d X_3 = 0$.  The error on
the derivative of $\d (m_U-m_I)/\d X_3$ in Fig.~\ref{fig:slUI_vs_RV_DATA} is
0.04, as evaluated from the spread of the simulation samples, meaning that the
significance of the rejection of $R_V = 3.1$ (and higher) is actually 3.5
standard deviations.

\subsection{Deriving $R_{V}$ from the grey fluctuation}

\begin{figure}
  \includegraphics[width=\columnwidth]{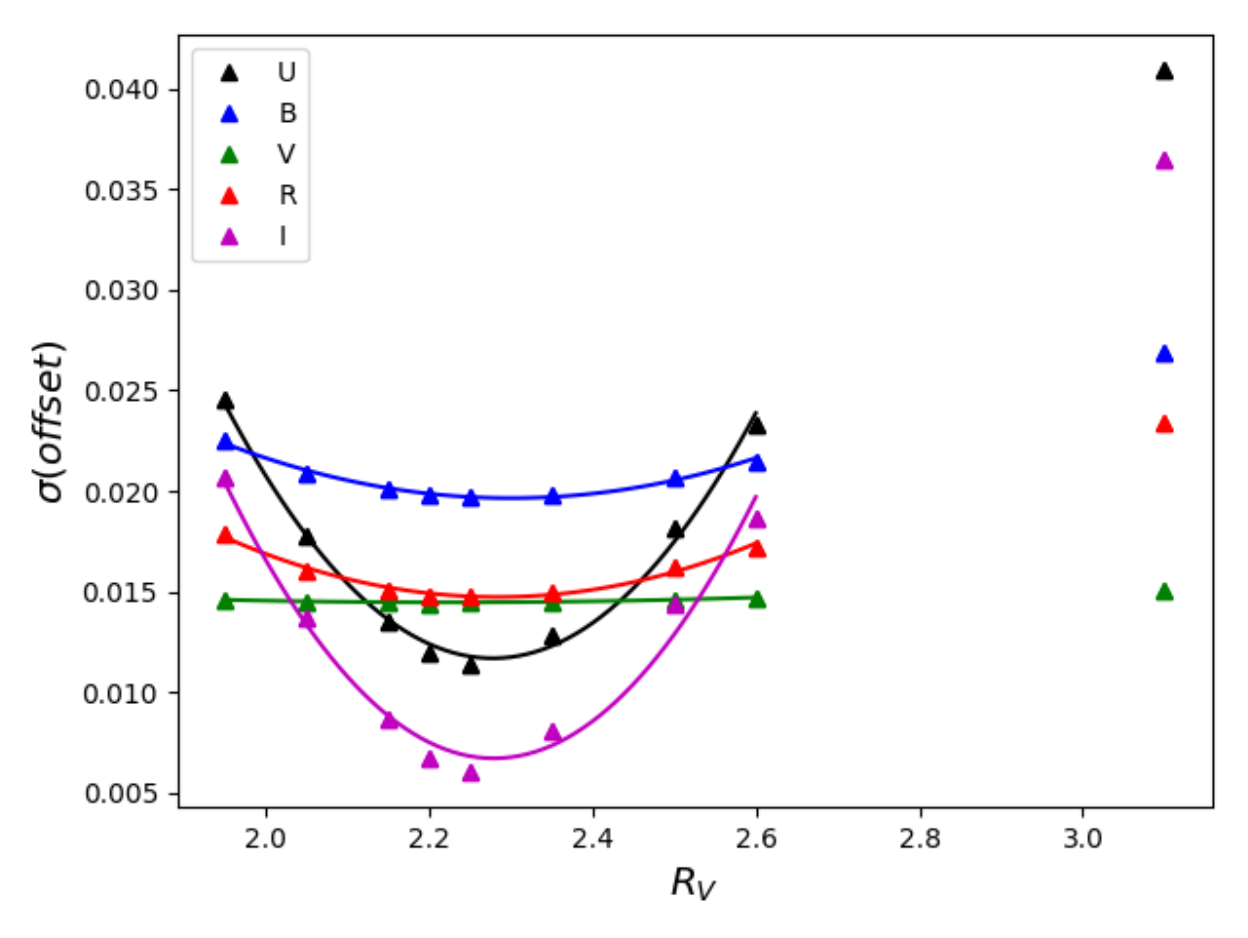}
  \caption{RMS of the offsets (with respect to $m_{\text{grey}}$) of
    the corrected $UBVRI$ magnitudes as a function of $R_V$.}
  \label{fig:offsets-filters}
\end{figure}

We mention in the previous section that the fluctuation of the magnitudes around
the common `grey' fluctuation of each SN~Ia $i$,
$m_{\text{grey}}(i) = \left( m_U^{\text{corr}}(i) + m_B^{\text{corr}}(i) +
  m_V^{\text{corr}}(i) + m_R^{\text{corr}}(i) + m_I^{\text{corr}}(i) \right)/5$,
was remarkably small.  This observation can be turned around and used to derive
$R_V$: in each bandpass $F$, we compute the RMS $\sigma_F$ of the dispersion of
the corrected magnitudes around $m_{\text{grey}}$.  We show how this dispersion
varies with $R_V$ in Fig.~\ref{fig:offsets-filters}: a minimum is reached for
$R_V = 2.277$, which confirms the previous result.  The dispersion in each
$UBVRI$ bandpass with respect to the average offset over all filters is
remarkably small, that is 0.0114, 0.0197, 0.0145, 0.0148, and 0.0060,
respectively, which is everywhere smaller than~0.02.  This result quantifies the
level at which the remaining magnitude fluctuation is grey, and also confirms
the validity of the modelling implemented (up to this `unaccounted' grey
fluctuation).

\section{Simulation}
\label{sec:simulation}

The previous results can be summarised by ten numbers: the five extinction
corrections to the extinction formula in the five bandpasses used for our
analysis, the three intrinsic colour ratios, the extinction scale $s$, and the
extinction parameter $R_V$.  These are the `effective' values found within our
algorithms, but not necessarily the `true' values.  The biases and the errors
are evaluated below based on a simulation that mimics the observations.  We use
the values of the parameters tuned to describe the data to generate the events,
and then proceed with the same analysis.  The differences from the reconstructed
values allow us to find the methodological biases, and the scatter among the
different generated samples helps us find the statistical error.

\subsection{Generation of colours with their error}
\label{sec:colors}

We generate nine samples of 165 SNe~Ia $UBVRI$ magnitudes.  The generated
intrinsic $I^g_1$ and extinction $X^g_1$ components of colour $1 \equiv U-V$
follow the observed distribution in the data after smoothing by a sliding local
averaging (to avoid counting the statistical fluctuations twice): in each
histogram bin, the data distribution of $I_1$ and $X_1$ was replaced by the
average of three adjacent bins.  Whenever the bin statistics was below four
events, the average of five bins was used.  The other intrinsic colours
$I^g_{k}$ obey $I^g_{k} = \gamma_{k1} I^g_1$, and the extinction colours
$X^{g}_{k}$ follow from $X^{g}_{k} = \delta_{k1}X^{g}_{1}$.  The coefficients
$\delta_{ij}$ used in the generation are the ones found in the data analysis;
that is, they include the correction to the \citetalias{Fitzpatrick1999} formula
described in Table~\ref{tab:extinctionoffsets}.  The simulation shows that the
analysis does recover the input corrections, with a small bias.  As the
measurement errors are included in the observed data distribution, they should
have been removed from the generation in the simulation.  We instead introduced
a single simulation scaling factor of the $I^g_{1}$ distribution, tuned so that
the reconstructed intrinsic components should match the observations once the
analysis is applied; it depends on the simulated sample, but its difference from
unity never exceeds~0.01.  This effect is expected to be smaller for the
extinction colours $X_i$, which have a larger range, and the observed
distribution of the data has been used.

\subsection{Colour noise}
\label{sec:colour-noise}

The noise is derived from the residuals observed in the data, and it combines
measurement and modelling errors.  The residuals $\epsilon_i$ measured from the
data in four colours (i.e. $2 \equiv U-R$, $5 \equiv B-R$, $7 \equiv V-R$,
$9 \equiv R-I$) are used to derive a $4\times 4$ covariance matrix
$\moy{\epsilon_i \epsilon_j}$.  The square roots of the dominant eigenvalues are
$\sigma_1 = 0.026525$, $\sigma_{2} = 0.017947$; the other eigenvalues are
negligible ($\sim 10^{-4}$ and $\sim 3\,10^{-5}$).  The Gaussian noise of the
two eigenvectors is then projected onto the four bandpasses to obtain a
simulated `colour noise' for the four colours.  We find that this noise has to
be multiplied by 1.004 in the simulation to reproduce the observed residuals.
This effect may arise from the contribution of modelling errors to the data,
while the model is exact in the simulation.

\subsection{Generation of magnitudes}
\label{sec:magnitudes}

The $R$ magnitude is used as the reference.  A correction reproducing its
(small) correlation with the intrinsic colour is added.  The magnitudes in the
$UBVI$ bandpasses are obtained by adding the $U-R$, $B-R$, $V-R$, and $I-R$
colours, as found from their intrinsic and extinction components, and including
the `colour noise'.  Finally, a Gaussian grey magnitude fluctuation with
$\sigma(grey) = 0.12$ is added, though its value is irrelevant for all the
results of the present study.  The generated magnitudes of event $n$ in bandpass
$R$ and $U$ are therefore described as
\begin{align}
  \label{eq:generation_magnitudes_R}
  m_R(n) &= z_{R1}I_1(n) + R_{R1}X_1(n) + \text{grey}(n),\\
  \label{eq:generation_magnitudes_U}
   m_U(n) &= m_R(n) + \gamma_{21}I_1(n) + \delta_{21} X_1(n) +
  \sigma_1 u_1(n)\,n_{21} + \sigma_2u_2(n)\,n_{22},
\end{align}
where $u_1(n)$ and $u_2(n)$ are centred random Gaussian variables with a
standard deviation of unity, $n_{21}$ and $n_{22}$ are the components of the
corresponding eigenvectors on the $U-R$ colour, and $z_{R1}$ is the observed
correlation between the $R$ magnitude and the intrinsic colour~1.  The generated
magnitudes are processed by the same algorithms as the data (after \ew{Ca} and
\ew{Si} corrections), providing reconstructed values for $I_i$ and $X_i$,
reconstructed extinction corrections, reconstructed intrinsic couplings, and a
reconstructed $R_V$.

\subsection{Simulated and observed distributions}
\label{sec:errors_in_simulation}

\begin{figure}
  \includegraphics[width=\columnwidth]{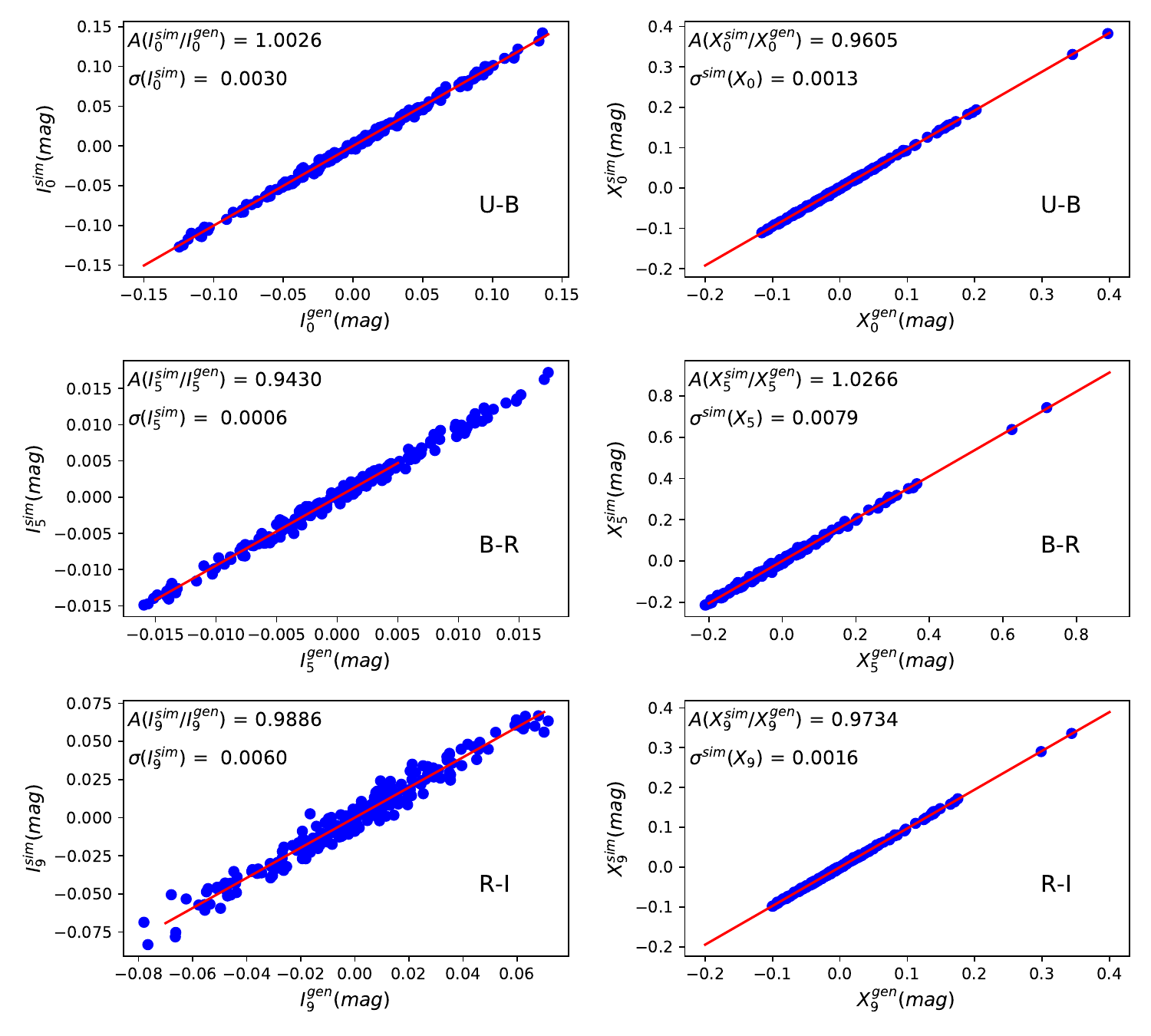}
  \caption{Comparison of intrinsic and extinction components (mag) in
    generation and reconstruction for colours $c_0$, $c_5$, and $c_9$.  The small
    differences found justify using the (smoothed) observed distributions of
    $I_1$ and $X_1$ as input.}
  \label{fig:Int_Intgen}
\end{figure}

\begin{figure}
  \includegraphics[width=\columnwidth]{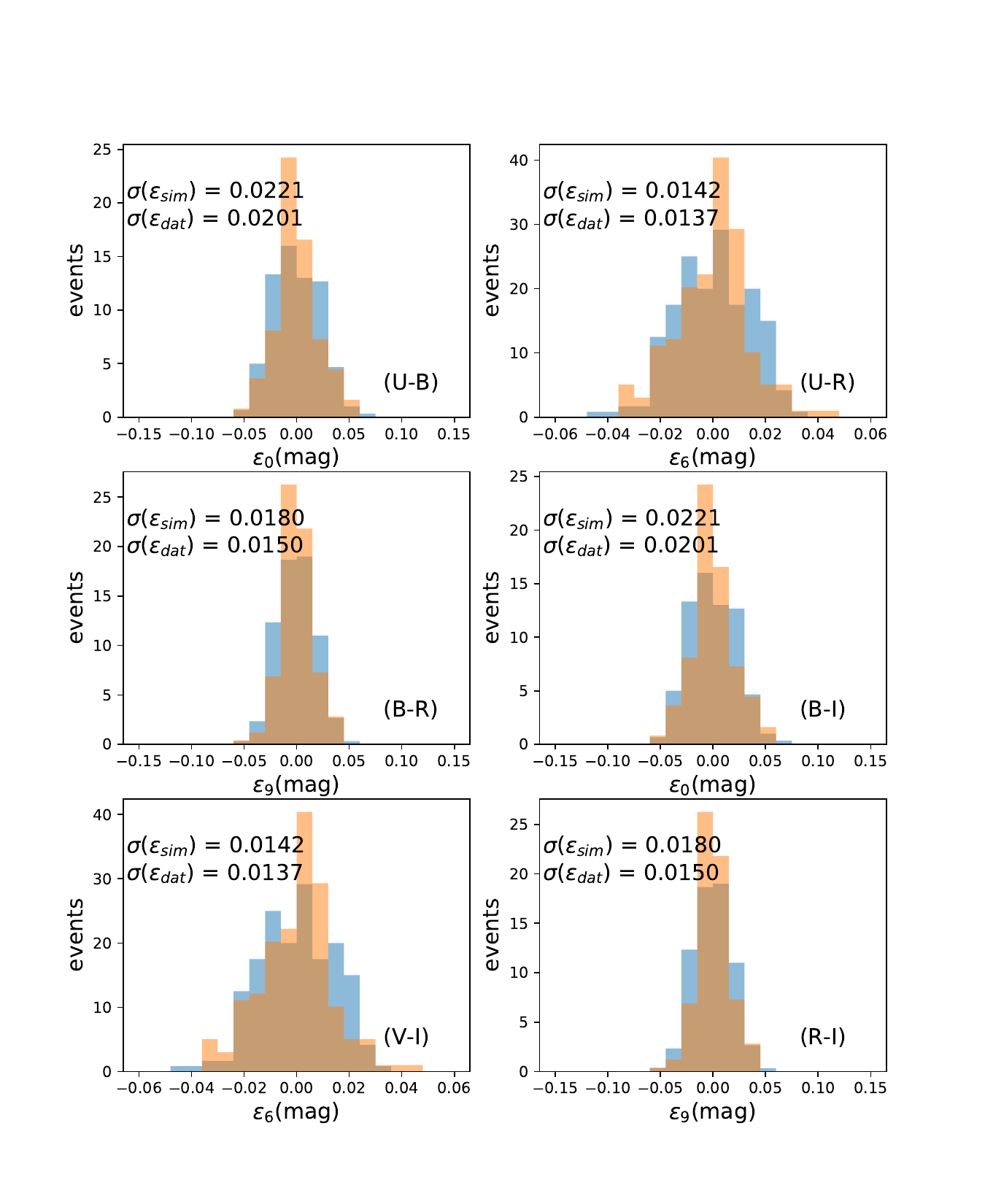}
  \caption{Residuals in observations and simulations for colours $c_0$, $c_2$,
    $c_5$, $c_6$, $c_8$, and $c_9$ ($R_V = 2.25$).  The RMS of the residuals of
    all colours are given in Table~\ref{tab:colour_error}.  The blue shade is the
    simulated distribution, the orange shade the data.}
  \label{fig:hist_diffcolsim}
\end{figure}

As we want to evaluate the errors and biases of the data analysis, with the help
of the simulation, both the distributions and the errors must be similar in the
observations and the simulation.  The $(I_i, X_i)$ obtained through the colour
analysis (as in the data), and the generated $(I^g_i, X^g_i)$ values are
compared for three colours in Fig.~\ref{fig:Int_Intgen} for the simulated sample
1 and in Table~\ref{tab:colour_error}.  There is a very small scatter between
the generated and reconstructed colour components, and their ratio is close to
unity.  This scatter varies from 0.002 to 0.006 for $I_i$, and from 0.001 to
0.006 for $X_i$.  The systematic trend in the ratio of measured and generated
colour is partially statistical and fluctuates from one sample to another in a
range of 0.02, which justifies using the observed distributions as an input for
the simulation.  The scatter in Fig.~\ref{fig:Int_Intgen} is surprisingly
smaller than the RMS of the observed residuals (in average 0.0176~mag).  This is
a consequence of the averaging over different colours performed in
Eq.~\ref{eq:XandI}.  The simulated accuracy is tuned to reproduce the observed
residuals by the introduction of two ad hoc factors in the noise and the
rescaling of the distribution of $I_1$ (of colour $U-V$) used in the generation.
The distribution of the residuals is shown in Fig.~\ref{fig:hist_diffcolsim} and
Table~\ref{tab:colour_error}.

\begin{figure}
  \includegraphics[width=\columnwidth]{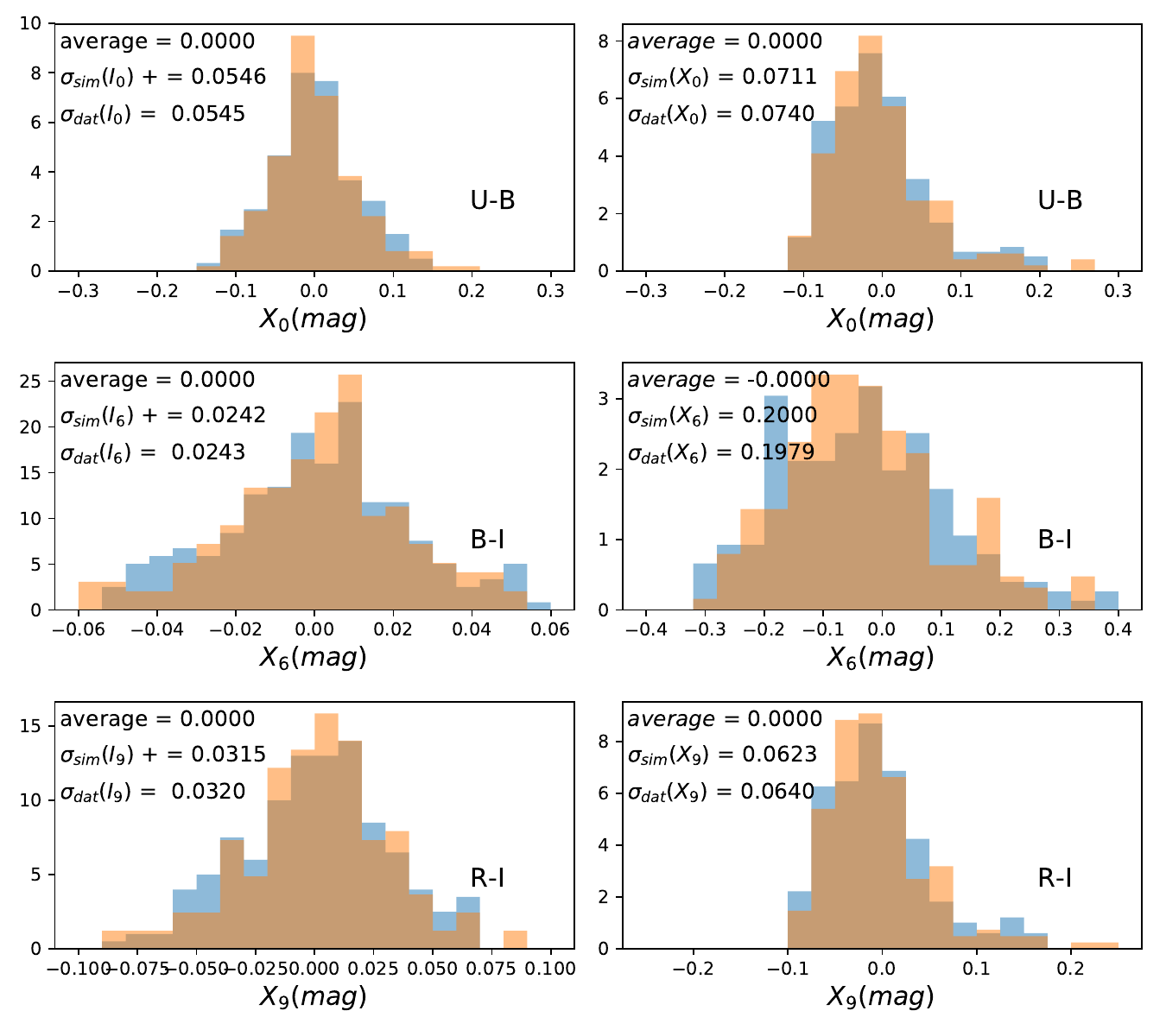}
  \caption{Distributions of intrinsic and extinction components of $c_0$, $c_6$,
    and $c_9$ in observation and simulation for $R_V = 2.25$ (mag).  The
    extinction is asymmetric, with an extended tail to high values.  The
    intrinsic part is symmetric.  The RMS for all colours is given in
    Table~\ref{tab:colour_error}.  The blue shade is the simulated distribution,
    the orange shade the data.}
  \label{fig:hist_Int_ext_DATA_sim}
\end{figure}

For the same three colours ($c_0$, $c_5$, $c_9$), the intrinsic and extinction
colour distributions for the data and one of the simulated samples are shown in
Fig.~\ref{fig:hist_Int_ext_DATA_sim}.  The results for all colours are
summarised in Table~\ref{tab:colour_error}.  The residuals in the data and the
simulation are compared in Fig.~\ref{fig:hist_diffcolsim} and
Table~\ref{tab:colour_error}.  The intrinsic and extinction colours in the
simulation and observation agree, and the trends of the errors are also well
reproduced by the simulation.

\begin{table*}
  \caption{Range of the intrinsic and extinction colours for observation,
    simulation (seed2), and residuals (mag).}
  \label{tab:colour_error}
  \centering
  \begin{tabular}{c c c c c c c c}
    \hline\hline
    Index & Colour & $\sigma(I_i)$ & $\sigma(I_i)$
    & $\sigma(X_i)$ & $\sigma(X_i)$ & $\sigma(\epsilon_i)$ & $\sigma(\epsilon_i)$ \\
          && Sim (mag) & Obs (mag) & Sim (mag) & Obs (mag) & Sim (mag) & Obs (mag) \\
    \hline
    0 & $U-B$ & 0.0544 & 0.0545 & 0.0681 & 0.0740 & 0.0216 & 0.0201 \\
    1 & $U-V$ & 0.0701 & 0.0692 & 0.1473 & 0.1560 & 0.0084 & 0.0114 \\
    2 & $U-R$ & 0.0625 & 0.0610 & 0.1996 & 0.2080 & 0.0169 & 0.0154 \\
    3 & $U-I$ & 0.0303 & 0.0304 & 0.2684 & 0.2719 & 0.0097 & 0.0089 \\
    4 & $B-V$ & 0.0149 & 0.0152 & 0.0791 & 0.0818 & 0.0250 & 0.0295 \\
    5 & $B-R$ & 0.0060 & 0.0069 & 0.1316 & 0.1339 & 0.0300 & 0.0250 \\
    6 & $B-I$ & 0.0239 & 0.0243 & 0.2003 & 0.1979 & 0.0141 & 0.0137 \\
    7 & $V-R$ & 0.0082 & 0.0082 & 0.0524 & 0.0519 & 0.0230 & 0.0248 \\
    8 & $V-I$ & 0.0390 & 0.0401 & 0.1210 & 0.1160 & 0.0094 & 0.0127 \\
    9 & $R-I$ & 0.0296 & 0.0320 & 0.0686 & 0.0640 & 0.0178 & 0.0150 \\
    \hline
  \end{tabular}
\end{table*}

\subsection{Errors on the extinction formula corrections}
\label{sec:simul_corrections}

We rely on the nine simulated samples to control the bias in the measurements
arising from the implemented algorithms, as well as the errors on the extracted
parameters: extinction corrections $\Dex_F$, the value of $R_V$, and the
extinction scale $s_X$.  The analysis was performed on each of the samples, on
the same grid of values of $R_V$.  The mean value of the reconstructed
extinction corrections and their dispersion is given in
Table~\ref{tab:extinction_corrections}.

\begin{table*}
  \caption{Extinction correction for $R_{V} = 2.25$.}
  \label{tab:extinction_corrections}
  \centering
  \begin{tabular}{c r r r r r}
    \hline\hline
    Filter
    & $\Dex_F$ obs & $\Dex_F$ sim & bias & $\sigma(\Dex_F)$ & $\Dex_F$ final \\
    & (= generation, mag) & (average, mag) & (average, mag) & (mag) & (mag) \\
    \hline
    $U$ & $ 0.0675$ & $ 0.0675$ & $ 0.0   $ & $0.0   $ & $ 0.0675$ \\
    $B$ & $-0.0415$ & $-0.0351$ & $ 0.0064$ & $0.0168$ & $-0.0479$ \\
    $V$ & $-0.0435$ & $-0.0433$ & $ 0.0002$ & $0.0120$ & $-0.0437$ \\
    $R$ & $ 0.1015$ & $ 0.0733$ & $-0.0282$ & $0.0342$ & $ 0.1297$ \\
    $I$ & $-0.1109$ & $-0.1390$ & $ 0.0280$ & $0.0498$ & $-0.0829$ \\
    \hline
  \end{tabular}
\end{table*}

When comparing to Table~\ref{tab:extinctioncoeff}, it is seen that the
corrections to the extinction are statistically significant, rising from 1 to 10
percent over the visible spectrum once the bias is taken into account.  The
algorithm used in deriving the corrections is able to extract their value, with
some biases.  Accepting the biases found in the simulation, our final values for
the corrections to the extinction coefficients are given in the last column of
Table~\ref{tab:extinction_corrections} for $R_V = 2.25$.
The error on the bias correction is $\sigma(\Dex_F)/3$ as nine samples are
generated.

\subsection{Errors on the intrinsic coupling coefficients $\gamma$}
\label{sec:errors-intrinsic}

For each simulation sample and each value of $R_V$ in the grid, the three
intrinsic coupling coefficients $\gamma_{10}$, $\gamma_{20}$, and $\gamma_{30}$,
are evaluated as in the data analysis, that is by minimising the scatter in the
plot comparing the input values to the results.  The standard deviation of the
three coefficients from one sample to the next is respectively 0.0075, 0.0061,
and 0.0041.  The values averaged over all samples allow us to estimate the bias
arising from the analysis; it is found to be of the same order as the
statistical error.  The largest contribution to the error is by far the one
discussed in Sect.~\ref{subsec:gamma}, which arises from different choices of
auxiliary colours, and is given in Table~\ref{tab:gammaij}.  The statistical
error will be added into the final uncertainty.

\subsection{Errors on the extinction parameter $R_V$}
\label{sec:errors-Rv}

The measurement of $R_V$ is also performed on the simulation.  The value in the
generation was $R_V = 2.20$, and the average value found is
$R_V = 2.284 \pm 0.112$.  We correct the result obtained in the data analysis,
of 2.265, to $R_V = 2.265-0.084 = 2.181 \pm 0.117$ (the error is increased to
account for the uncertainty on the bias correction).  As the values found in the
four filters $BVRI$ are almost fully correlated, we use the error on $R_V$
derived from the $U-I$ colour.  We add the impact of an estimated systematic
error in the measurement of colours and other corrections (earth atmosphere,
galactic extinction, host galaxy subtraction, silicon and calcium contribution).
As we only use differences between mean and measured colours, many instrumental
systematic errors cancel out.  To evaluate the impact on $R_V$, all colours were
modified by a shift proportional to their value and reaching 0.005 mag over
their range of variation.  The outcome is a change in $R_V$ by 0.05.  The final
result for the parameter $R_V$ (including the error on the bias) is then
\begin{equation}
  \label{RV_final}
  R_V = 2.181 \pm 0.117\;(\text{stat.}) \pm 0.050\;(\text{syst.}).
\end{equation}

\section{No calcium or silicon line corrections}
\label{sec:noCaSi}

\begin{figure}
  \includegraphics[width=\columnwidth]{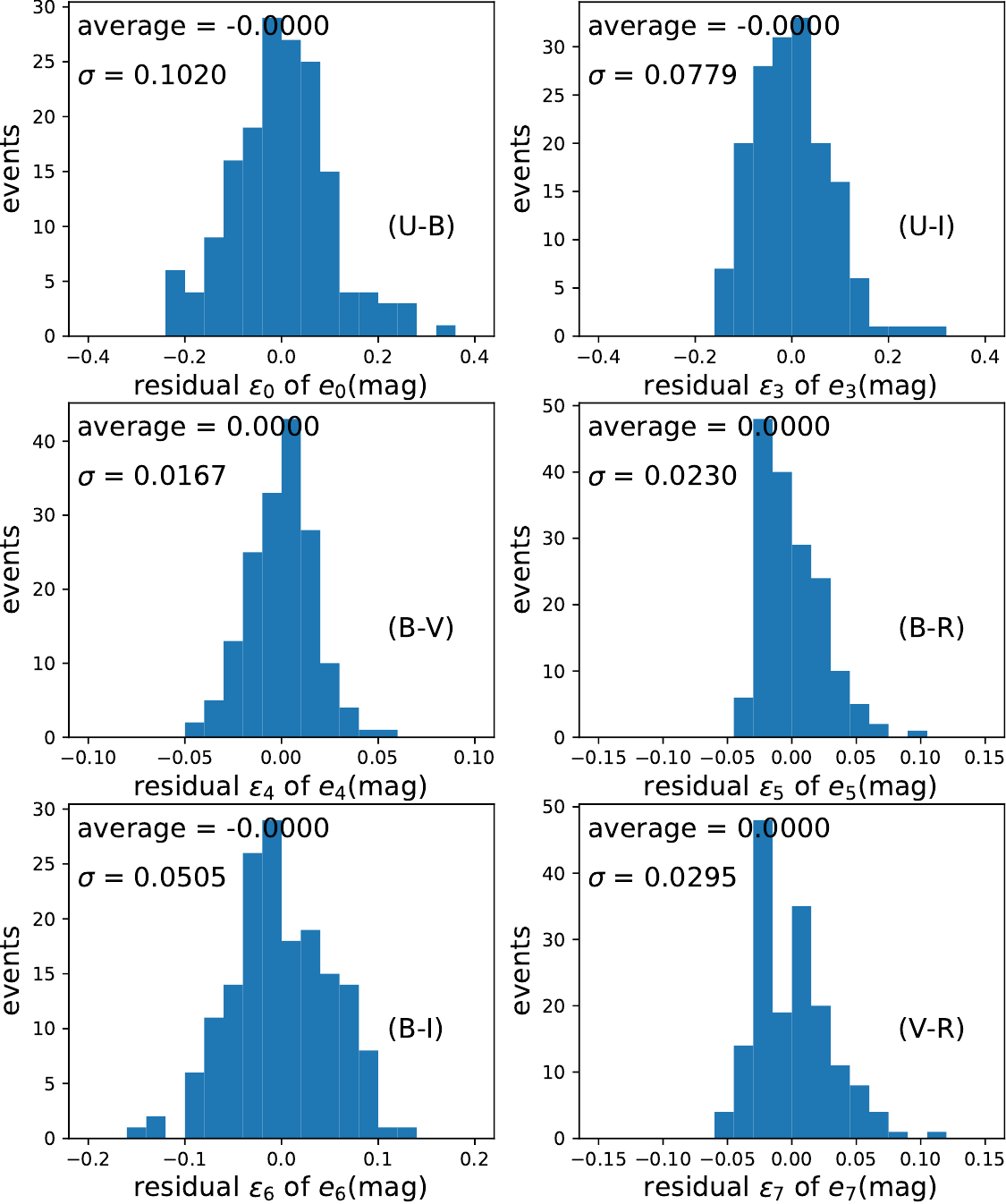}
  \caption{Distributions of six `full' intrinsic colours including \SiII{} and
    \CaII~contributions.}
  \label{fig:hist_Intfull}
\end{figure}

We tried to reproduce our analysis without performing the initial correction for
the \ew{Si} and \ew{Ca} spectral lines.  The intrinsic colours are expected to
be much larger than in the previous analysis.  They reach 0.10 in colours
($U-B$, $U-V$, $U-R$).  The mean colour residual is then 0.0402 (instead of
0.0176 with the colour corrections from \ew{Ca} and \ew{Si}).  The square roots
of the dominant eigenvalues of the covariance matrix from the residuals are
0.0724 and 0.0296, meaning that it would be necessary to add two spectral
coordinates for each SN in order to reduce these residuals, instead of
tolerating these contributions as `noise' as is done in the present work.  The
two eigenvectors are $(0.5842, 0.5974, 0.4373, -0.3324)$ and
$(0.5972, -0.7405, 0.2908, 0.1013)$.  In the absence of \ew{Si} and \ew{Ca}
corrections, the model we used fails to describe the observations, as we are
missing key spectral information.  Instead, the approach we take is to add back
the colour contributions of \SiII~$\lambda$4131 and \CaII~$\lambda$3945 ---which
are subtracted in Sect.~\ref{sec:CaSi}--- to obtain a `full' intrinsic component
of colour variability; this is shown in Fig.~\ref{fig:hist_Intfull}.

The range of the full intrinsic colour component is given in
Table~\ref{tab:Intfull} for all colours.  The small size of the intrinsic part
of $e_4 \equiv B-V$ justifies using it as a simplified indicator of extinction
for SNe~Ia.

\begin{table}
  \caption{Range of the full intrinsic colour component and correlations between
    the full calcium--silicon colour correction $\DCaSi$ and the SALT $x_1$
    variable ($R_{V} = 2.25$).}
  \label{tab:Intfull}
  \centering
  \begin{tabular}{c c c r c}
    \hline\hline
    Index & colour & $\sigma_{\text{full}}$ & $\d\DCaSi/\d x_1$ & $\sigma(\DCaSi)$ \\
    \hline
    0 & $U-B$  & 0.1020 &  $-0.0300$ & 0.0808\\
    1 & $U-V$  & 0.1088 &  $-0.0246$ & 0.0802 \\
    2 & $U-R$  & 0.1170 &  $-0.0482$ & 0.0873 \\
    3 & $U-I$  & 0.0780 &  $-0.0495$ & 0.0518 \\
    4 & $B-V$  & 0.016 &   0.0054    & 0.0035 \\
    5 & $B-R$  & 0.0230 & $-0.0182$  & 0.0121 \\
    6 & $B-I$  & 0.0505 & $-0.0194$  & 0.0397 \\
    7 & $V-R$  & 0.0295 & $-0.0235$  & 0.0155 \\
    8 & $V-I$  & 0.0627 & $-0.0248$  & 0.0411 \\
    9 & $R-I$  & 0.0519 & $-0.0013$  & 0.0407 \\
    \hline
  \end{tabular}
\end{table}

\begin{figure}%
  \includegraphics[width=\columnwidth]{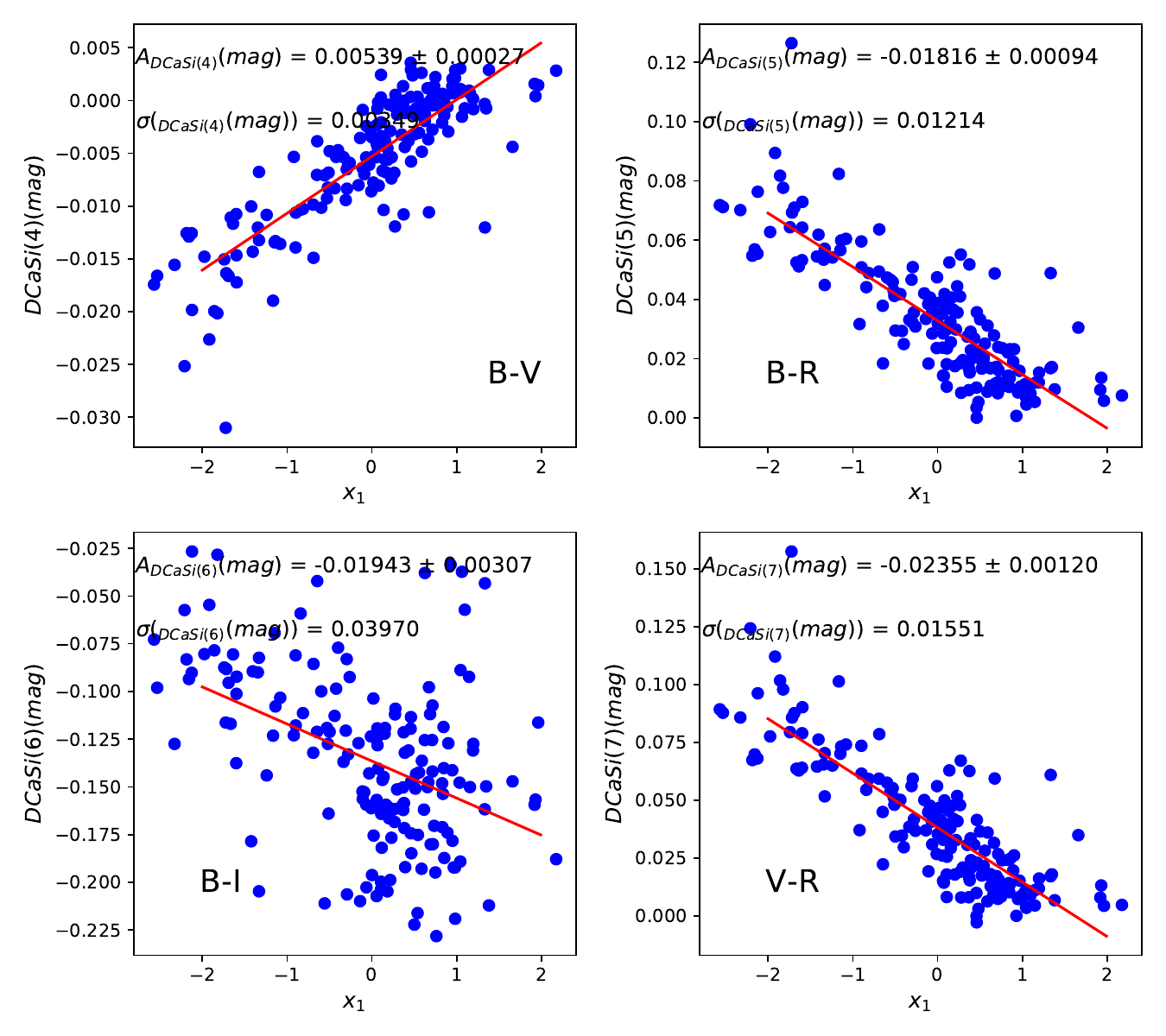}
  \caption{Correlations between the combined calcium--silicon correction
    $\DCaSi$ and the SALT2 variable $x_1$ for four colours.}
  \label{fig:DCaSi_vs_xx1}
\end{figure}

The standard choice of colour $4 \equiv B-V$ as an extinction indicator is
fortunate, as the intrinsic content happens to be small.  However, a large part
of the calcium and silicon correction can be recovered, with information
extracted from the light curve.  The SALT2 stretch variable $x_1$, as described
in~\cite{Betoule2014}, is strongly correlated to the combined calcium--silicon
corrections $\DCaSi(i) = \sigma_i\ew{Si} + \kappa_i \ew{\Ca}$, as shown in
Fig.~\ref{fig:DCaSi_vs_xx1} and in Table~\ref{tab:Intfull} for all colours.  The
dispersion of the full intrinsic colour can be compared to the results of
previous determinations: $\sigma_{BV}^{\text{int}} = 0.049$ \citep{Jha2007},
$\sigma_{BV}^{\text{int}} = 0.03 \pm 0.01$;
$\sigma_{VR}^{\text{int}} = 0.04 \pm 0.01$,
$\sigma_{RI}^{\text{int}} = 0.06 \pm 0.01$ \citep{Nobili2008},
$\sigma_{BV}^{\text{int}} = 0.065 \pm 0.008$ \citep{Mandel2017},
$\sigma_{BV}^{\text{int1}} = 0.076 \pm 0.011$,
$\sigma_{BV}^{\text{int2}} = 0.043 \pm 0.013$ \citep[2populations/decline
rate,][]{Wojtak2023}.  The values of $\sigma_{BV}^{\text{int}}$ found in this
work for the two populations of the $H_{\alpha}$ subsample (galaxies with high
and low $H_{\alpha}$ signal, as discussed in item 10 of
Sect.~\ref{sec:conclusions}) are: $\sigma_{BV}^{\text{int1}} = 0.0127$,
$\sigma_{BV}^{\text{int2}} = 0.0149 $.

The smallest scatter of $\DCaSi$ is found for colour $4 \equiv B-V$, with
$\sigma = 0.00349$.  This provides an estimate of the contribution of
measurement errors to colours, all colours being processed in the same way.
Extra physical variabilities worsen the evaluation of the $\DCaSi$ contribution
of other colours.

\section{Conclusions}
\label{sec:conclusions}

The focus of this work is the extraction for each SN of our sample of extinction
and intrinsic colour components, using the extinction formula for dust in our
galaxy as leverage.  This colour analysis is well adapted to the determination of
the reddening, but does not make use of the full power of the spectral
information for the intrinsic part.  The linearity that we find in the
correlations between different colours is nevertheless an interesting feature by
itself.
\begin{enumerate}
\item The model presented in this work assumes the factorisation of the
  extinction by the host galaxy, without any correlation with the intrinsic
  properties of the SN.  We acknowledge that the presence of circumstellar dust
  might invalidate this assumption.

\item The modelling of colour fluctuations requires the inclusion of four
  corrections to the extinction formula in the $BVRI$ bandpasses, three
  intrinsic coefficients correlating the different colours, and one value for
  $R_V$.  An additional scale parameter for the extinction should be set to
  unity.  The colour accuracy achieved varies from 0.011 to 0.028~mag depending
  on the colour.  All colours discussed in this work are `rest-frame' colours.
  Such corrections could arise from different average dust contents in host
  galaxies and ours.  One of the assumptions of the model (factorisation of the
  extinction, linearity of the correlations of intrinsic colours) could be
  wrong, though plausible, but we are surprised by the large difference found in
  the corrections to the R and I bandpasses $\Dex_R$ and $\Dex_I$ shown in
  Table~\ref{tab:extinctioncoeff} and observed directly in
  Fig.~\ref{fig:diffcol_vs_colx_nocorr}.  The simulation confirms that the
  algorithm we use is indeed sensitive to these small corrections.  The errors
  arising from the linear approximation cannot account for the size of the
  effects found, and additional intrinsic contributions are extremely unlikely
  to induce a correlation with extinction as strong as the one observed in
  Fig.~\ref{fig:diffcol_vs_colx_nocorr}.

\item A single extra-intrinsic contribution beyond \SiII~$\lambda$4131~\AA and
  \CaII~$\lambda$3945~\AA\ is needed to account for the colour correlations
  within the accuracy described immediately above We can then extract intrinsic
  and extinction colour components for each SN.

\item We show that the remaining magnitude fluctuation of 0.13~mag is
  independent of the five $UBVRI$ bandpasses to an accuracy of better than
  0.02~mag, reaching 0.007~mag for the $I$ bandpass.  There is no attempt to
  minimise this magnitude dispersion directly in our procedure.

\item When the extinction formula of \citetalias{Fitzpatrick1999} is used as
  leverage to extract this extra intrinsic (one-dimensional) component, and the
  associated extinction component, the intrinsic coupling coefficients between
  the different colours are consistent over the full sample of SNe~Ia at the
  level of 0.01--0.05.  The values of the three independent coefficients are
  (for $R_V = 2.25$)
  \begin{equation}
    \begin{gathered}
      \gamma_{10} = 1.265 \pm 0.075, \\
      \gamma_{20} = 1.116 \pm 0.061, \\
      \gamma_{30} = 0.539 \pm 0.041.
    \end{gathered}
    \label{eq:6}
  \end{equation}
  The errors quoted include the systematic errors from the colour choice, and
  the statistical error derived from the simulation.  The extra `intrinsic'
  content of $U-I$ is half the size of $U-B$ or $U-V$, with a large uncertainty.

\item After correcting for the analysis bias found in the simulation, the
  optimal {mean} value of the \citetalias{Fitzpatrick1999} parameter $R_V$ over
  our sample is found to be
  \begin{equation}
    R_V  = 2.218  \pm  0.117\;(\text{stat.}) \pm 0.050\;(\text{syst.}).
    \label{eq:7}
  \end{equation}
  This value is obtained within our specific treatment of the extinction scale,
  which is not universally adopted.  The usual `operational' definition
  $R_V = A_V/E(B-V) = R_{V4}$ then leads to $R_{V4} = 2.7$ according to
  Table~\ref{tab:extinctioncoeff}, which includes the effect of our different
  filter definition.

\item The associated scaling factor for the extinction component is measured to
  be $ 1.001\pm 0.044$, and, as desired for consistency, is compatible with
  unity.

\item A previous SNfactory~analysis in \citetalias{Chotard2011} found
  $R_V = 2.8 \pm 0.30$, with a different method and a substantially different
  data set.  A few improvements are brought here, namely the rescaling of the
  extinction coefficients, ensuring consistency with the $B$ and $V$ SN spectra;
  the presence of an `extra-intrinsic' colour component; and we avoid an
  arbitrary extinction parameter for each SN ($A_V$).  The presence of the grey
  fluctuation was ignored by \citetalias{Chotard2011}, and the extraction of the
  extinction colour component was not explicit.  The $R_V$ value that we obtain
  is dependent on the method used, with an emphasis on colour correlations
  rather than a minimisation of Hubble residuals.  The smaller number of
  parameters involved and the accuracy of the colour description is an argument
  in favour of the present method.

\item Within the accuracy of the present observations, we see no hint of a
  variation of the parameter $R_V$ from one SN to another.  As we find
  corrections to the extinction formula to be necessary, for all values of $R_V$
  in the range 1.95 to 2.60, the meaning of this parameter is slightly blurred.
  Simulations have been performed with samples including a Gaussian distribution
  of the value of $R_V$ around the value of 2.20 and with an RMS of~0.5.  The
  results do not show any significant change, and we would thus be insensitive
  to such a fluctuation.

\item Many instances have been given in the past of a correlation between the
  star formation rate \citep[H$\alpha$ signal,][]{Rigault2013} or the host
  galaxy mass \citep{Hamuy2000,Kelly2010} and the absolute magnitudes of SNe, as
  well as their intrinsic properties, suggesting two populations depending on
  their age.  We checked the impact on our analysis of including a sample of 85
  SNe from the SNfactory collaboration common to the \cite{Rigault2013} sample.
  A striking difference is seen in the amount of extinction of the two groups.
  The young SNe (high H$\alpha$ signal of the host galaxy) have significantly
  more dust, with $\moy{X_3} = 0.0124 \pm 0.036$, whilst the older ones (low
  galactic H$\alpha$ signal) have less extinction and
  $\moy{X_3} = -0.0577 \pm 0.025$.  Our corrections slightly reduce the
  discrepancy in absolute magnitude of the two samples from 0.095 to
  0.087~mag.  The statistics of the two groups (47 and 37 SNe) are too small to
  evaluate a difference in the associated $R_V$.
\end{enumerate}
The colour model used here achieves an impressive level of accuracy in the
description of the 'grey' magnitude fluctuation, which may help interpret the
effect.  The accuracy reached on the (smaller) intrinsic colour might be improved
by using additional spectral information, in particular
\SiII~$\lambda$6355~\AA{}, which is better measured.  This will require dealing
with the additional complexity arising from non-linear effects.

\begin{acknowledgements}
  We thank the Nearby Supernova Factory collaboration for providing access to
  its data.  The contribution of N.~Chotard was crucial in the processing of the
  data, the estimate of the magnitudes, and the evaluation of the equivalent
  widths used in this analysis.  We are grateful to the technical and scientific
  staff of the University of Hawaii 2.2~m telescope, to the Palomar Observatory,
  and to the High Performance Research and Educational Network (HPWREN) for
  their assistance in obtaining these data.  We also thank the people of Hawaii
  for the access to Mauna Kea.  We thank D.~Birchall for observing assistance.
  This work was supported by the Director, Office of Science, Office of High
  Energy Physics of the US Department of Energy under contract
  DE-AC02-05CH11231.  This work was supported in France by CNRS/IN2P3,
  CNRS/INSU, and PNC.  Support in Germany was provided by the DFG through TRR33
  ``The Dark Universe'', and in China from Tsinghua University grant 985 and
  NSFC grant 11173017.  Some results were obtained using resources and support
  from the National Energy Research Scientific Computing Center, supported by
  the Director, Office of Science, Office of Advanced Scientific Computing
  Research, of the US Department of Energy under contract DE-AC02-05CH11231.
  HPWREN is funded by National Science Foundation Grant ANI-0087344, and the
  University of California, San Diego.  We thank G.~Aldering and A.~Kim for
  their remarks, which have helped clarify the present text.  We thank
  particularly the referee for his careful reading and numerous useful questions
  and comments.  The authors take responsibility for the remaining inadequacies.
\end{acknowledgements}

%
%

\bibliographystyle{aa}          
\bibliography{aa_extinction}    

\begin{thebibliography}{43}
\expandafter\ifx\csname natexlab\endcsname\relax\def\natexlab#1{#1}\fi

\bibitem[{Aldering {et~al.}(2020)Aldering, Antilogus, Aragon, Bailey, Baltay,
  Bongard, Boone, Buton, Chotard, Copin, Dixon, Fakhouri, Feindt, Fouchez,
  Gangler, Hayden, Hillebrandt, Kim, Kowalski, K{\"u}sters, L{\'e}get, Lin,
  Lombardo, Mondon, Nordin, Pain, Pecontal, Pereira, Perlmutter, Ponder,
  Pruzhinskaya, Rabinowitz, Rigault, Rubin, Runge, Saunders, Says, Smadja,
  Suzuki, Tao, Taubenberger, Thomas, Vincenzi, Weaver, \& {Nearby Supernova
  Factory Collaboration}}]{Aldering2020}
Aldering, G., Antilogus, P., Aragon, C., {et~al.} 2020, Research Notes of the
  American Astronomical Society, 4, 63

\bibitem[{Aldering {et~al.}(2006)Aldering, Antilogus, Bailey, Baltay, Bauer,
  Blanc, Bongard, Copin, Gangler, Gilles, Kessler, Kocevski, Lee, Loken,
  Nugent, Pain, P{\'e}contal, Pereira, Perlmutter, Rabinowitz, Rigaudier,
  Scalzo, Smadja, Thomas, Wang, \& Weaver}]{Aldering2006}
Aldering, G., Antilogus, P., Bailey, S., {et~al.} 2006, ApJ, 650, 510

\bibitem[{{Amanullah} {et~al.}(2014){Amanullah}, {Goobar}, {Johansson},
  {Banerjee}, {Venkataraman}, {Joshi}, {Ashok}, {Cao}, {Kasliwal}, {Kulkarni},
  {Nugent}, {Petrushevska}, \& {Stanishev}}]{Amanullah2014}
{Amanullah}, R., {Goobar}, A., {Johansson}, J., {et~al.} 2014, ApJL, 788, L21

\bibitem[{{Amanullah} {et~al.}(2015){Amanullah}, {Johansson}, {Goobar},
  {Ferretti}, {Papadogiannakis}, {Petrushevska}, {Brown}, {Cao}, {Contreras},
  {Dahle}, {Elias-Rosa}, {Fynbo}, {Gorosabel}, {Guaita}, {Hangard}, {Howell},
  {Hsiao}, {Kankare}, {Kasliwal}, {Leloudas}, {Lundqvist}, {Mattila}, {Nugent},
  {Phillips}, {Sandberg}, {Stanishev}, {Sullivan}, {Taddia}, {{\"O}stlin},
  {Asadi}, {Herrero-Illana}, {Jensen}, {Karhunen}, {Lazarevic}, {Varenius},
  {Santos}, {Sridhar}, {Wallstr{\"o}m}, \& {Wiegert}}]{Amanullah2015}
{Amanullah}, R., {Johansson}, J., {Goobar}, A., {et~al.} 2015, MNRAS, 453, 3300

\bibitem[{{Bessell} \& {Murphy}(2012)}]{Bessellfilters}
{Bessell}, M. \& {Murphy}, S. 2012, PASP, 124, 140

\bibitem[{Betoule {et~al.}(2014)Betoule, Kessler, Guy, Mosher, Hardin, Biswas,
  Astier, {El-Hage}, Konig, Kuhlmann, Marriner, Pain, Regnault, Balland,
  Bassett, Brown, Campbell, Carlberg, {Cellier-Holzem}, Cinabro, Conley,
  D'Andrea, DePoy, Doi, Ellis, Fabbro, Filippenko, Foley, Frieman, Fouchez,
  Galbany, Goobar, Gupta, Hill, Hlozek, Hogan, Hook, Howell, Jha, Le~Guillou,
  Leloudas, Lidman, Marshall, M{\"o}ller, Mour{\~a}o, Neveu, Nichol, Olmstead,
  {Palanque-Delabrouille}, Perlmutter, Prieto, Pritchet, Richmond, Riess,
  {Ruhlmann-Kleider}, Sako, Schahmaneche, Schneider, Smith, Sollerman,
  Sullivan, Walton, \& Wheeler}]{Betoule2014}
Betoule, M., Kessler, R., Guy, J., {et~al.} 2014, A\&A, 568, A22

\bibitem[{Boone {et~al.}(2021)Boone, Aldering, Antilogus, Aragon, Bailey,
  Baltay, Bongard, Buton, Copin, Dixon, Fouchez, Gangler, Gupta, Hayden,
  Hillebrandt, Kim, Kowalski, K{\"u}sters, L{\'e}get, Mondon, Nordin, Pain,
  Pecontal, Pereira, Perlmutter, Ponder, Rabinowitz, Rigault, Rubin, Runge,
  Saunders, Smadja, Suzuki, Tao, Taubenberger, Thomas, \& Vincenzi}]{Boone2021}
Boone, K., Aldering, G., Antilogus, P., {et~al.} 2021, ApJ, 912, 70

\bibitem[{{Borkowski} {et~al.}(2009){Borkowski}, {Blondin}, \&
  {Reynolds}}]{Borkowski2009}
{Borkowski}, K.~J., {Blondin}, J.~M., \& {Reynolds}, S.~P. 2009, ApJL, 699, L64

\bibitem[{{Bronder} {et~al.}(2008){Bronder}, {Hook}, {Astier}, {Balam},
  {Balland}, {Basa}, {Carlberg}, {Conley}, {Fouchez}, {Guy}, {Howell}, {Neill},
  {Pain}, {Perrett}, {Pritchet}, {Regnault}, {Sullivan}, {Baumont}, {Fabbro},
  {Filliol}, {Perlmutter}, \& {Ripoche}}]{Bronder2008}
{Bronder}, T.~J., {Hook}, I.~M., {Astier}, P., {et~al.} 2008, A\&A, 477, 717

\bibitem[{{Brout} \& {Scolnic}(2021)}]{Brout2021}
{Brout}, D. \& {Scolnic}, D. 2021, ApJ, 909, 26

\bibitem[{Burns {et~al.}(2014)Burns, Stritzinger, Phillips, Hsiao, Contreras,
  Persson, Folatelli, Boldt, Campillay, Castell{\'o}n, Freedman, Madore,
  Morrell, Salgado, \& Suntzeff}]{Burns2014}
Burns, C.~R., Stritzinger, M., Phillips, M.~M., {et~al.} 2014, ApJ, 789, 32

\bibitem[{Cardelli {et~al.}(1989)Cardelli, Clayton, \& Mathis}]{CCM1989}
Cardelli, J.~A., Clayton, G.~C., \& Mathis, J.~S. 1989, ApJ, 345, 245

\bibitem[{Chotard(2011)}]{ChotardPhD}
Chotard, N. 2011, {Th\`ese de doctorat}, Universit\'e Claude Bernard - Lyon I,
  {Institut de Physique Nucl\'eaire de Lyon}

\bibitem[{Chotard {et~al.}(2011)Chotard, Gangler, Aldering, Antilogus, Aragon,
  Bailey, Baltay, Bongard, \& Buton}]{Chotard2011}
Chotard, N., Gangler, E., Aldering, G., {et~al.} 2011, A\&A, 529, L4

\bibitem[{{Ferretti} {et~al.}(2017){Ferretti}, {Amanullah}, {Bulla}, {Goobar},
  {Johansson}, \& {Lundqvist}}]{Ferretti2017}
{Ferretti}, R., {Amanullah}, R., {Bulla}, M., {et~al.} 2017, ApJL, 851, L43

\bibitem[{Fitzpatrick(1999)}]{Fitzpatrick1999}
Fitzpatrick, E.~L. 1999, PASP, 111, 63

\bibitem[{{Fitzpatrick} \& {Massa}(2005)}]{FM05}
{Fitzpatrick}, E.~L. \& {Massa}, D. 2005, AJ, 130, 1127

\bibitem[{{Guy} {et~al.}(2007){Guy}, {Astier}, {Baumont}, {Hardin}, {Pain},
  {Regnault}, {Basa}, {Carlberg}, {Conley}, {Fabbro}, {Fouchez}, {Hook},
  {Howell}, {Perrett}, {Pritchet}, {Rich}, {Sullivan}, {Antilogus}, {Aubourg},
  {Bazin}, {Bronder}, {Filiol}, {Palanque-Delabrouille}, {Ripoche}, \&
  {Ruhlmann-Kleider}}]{Guy2007}
{Guy}, J., {Astier}, P., {Baumont}, S., {et~al.} 2007, \aap, 466, 11

\bibitem[{Hamuy {et~al.}(1996)Hamuy, Phillips, Suntzeff, Schommer, Maza, \&
  Aviles}]{Hamuy1996}
Hamuy, M., Phillips, M.~M., Suntzeff, N.~B., {et~al.} 1996, AJ, 112, 2391

\bibitem[{{Hamuy} {et~al.}(2000){Hamuy}, {Trager}, {Pinto}, {Phillips},
  {Schommer}, {Ivanov}, \& {Suntzeff}}]{Hamuy2000}
{Hamuy}, M., {Trager}, S.~C., {Pinto}, P.~A., {et~al.} 2000, AJ, 120, 1479

\bibitem[{Huang {et~al.}(2017)Huang, Raha, Aldering, Antilogus, Bailey, Baltay,
  Barbary, Baugh, Boone, Bongard, Buton, Chen, Chotard, Copin, Fagrelius,
  Fakhouri, Feindt, Fouchez, Gangler, Hayden, Hillebrandt, Kim, Kowalski,
  Leget, Lombardo, Nordin, Pain, Pecontal, Pereira, Perlmutter, Rabinowitz,
  Rigault, Rubin, Runge, Saunders, Smadja, Sofiatti, Stocker, Suzuki,
  Taubenberger, Tao, Thomas, \& {The Nearby Supernova Factory}}]{Huang2017}
Huang, X., Raha, Z., Aldering, G., {et~al.} 2017, ApJ, 836, 157

\bibitem[{Jha {et~al.}(2007)Jha, Riess, \& Kirshner}]{Jha2007}
Jha, S., Riess, A.~G., \& Kirshner, R.~P. 2007, ApJ, 659, 122

\bibitem[{Kelly {et~al.}(2010)Kelly, Hicken, Burke, Mandel, \&
  Kirshner}]{Kelly2010}
Kelly, P.~L., Hicken, M., Burke, D.~L., Mandel, K.~S., \& Kirshner, R.~P. 2010,
  ApJ, 715, 743

\bibitem[{Lantz {et~al.}(2004)Lantz, Aldering, Antilogus, Bonnaud, Capoani,
  Castera, Copin, Dubet, Gangler, Henault, Lemonnier, Pain, Pecontal, Pecontal,
  \& Smadja}]{Lantz2004}
Lantz, B., Aldering, G., Antilogus, P., {et~al.} 2004, in Proceedings of the
  {{SPIE}}, Vol. 5249, Optical {{Design}} and {{Engineering}}, 146--155

\bibitem[{L{\'e}get {et~al.}(2020)L{\'e}get, Gangler, Mondon, Aldering,
  Antilogus, Aragon, Bailey, Baltay, Barbary, Bongard, Boone, Buton, Chotard,
  Copin, Dixon, Fagrelius, Feindt, Fouchez, Hayden, Hillebrandt, Kim, Kowalski,
  Kuesters, Lombardo, Lin, Nordin, Pain, Pecontal, Pereira, Perlmutter, Ponder,
  Pruzhinskaya, Rabinowitz, Rigault, Runge, Rubin, Saunders, Says, Smadja,
  Sofiatti, Suzuki, Taubenberger, Tao, \& Thomas}]{Leget2020}
L{\'e}get, P.-F., Gangler, E., Mondon, F., {et~al.} 2020, A\&A, 636, A46

\bibitem[{{Lira} {et~al.}(1998){Lira}, {Suntzeff}, {Phillips}, {Hamuy}, {Maza},
  {Schommer}, {Smith}, {Wells}, {Avil{\'e}s}, {Baldwin}, {Elias},
  {Gonz{\'a}lez}, {Layden}, {Navarrete}, {Ugarte}, {Walker}, {Williger},
  {Baganoff}, {Crotts}, {Rich}, {Tyson}, {Dey}, {Guhathakurta}, {Hibbard},
  {Kim}, {Rehner}, {Siciliano}, {Roth}, {Seitzer}, \& {Williams}}]{Lira1998}
{Lira}, P., {Suntzeff}, N.~B., {Phillips}, M.~M., {et~al.} 1998, AJ, 115, 234

\bibitem[{{Mandel} {et~al.}(2017){Mandel}, {Scolnic}, {Shariff}, {Foley}, \&
  {Kirshner}}]{Mandel2017}
{Mandel}, K.~S., {Scolnic}, D.~M., {Shariff}, H., {Foley}, R.~J., \&
  {Kirshner}, R.~P. 2017, ApJ, 842, 93

\bibitem[{{Nagao} {et~al.}(2017){Nagao}, {Maeda}, \& {Yamanaka}}]{Nagao2017}
{Nagao}, T., {Maeda}, K., \& {Yamanaka}, M. 2017, ApJ, 835, 143

\bibitem[{{Nobili} \& {Goobar}(2008)}]{Nobili2008}
{Nobili}, S. \& {Goobar}, A. 2008, A\&A, 487, 19

\bibitem[{Nugent {et~al.}(1995)Nugent, Phillips, Baron, Branch, \&
  Hauschildt}]{Nugent1995}
Nugent, P., Phillips, M., Baron, E., Branch, D., \& Hauschildt, P. 1995, ApJL,
  455, L147

\bibitem[{O'Donnell(1994)}]{ODonnell1994}
O'Donnell, J.~E. 1994, ApJ, 422, 158

\bibitem[{Perlmutter {et~al.}(1999)Perlmutter, Aldering, Goldhaber, Knop,
  Nugent, Castro, Deustua, Fabbro, Goobar, Groom, Hook, Kim, Kim, Lee, Nunes,
  Pain, Pennypacker, Quimby, Lidman, Ellis, Irwin, McMahon, {Ruiz-Lapuente},
  Walton, Schaefer, Boyle, Filippenko, Matheson, Fruchter, Panagia, Newberg,
  Couch, \& {Supernova Cosmology Project}}]{Perlmutter1999}
Perlmutter, S., Aldering, G., Goldhaber, G., {et~al.} 1999, ApJ, 517, 565

\bibitem[{Phillips {et~al.}(1999)Phillips, Lira, Suntzeff, Schommer, Hamuy, \&
  Maza}]{Phillips1999}
Phillips, M.~M., Lira, P., Suntzeff, N.~B., {et~al.} 1999, AJ, 118, 1766

\bibitem[{{Rieke} \& {Lebofsky}(1985)}]{Rieke1985}
{Rieke}, G.~H. \& {Lebofsky}, M.~J. 1985, ApJ, 288, 618

\bibitem[{Riess {et~al.}(1998)Riess, Filippenko, Challis, Clocchiatti, Diercks,
  Garnavich, Gilliland, Hogan, Jha, Kirshner, Leibundgut, Phillips, Reiss,
  Schmidt, Schommer, Smith, Spyromilio, Stubbs, Suntzeff, \& Tonry}]{Riess1998}
Riess, A.~G., Filippenko, A.~V., Challis, P., {et~al.} 1998, AJ, 116, 1009

\bibitem[{Rigault {et~al.}(2013)Rigault, Copin, Aldering, Antilogus, Aragon,
  Bailey, Baltay, Bongard, Buton, Canto, {Cellier-Holzem}, Childress, Chotard,
  Fakhouri, Feindt, Fleury, Gangler, Greskovic, Guy, Kim, Kowalski, Lombardo,
  Nordin, Nugent, Pain, P{\'e}contal, Pereira, Perlmutter, Rabinowitz, Runge,
  Saunders, Scalzo, Smadja, Tao, Thomas, \& Weaver}]{Rigault2013}
Rigault, M., Copin, Y., Aldering, G., {et~al.} 2013, A\&A, 560, 66

\bibitem[{Saunders {et~al.}(2018)Saunders, Aldering, Antilogus, Bailey, Baltay,
  Barbary, Baugh, Boone, Bongard, Buton, Chen, Chotard, Copin, Dixon,
  Fagrelius, Fakhouri, Feindt, Fouchez, Gangler, Hayden, Hillebrandt, Kim,
  Kowalski, K{\"u}sters, Leget, Lombardo, Nordin, Pain, Pecontal, Pereira,
  Perlmutter, Rabinowitz, Rigault, Rubin, Runge, Smadja, Sofiatti, Suzuki, Tao,
  Taubenberger, Thomas, Vincenzi, \& Nearby Supernova~Factory}]{Saunders2018}
Saunders, C., Aldering, G., Antilogus, P., {et~al.} 2018, ApJ, 869, 167

\bibitem[{Scalzo {et~al.}(2010)Scalzo, Aldering, Antilogus, Aragon, Bailey,
  Baltay, Bongard, Buton, Childress, Chotard, Copin, Fakhouri, {Gal-Yam},
  Gangler, Hoyer, Kasliwal, Loken, Nugent, Pain, P{\'e}contal, Pereira,
  Perlmutter, Rabinowitz, Rau, Rigaudier, Runge, Smadja, Tao, Thomas, Weaver,
  \& Wu}]{Scalzo2010}
Scalzo, R.~A., Aldering, G., Antilogus, P., {et~al.} 2010, ApJ, 713, 1073

\bibitem[{{Schlafly} {et~al.}(2016){Schlafly}, {Meisner}, {Stutz},
  {Kainulainen}, {Peek}, {Tchernyshyov}, {Rix}, {Finkbeiner}, {Covey}, {Green},
  {Bell}, {Burgett}, {Chambers}, {Draper}, {Flewelling}, {Hodapp}, {Kaiser},
  {Magnier}, {Martin}, {Metcalfe}, {Wainscoat}, \& {Waters}}]{Schlafly2016}
{Schlafly}, E.~F., {Meisner}, A.~M., {Stutz}, A.~M., {et~al.} 2016, ApJ, 821,
  78

\bibitem[{{Schlafly} {et~al.}(2017){Schlafly}, {Peek}, {Finkbeiner}, \&
  {Green}}]{Schlafly2017}
{Schlafly}, E.~F., {Peek}, J.~E.~G., {Finkbeiner}, D.~P., \& {Green}, G.~M.
  2017, ApJ, 838, 36

\bibitem[{Thorp {et~al.}(2021)Thorp, Mandel, Jones, Ward, \&
  Narayan}]{Thorp2021}
Thorp, S., Mandel, K.~S., Jones, D.~O., Ward, S.~M., \& Narayan, G. 2021,
  MNRAS, 508, 4310

\bibitem[{Wang {et~al.}(2003)Wang, Goldhaber, Aldering, \&
  Perlmutter}]{Wang2003}
Wang, L., Goldhaber, G., Aldering, G., \& Perlmutter, S. 2003, ApJ, 590, 944

\bibitem[{{Wojtak} {et~al.}(2023){Wojtak}, {Hjorth}, \& {Osman
  Hjortlund}}]{Wojtak2023}
{Wojtak}, R., {Hjorth}, J., \& {Osman Hjortlund}, J. 2023, arXiv e-prints,
  arXiv:2302.01906

\end{thebibliography}

\appendix

\section{Linearisation of the extinction formula}
\label{sec:Annex}

The observed photon-count distribution was given in Eq.~\ref{eq:extinctionlaw}
where the function $\phi(\lambda)$ is provided by \cite{CCM1989, ODonnell1994,
  Fitzpatrick1999, FM05}.  To simplify the computations, we assume the
extinction $E(B-V)$ is small enough to allow a linear approximation under the
exponentiation.
\begin{equation}
  10^{-0.4 \phi(\lambda, R_{V})\,E(B-V)}
  \approx 1 - 0.4\ln 10\,\phi(\lambda, R_{V})\,E(B-V).
\end{equation}
We define $A(\lambda, R_{V}) = \phi(\lambda, R_{V})\,E(B-V)$, the observed
spectrum (photon counts) is then
\begin{equation}
  \label{eq:extinctionlambda}
  m(\lambda) = -2.5\log_{10} s(\lambda) = -2.5\log_{10} s_0(\lambda) + A(\lambda, R_{V}).
\end{equation}
When the unextincted SN spectrum $s_0(\lambda)$ is integrated over a spectral
band, such as $B$, with transmission $T(\lambda)$, the rest frame flux is:
\begin{equation}
  \label{eq:extinctionspectrum}
  s_B = \int_B s_0(\lambda) T(\lambda)\,10^{-0.4\,\phi(\lambda)} \d\lambda.
\end{equation}
In our synthetic photometry, unit-transmission top-hat filters are used, and so
we omit $T(\lambda)$.  Using the previous linear approximation:
\begin{align}
  \label{eq:extinctionfiltermag}
  m_{B}^{\text{obs}} &= m_{B}^{\text{true}}  + A_B \nonumber\\
  &= -2.5 \log_{10} \left(
    \int_B s_0(\lambda) \left[1 - 0.4 \ln 10\,A(\lambda, R_{V}) \right]
    \d\lambda \right). 
\end{align}
The logarithm can be expanded to give
\begin{equation}
  \label{eq:extinctionfilterAnnex}
  A_B = m_{B}^{\text{obs}} -m_{B}^{\text{true}} = \frac{%
    \int_{B} s_{0}(\lambda) A(\lambda, R_{V})\,\d\lambda}{%
    \int_{B} s_{0}(\lambda)\,\d\lambda}.
\end{equation}
In this approximation, the extinction is simply weighted by the spectrum over
the filter bandpass.  The maximal value of the extinction component $X_4$ is
0.4, and the error
$\delta R_{F4} = R_{F4}^{\text{linear}} - R_{F4}^{\text{exact}}$ arising from
the linear approximation in the evaluation of the extinctions coefficient
$R_{F4}$ in the different (top-hat) filters is given in
Table~\ref{tab:linearcoeff}.  The magnitude correction is obtained by
multiplying the coefficients $R_{F4}$ by $X_4$, and so the error on this
correction never exceeds 0.005~mag, and cannot account for the effects observed
in Fig.~\ref{fig:diffcol_vs_colx_nocorr}.

\begin{table*}
  \caption{Error on the bandpass extinction coefficients arising from the linear
    approximation for $X_4 = 0.4$}.
  \label{tab:linearcoeff}
  \begin{tabular}{l c c c c c}
    \hline\hline
    Filter & $U$ & $B$ &  $V$ & $R$ & $I$ \\
    $\delta R'_{F4}$  & $-0.00976$ & $-0.00870$ & $-0.01147$ & $-0.00494$ & $-0.00195$ \\
    \hline
  \end{tabular}
\end{table*}
\end{document}